%% file: get_kernel.tex
\documentclass[english, journal]{IEEEtran}

\usepackage{enumitem}
\usepackage{amsthm}
\usepackage{hyperref}
\usepackage[T1]{fontenc}
\usepackage[utf8]{inputenc}
\usepackage{verbatim}
\usepackage{float}
\usepackage[cmex10]{amsmath}
\usepackage{amssymb}
\usepackage{graphicx}
\usepackage{epstopdf}
\usepackage{pgf,tikz,psfrag}
\usepackage{stmaryrd}
\usepackage{mathrsfs}
\usepackage{cite}
\usepackage{color}
\makeatletter

\usepackage{subfigure}
\makeatother

\newcommand{\cip}{\overset{\mathcal{P}}{\longrightarrow}}
\providecommand{\eref}[1]{\eqref{eq:#1}}  
\providecommand{\sref}[1]{Section~\ref{sec:#1}}
\providecommand{\fref}[1]{Figure~\ref{fig:#1}}

\DeclareMathOperator{\polylog}{polylog}
\DeclareMathOperator{\sign}{sign}

\newcommand*{\tran}{^{\mkern-1.5mu\mathsf{T}}}

\usepackage{dsfont}
\newcommand{\charfn}{\mathds{1}}

\usepackage{dsfont}
\newtheorem{thm}{Theorem}
\newtheorem{prop}{Proposition}
\newtheorem{lem}{Lemma}

\newtheorem{rem}{Remark}

\newtheorem{example}{Example}

\include{command}
\title{Universality Laws for High-Dimensional Learning with Random Features}

\author{
\IEEEauthorblockN{Hong Hu and Yue M. Lu,~\IEEEmembership{Senior Member,~IEEE}}
\thanks{
This work was supported by the Harvard FAS Dean's Fund for Promising Scholarship, and by the US National Science Foundation under grants CCF-1718698 and CCF-1910410. 

H. Hu was with the John A. Paulson School of Engineering and Applied Sciences, Harvard University, Cambridge, MA 02138, USA. He is now with Department of Statistics and Data Science, University of Pennsylvania, Philadelphia, PA 19104, USA (e-mail: huhong@wharton.upenn.edu).

Y. M. Lu is with the John A. Paulson School of Engineering and Applied Sciences, Harvard University, Cambridge, MA 02138, USA. (e-mail: yuelu@seas.harvard.edu). 
}
}
\date{}

\begin{document}


\maketitle

\input{abstract}


\input{introduction}

\input{proof_sketch}

\input{clt_new}

\input{conclusion}

\appendix

\input{appendix_truncation}
\input{appendix_covariance}
\input{appendix-concentration}

\input{appendix-boundedness}
\input{appendix-quadapprox}

\bibliographystyle{IEEEtran}
\bibliography{refs}

\end{document}

%% file: command.tex
\global\long\def\t#1{\widetilde{#1}}%
\global\long\def\abs#1{\left\lvert #1\right\rvert }%
\global\long\def\absb#1{\big\lvert #1\big\rvert }%
\global\long\def\norm#1{\lVert#1\rVert}%

\global\long\def\set#1{\left\{  #1\right\}  }%
\global\long\def\bydef{\overset{\text{def}}{=}}%
\global\long\def\tleq#1{\overset{#1}{\leq}}%

\global\long\def\EE{\mathbb{E}\,}%

\global\long\def\R{\mathbb{R}}%
\global\long\def\E{\mathbb{E}}%
\global\long\def\P{\mathbb{P}}%

\global\long\def\va{\boldsymbol{a}}%
\global\long\def\vb{\boldsymbol{b}}%
\global\long\def\ve{\boldsymbol{e}}%
\global\long\def\vf{\boldsymbol{f}}%
\global\long\def\vg{\boldsymbol{g}}%

\global\long\def\vh{\boldsymbol{h}}%
\global\long\def\vr{\boldsymbol{r}}%

\global\long\def\vu{\boldsymbol{u}}%
\global\long\def\vw{\boldsymbol{w}}%
\global\long\def\vx{\boldsymbol{x}}%
\global\long\def\vz{\boldsymbol{z}}%

\global\long\def\vtheta{\boldsymbol{\theta}}%
\global\long\def\vbeta{\boldsymbol{\beta}}%
\global\long\def\vone{\boldsymbol{1}}%

\global\long\def\mA{\boldsymbol{A}}%
\global\long\def\mB{\boldsymbol{B}}%
\global\long\def\mD{\boldsymbol{D}}%
\global\long\def\mF{\boldsymbol{F}}%
\global\long\def\mG{\boldsymbol{G}}%

\global\long\def\mH{\boldsymbol{H}}%
\global\long\def\mI{\boldsymbol{I}}%
\global\long\def\mM{\boldsymbol{M}}%
\global\long\def\mP{\boldsymbol{P}}%
\global\long\def\mR{\boldsymbol{R}}%
\global\long\def\mSig{\boldsymbol{\Sigma}}%

\global\long\def\mzero{\boldsymbol{0}}%

\global\long\def\calS{\mathcal{S}}%
\global\long\def\calN{\mathcal{N}}%
\global\long\def\calL{\mathcal{L}}%
\global\long\def\calM{\mathcal{M}}%
\global\long\def\calB{\mathcal{B}}%

\global\long\def\w{\omega}%
\global\long\def\veps{\varepsilon}%

\global\long\def\tw{\t w}%

\global\long\def\asconv{\overset{a.s.}{\longrightarrow}}%
\global\long\def\iid{\overset{i.i.d.}{\sim}}%

\global\long\def\sgl{\boldsymbol{\xi}}%
\global\long\def\bs{\backslash}%
\global\long\def\diag{\text{diag}}%

\global\long\def\argmin#1{\underset{#1}{\text{\ensuremath{\arg\min} }}}%

\DeclareMathOperator{\prox}{Prox}%
\global\long\def\cmtx{\mG}%
\global\long\def\cvec{\vg}%
\global\long\def\fmtx{\mF}%
\global\long\def\fvec{\vf}%

\global\long\def\loss{\ell}%
\global\long\def\regu{h}%
\global\long\def\test{\varphi}%
\global\long\def\kernel{\sigma}%

\global\long\def\wt{\vw^{*}}%
\global\long\def\loowt#1{\vw_{#1}^{*}}%
\global\long\def\qwt{\widetilde{\vw}}%

\global\long\def\ie{\emph{i.e.}}
\global\long\def\eg{\emph{e.g.}}

\global\long\def\terr{\mathcal{E}_\text{train}}
\global\long\def\gerr{\mathcal{E}_\text{gen}}
\global\long\def\terrA{\mathcal{E}_\text{train}(\mA)}
\global\long\def\terrB{\mathcal{E}_\text{train}(\mB)}
\global\long\def\gerrA{\mathcal{E}_\text{gen}(\mA)}
\global\long\def\gerrB{\mathcal{E}_\text{gen}(\mB)}
\global\long\def\gerrLim{\mathcal{E}_\text{gen}^\ast}

\global\long\def\scp{\tfrac{1}{\sqrt{p}}}
\global\long\def\mol{\zeta}
\global\long\def\btest{\varphi}
\global\long\def\bts{\psi}
\global\long\def\strunc{\Omega}
\global\long\def\wv{\beta}
\global\long\def\wvv{\vbeta}

\global\long\def\BP{\kappa_p}
\global\long\def\BPs{\kappa^2_p}
\global\long\def\BPt{\kappa^3_p}
\global\long\def\BPf{\kappa^4_p}

\global\long\def\ajmi{\tilde{a}_{j, \bs i}}

\global\long\def\linf#1{\left\Vert #1\right\Vert _{\infty}}%

\global\long\def\Tr{\text{Tr}}%

\global\long\def\fteacher{\theta_\text{teach}}
\global\long\def\fout{\theta_\text{out}}
\global\long\def\genvar{\rho}
\global\long\def\gencor{\pi}

\global\long\def\epn{m}%
\global\long\def\didx{t}%
\global\long\def\genfn{h}%

\global\long\def\Ecnd#1{\E_{\bs#1}}%
\global\long\def\Pcnd#1{\P_{\bs#1}}%
\global\long\def\Eone#1{\E_{#1}}%

\global\long\def\lsbd{L_S}%

%% file: abstract.tex

\begin{abstract}
We prove a universality theorem for learning with random features. Our result shows that, in terms of training and generalization errors, a random feature model with a nonlinear activation function is asymptotically equivalent to a surrogate linear Gaussian model with a matching covariance matrix. This settles a so-called Gaussian equivalence conjecture based on which several recent papers develop their results. Our method for proving the universality theorem builds on the classical Lindeberg approach. Major ingredients of the proof include a leave-one-out analysis for the optimization problem associated with the training process and a central limit theorem, obtained via Stein's method, for weakly correlated random variables.
\end{abstract} 

%% file: introduction.tex
\section{Introduction}

\subsection{Background and Motivation}
Consider a supervised learning problem with a collection of training samples $\set{\cvec_\didx, y_\didx}_{1 \le \didx \le n}$. We seek to learn a relationship between the input $\cvec_\didx \in \R^d$ and the output $y_\didx \in \R$ by fitting the training data on a parametric family of functions in the form of
\[
\big\{Y_{\vw}(\cvec) = \scp \vw\tran \mathcal{T}(\cvec): \vw \in \R^p\big\},
\]
where $\mathcal{T}: \R^d \mapsto \R^p$ is a (possibly nonlinear and stochastic) feature map. Each such function $Y_{\vw}(\cvec)$ is indexed by a weight vector $\vw$, and we choose the optimal $\vw$ by solving an optimization problem
\begin{equation}\label{eq:training}
\wt_{\mR} = \argmin{\vw \in \R^p} \textstyle\sum_{\didx = 1}^{n} \loss(\tfrac{1}{\sqrt{p}}\vr_{\didx}\tran \vw; y_\didx) + \sum_{j=1}^{p} \regu(w_j).
\end{equation}
Here, $\loss(x; y)$ is a loss function, $\regu(x)$ is a regularizer, and $\mR = [\vr_1, \vr_2, \ldots, \vr_n]\tran \in \R^{n \times p}$ denotes the matrix whose rows are the regressors used in \eref{training}, \ie,
\begin{equation}\label{eq:regressor}
\vr_\didx = \mathcal{T}(\cvec_\didx), \qquad 1 \le \didx \le n.
\end{equation}
Examples of the loss function include the squared loss [$\loss(x, y) = \tfrac{1}{2}(x-y)^2$] and the logistic loss [$\loss(x, y) = \log(1+e^{-yx})$]. The latter is often used in binary classification tasks, where the labels $y_\didx \in \set{\pm 1}$.

The supervised learning process described above has two main performance metrics: the \emph{training error}
\begin{equation}\label{eq:training_error}
\terr(\mR) = \frac{1}{p} \big\{\textstyle\sum_{\didx=1}^{n} \loss(\scp \vr_\didx\tran \wt_{\mR}; y_\didx) + \sum_{j=1}^{p}\regu(w^\ast_{\mR,j})\big\},
\end{equation}
which is simply a scaled version of the optimal value of \eref{training}, and the \emph{generalization error}
\begin{equation}\label{eq:gen_err}
\gerr(\mR) = \EE \big(y_\text{new} - {\fout}[\scp(\wt_{\mR})\tran \mathcal{T}(\cvec_\text{new})]\big)^2,
\end{equation}
where $\fout(\cdot)$ is some post-processing function (\eg, the sign function) and the expectation in \eref{gen_err} is taken over a fresh pair of samples $\set{\cvec_\text{new}, y_\text{new}}$ that are independent of the training data. To carry out theoretical analysis of the training and generalization errors, it is necessary to make some further assumptions on how the training samples $\set{\cvec_\didx, y_\didx}$ are generated. A classical model, which is also the one adopted in this work, is the so-called teacher-student framework. Specifically, we assume that $\cvec_\didx \overset{\text{i.i.d.}}{\sim} \mathcal{N}(0, \mI_d)$ and
\begin{equation}\label{eq:teacher}
y_\didx = \fteacher(\cvec_\didx\tran \sgl),
\end{equation}
where $\sgl \in \R^d$ is a fixed and unknown \emph{teacher vector}, and $\fteacher(\cdot)$ is an unknown function.

In this paper, we study a particular case of the above setting, known in the literature as the \emph{random feature model} \cite{rahimi2008random}. It corresponds to specializing the general regressors in \eref{regressor} to
\begin{equation}\label{eq:a_vec}
\vr_\didx = \va_\didx \bydef \kernel(\fmtx\tran \cvec_\didx),
\end{equation}
where $\fmtx \in \R^{d \times p}$ is a random feature matrix, and $\kernel: \R \mapsto \R$ is a nonlinear scalar activation function [\eg, $\kernel(x) = \tanh(x)$] applied to individual elements of $\mF\tran \cvec_\didx$. Alternatively, the model in \eref{a_vec} can be viewed as a two-layer neural network, with $\cvec_\didx$ being the input to the network, $\fmtx$ the weight matrix in the first layer, and $\kernel(x)$ the activation function. The optimization in \eref{training} (with $\set{\vr_\didx}$ replaced by $\set{\va_\didx}$) then corresponds to learning $\vw$, the second-layer weights of the network, with the first layer weights $\fmtx$ kept fixed.

The random feature model has received considerable attention in the last few years mainly due to its impressive performance and its connection to overparameterized neural networks \cite{rahimi2008random,daniely2016toward,daniely2017sgd,bach2017equivalence,jacot2018neural,belkin2018understand,liu2020random}.
Some of that attention has been directed towards analyzing the performance of this model in high-dimensional regimes. Developments along this line can be found in \emph{e.g.}, \cite{louart2018random,hastie2019surprises,mei2019generalization,montanari2019generalization,goldt2019modelling,gerace2020generalisation,Goldt2020Gaussian, ghorbani2019linearized,ba2019generalization,dhifallah2020precise}. In \cite{louart2018random,mei2019generalization}, the authors precisely characterized the training and generalization errors associated with a special case of \eref{training}, where the loss function $\loss(x; y) = \tfrac{1}{2}(x-y)^2$ and the regularization function is $\regu(x) = \tfrac{\lambda}{2} x^2$. This setting, known as ridge regression, has a closed-form solution. By studying a corresponding (kernel) random matrix, one can show that $\terr$ and $\gerr$ converge to well-defined deterministic limits as the number of training samples $n$ and the problem dimensions $d, p$ grow to infinity at fixed ratios. However, it is difficult to extend such analysis to more general (non-quadratic) loss and regularization functions for which no closed-form solution exists. In particular, the presence of the nonlinear activation function $\kernel(x)$ in \eref{a_vec} makes the regressors $\set{\va_t}$ in \eref{a_vec} non-Gaussian. This then prevents the direct application of analysis tools such as Gaussian min-max theorems (GMT) \cite{gordon1985some,thrampoulidis2018precise}, Gaussian width \cite{chandrasekaran2012convex}, or statistical dimensions \cite{amelunxen2014living}, as they have all been built for analyzing problems involving Gaussian vectors.


\subsection{The Gaussian Equivalence Conjecture}
Fortunately, it has been observed by many authors (see, \eg, \cite{hastie2019surprises,mei2019generalization,montanari2019generalization,goldt2019modelling, gerace2020generalisation,Goldt2020Gaussian,dhifallah2020precise,seddik2020random,Dhifallah2021inherent,Loureiro2021capturing}, and also \cite{cheng2013spectrum,pennington2017nonlinear,louart2018random} in the context of random kernel matrices) that the random feature model considered above should be asymptotically equivalent to a Gaussian model, where we set the regressors in \eref{regressor} to
\begin{equation}\label{eq:b_vec}
\vr_\didx = \vb_\didx \bydef \mu_0 \vone + \mu_1 \fmtx\tran \cvec_\didx + \mu_2 \vz_\didx.
\end{equation}
Here, $\vone$ denotes an all-one vector in $\R^p$, $\vz_\didx \overset{\text{i.i.d.}}{\sim} \mathcal{N}(0, \mI_p)$ is independent of $\cvec_\didx$, and $\mu_0, \mu_1, \mu_2$ are three constants defined as follows. Let $z$ be a standard Gaussian random variable, then
\begin{equation}\label{eq:mu_constants}
\begin{aligned}
   \mu_0 &= \EE[\kernel(z)], \quad \mu_1 = \EE[z\kernel(z)]\quad \text{and} \\ \quad \mu_2 &= (\EE[\kernel^2(z)] - \mu_0^2 - \mu_1^2)^{1/2}. 
\end{aligned}
\end{equation}
In what follows, we shall refer to the setting where the regressors are $\set{\va_\didx}$ in \eref{a_vec} as the \emph{nonlinear feature model}, and refer to the one using $\set{\vb_\didx}$ in \eref{b_vec} as the \emph{linear Gaussian model}. Let
\begin{equation}\label{eq:mat_AB}
\mA = [\va_1, \va_2, \ldots, \va_n]\tran \quad \text{and} \quad \mB = [\vb_1, \vb_2, \ldots, \vb_n]\tran.
\end{equation}
The optimal weight vectors, the training and the generalization errors of these two formulations can then be written as $\wt_{\mA}, \wt_{\mB}$, $\terrA, \terrB$, and $\gerrA, \gerrB$, respectively.

Roughly speaking, the \emph{Gaussian equivalence conjecture} states that, under certain conditions on the feature matrix $\fmtx$, we have
\begin{equation}\label{eq:asymp_equiv}
\terrA \approx \terrB \quad \text{and} \quad \gerrA \approx \gerrB \quad \text{as } p \to \infty.
\end{equation}

\begin{example}
We illustrate this conjecture with two numerical examples. \fref{quadratic} shows the training and generalization errors of a regression problem, where $\fteacher(x) = \fout(x) = x$ and $\kernel(x) = \max(x, 0)$ is the ReLU function. The feature matrix $\fmtx$ in \eref{a_vec} is chosen to be a random matrix with i.i.d. normal entries drawn from $\mathcal{N}(0,1/d)$. To find the optimal weight vector in \eref{training}, we use a quadratic loss $\loss(x; y) = \tfrac{1}{2}(x-y)^2$ and a ridge regularizer $\regu(x) = \frac{\lambda}{2}x^2$. We can see from the simulation results that, even at a moderate problem size ($d=200$ and $n = 600$), the training and generalization errors under the nonlinear feature model and the corresponding linear Gaussian model are already very close. Moreover, they match the analytical predictions developed for the Gaussian model \cite{dhifallah2020precise}. The same phenomenon can also been observed in \fref{logistic}, where we consider a binary classification problem with $\fteacher(x) = \fout(x) = \sign(x)$ and $\kernel(x) = \tanh(x)$. The loss function here is the logistic loss $\loss(x, y) = \log(1+e^{-yx})$, and the regularizer is $\regu(x) = \frac{\lambda}{2}x^2$.
\end{example}

\begin{figure}[t]
\centering
\subfigure[linear regression]{\label{fig:quadratic}
\includegraphics[scale=0.45]{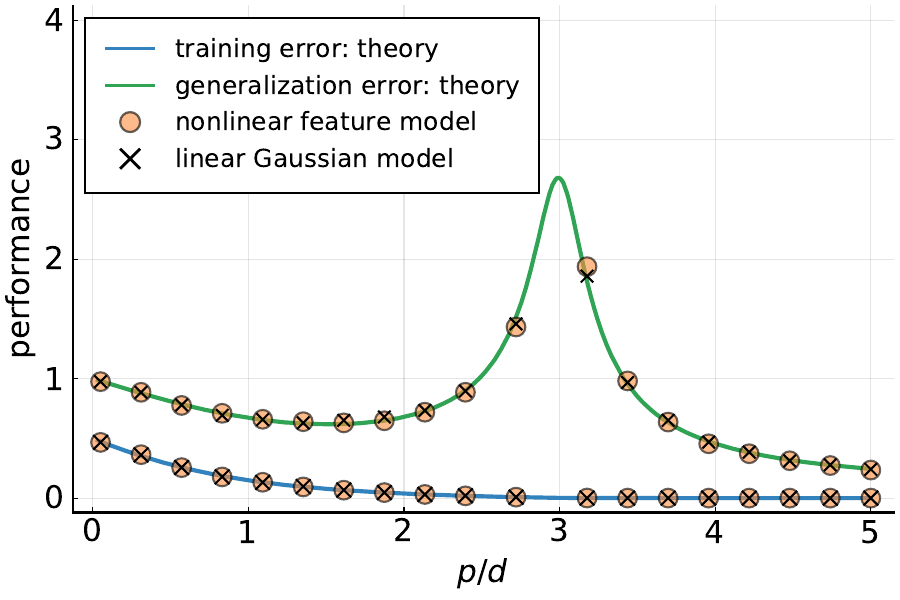}}
\hspace{6ex}
\subfigure[binary classification]{\label{fig:logistic}
\includegraphics[scale=0.45]{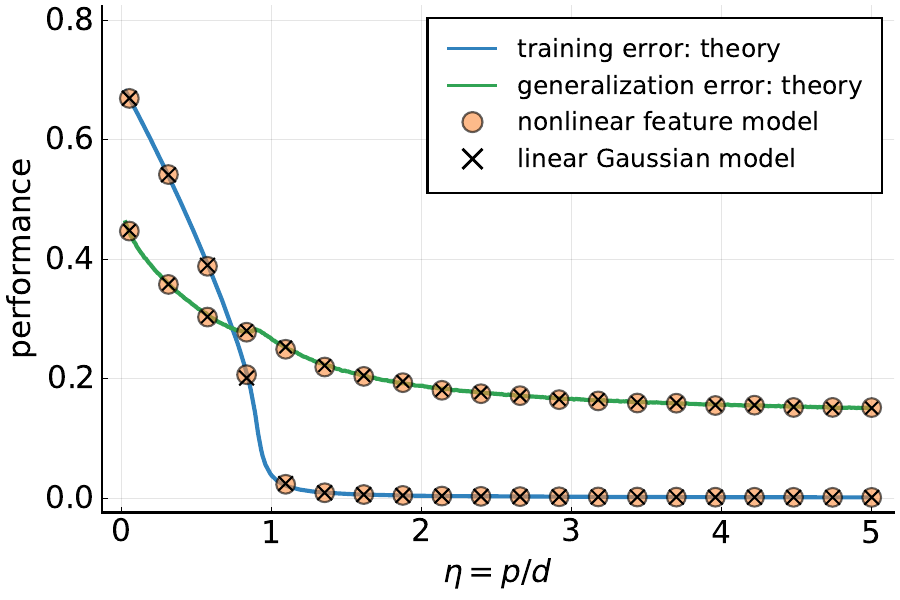}}
\caption{Numerical simulations to demonstrate the asymptotic Gaussian equivalence stated in \eref{asymp_equiv}. (a) A regression problem, where the activation function $\kernel(x) = \max(x, 0)$, the loss function $\ell(x; y)$ is the quadratic function, and $\fteacher(x) = \fout(x) = x$. (b) A binary classification problem, where $\kernel(x) = \tanh(x)$, $\ell(x; y)$ is the logistic loss, and $\fteacher(x) = \fout(x) = \sign(x)$. In both cases, we set $d=200$ and $n=600$, and vary the values of $p$. The simulation results are averaged over 100 independent trials, and the theoretical curves are the analytical predictions \cite{dhifallah2020precise} developed for the Gaussian model.}\label{fig:universality_sim}
\end{figure}

That the nonlinear feature model and the linear Gaussian model can be asymptotically equivalent has a simple intuitive explanation. Under certain conditions on the random feature matrix $\fmtx$, one can show that the random vectors $\va_\didx$ in \eref{a_vec} and $\vb_\didx$ in \eref{b_vec} have asymptotically matching first and second moments. (See Appendix~\ref{appendix:covariance} for details.) Thus, the asymptotic equivalence in \eref{asymp_equiv} points to the emergence of a \emph{universality phenomenon} that is inherent in many large random systems: The macroscopic behaviors of such systems only depend on a few key parameters (the first two moments of $\va_\didx$ and $\vb_\didx$ in our case), whereas the microscopic structures of the systems (\emph{i.e.}, the exact probability distributions of $\va_\didx$ and $\vb_\didx$) are irrelevant.

Notice that the surrogate Gaussian formulation is much more amenable to theoretical analysis, as it only involves Gaussian vectors $\set{\vb_\didx}$. Indeed, based on the Gaussian equivalence conjecture, the authors of \cite{montanari2019generalization} provided a precise asymptotic characterization of maximum-margin linear classifiers in the overparameterized regime using Gaussian min-max theorems \cite{gordon1985some,thrampoulidis2018precise}. The performance of the linear Gaussian model under more general settings, where one uses generic convex loss functions and ridge regularization in \eref{training}, was studied in \cite{gerace2020generalisation} by using the non-rigorous replica method \cite{mezard1987spin} from statistical physics. More recently, these replica predictions have been rigorously proved in \cite{dhifallah2020precise}.

\subsection{Main Contributions}

The main contribution of this paper is to prove the aforementioned Gaussian equivalence conjecture. Our results are based on the following technical assumptions.

\newcounter{assumptions}
\begin{enumerate}[label={(A.\arabic*)}]
\item \label{a:Gaussian} The latent input vectors $\cvec_\didx \overset{\text{i.i.d.}}{\sim} \mathcal{N}(0, \mI_d)$ in \eref{a_vec} and \eref{b_vec}.

\item \label{a:dimensions} The dimension of the latent input vectors $\cvec_\didx$ (denoted by $d$), the dimension of the regression vectors (denoted by $p$), and the number of training samples (denoted by $n$) tend to infinity at fixed ratios. Specifically, $n / d \to \alpha > 0$ and $p / d \to \eta > 0$ as $d \to \infty$.

\item \label{a:sgl} The unknown teacher vector $\sgl$ in \eref{teacher} is deterministic, with $\norm{\sgl} = 1$.

\item \label{a:loss} The loss function $\loss(x; y)\ge 0$ for all $x, y$, and it is convex with respect to its first variable $x$. The third partial derivative of $\loss(x; y)$ with respect to $x$ exists. Moreover, there exist constants $C>0$ and $K_1 \in\mathbb{Z}^{+}$ such that
\[
\abs{\loss'''(x; \fteacher[s])} \le C(1+|s|^{K_1}), \quad \text{for all }x \in \R,
\]
and
\begin{align*}
    &\max\set{\abs{\loss(0; \fteacher[s])}, \abs{\loss'(0; \fteacher[s])}, \abs{\loss''(0; \fteacher[s])}} \\
    &\le C(1+|s|^{K_1}),
\end{align*}
where $\fteacher(\cdot)$ is the function in \eref{teacher}.

\item \label{a:regu} The regularizer $h(\cdot)$ in \eref{training} is strongly convex with parameter $\lambda > 0$. In addition, $h'''(x)$ exists, and it is uniformly bounded over $x \in \R$.

\item \label{a:kernel} The activation function $\kernel(\cdot)$ is an odd function, with bounded first, second, and third derivatives.

\item \label{a:fout} The function $\fout(x)$ in \eref{gen_err} is differentiable except at a finite number of points $\set{x_1, x_2, \ldots, x_L}$. Moreover, there exist constants $C>0$ and $K_2 \in \mathbb{Z}^+$ such that
\[
\max\set{\abs{\fteacher(x)}, \abs{\fout(x)}} \le C(1 + \abs{x}^{K_2}), ~\text{for all }x \in \R
\]
and
\[
\abs{\fout'(x)} \le C(1 + \abs{x}^{K_2}),\quad \text{for } x \not\in \set{x_1, x_2, \ldots, x_L}.
\]

\item \label{a:feature} The columns of the feature matrix $\fmtx = [\vf_1, \vf_2, \ldots, \vf_p]$ are independent Gaussian random vectors: $\vf_i \overset{\text{i.i.d.}}{\sim} \mathcal{N}(\mzero, \tfrac{1}{d}\mI_d)$ for $1 \le i \le p$. Moreover, $\fmtx$ is independent of the latent input variables $\set{\cvec_\didx}$.
\setcounter{assumptions}{\value{enumi}}
\end{enumerate}

\begin{rem}
We can verify that the conditions in Assumption \ref{a:loss} are satisfied by the quadratic loss function, the logistic loss function, and by any $\fteacher(s)$ that grows no faster than some polynomial of $\abs{s}$ as $\abs{s} \to \infty$. Possible ways to generalize our results to non-differentiable loss functions (\emph{e.g.} the hinge loss) will be discussed in \sref{conclusion}. To simplify our analysis, we require in Assumption~\ref{a:kernel} that the activation function $\kernel(x)$ be odd, which then implies that $\mu_0 = 0$ in \eref{mu_constants}. This is merely a limitation of our current results, and the asymptotic equivalence in \eref{asymp_equiv} is expected to hold for more general activation functions [such as the ReLU function as shown in \fref{quadratic}]. Yet another limitation of our work is the Gaussian assumption on the feature vectors in Assumption~\ref{a:feature}. With some extra effort (mostly on generalizing the concentration inequalities in Appendix~\ref{appendix:Concentration_Lipschitz}), our proof can be easily extended to cases where the columns of the feature matrix are independent sub-Gaussian random vectors. However, we expect that the majority of our proof technique should work for \emph{deterministic} feature matrices that satisfy the conditions in \eref{BP} and \eref{fnorm}. We will elaborate on this point in \sref{conclusion} and pinpoint the one technical difficulty that prevents us from working with deterministic matrices.
\end{rem}

To state the results of our main theorem, we first introduce a perturbed version of the optimization problem in \eref{training}:
\begin{equation}\label{eq:gen_opt}
\begin{aligned}
    \Phi_{\mR}(\tau_1, \tau_2) \bydef& \inf_{\vw \in \R^p} \big\{\textstyle\sum_{\didx=1}^{n} \loss(\tfrac{1}{\sqrt{p}}\vr_\didx\tran \vw; y_\didx) + \sum_{j = 1}^p \regu(w_j) \\
    &\hspace{2.5em}+ {\tau_1} (\vw\tran \mSig \vw) + \tau_2 (\sqrt{p} \mu_1\sgl\tran \fmtx \vw)\big\},
\end{aligned}
\end{equation}
where $\tau_1, \tau_2$ are two parameters, $\sgl$ is the teacher vector in \eref{teacher}, and
\begin{equation}\label{eq:Sig}
\mSig \bydef \mu_1^2 \fmtx\tran \fmtx + \mu_2^2 \mI_p.
\end{equation}
Note that $\tfrac{1}{p}\Phi_{\mA}(0, 0)$ and $\tfrac{1}{p}\Phi_{\mB}(0, 0)$ [with the regressor matrix $\mR$ specialized to $\mA$ and $\mB$ in \eref{mat_AB}] are exactly the training errors associated with the feature and Gaussian formulations, respectively. The two extra terms $\tau_1 (\vw\tran \mSig \vw)$ and $\tau_2 (\sqrt{p} \mu_1\sgl\tran \fmtx \vw)$ in \eref{gen_opt} will be needed in our analysis of the generalization error. In particular, we shall consider different values of $\tau_1, \tau_2$ such that
\begin{equation}\label{eq:tau12}
\abs{\tau_1} \le \tau^\ast \bydef \frac{\lambda/4}{\mu_1^2 (1+2\sqrt{\eta})^2 + \mu_2^2} \quad \text{and} \quad \abs{\tau_2} \le 1.
\end{equation}

\begin{rem}\label{rem:convex}
The bound $\tau^\ast$ requires some explanation. At first glance, the possibility that $\tau_1$ can take negative values is worrisome, as $\tau_1 \vw\tran \mSig \vw$ will then be a \emph{concave} function of $\vw$. This concave term, however, will (most likely) not change the convexity of the overall objective function in \eref{gen_opt}. To see this, we recall from Assumption~\ref{a:feature} that $\fmtx\tran \fmtx$ has a Wishart distribution and thus its spectral norm is bounded with high probability. Specifically, it is easy to show (see Appendix~\ref{appendix:matrix}) that
\[
\P(\norm{\fmtx} \ge 1+2\sqrt{\eta}) \le 2 e^{-c p},
\]
where $\eta = p/d$ and $c$ is some positive constant. By Assumption~\ref{a:regu}, the regularizer $h(x)$ is strongly convex with parameter $\lambda > 0$. It follows that, with $\tau_1 \ge -\tau^\ast$, the overall objective function of \eref{gen_opt} is $\frac{\lambda}{2}$-strongly convex with probability at least $1 - 2e^{-cp}$,
\end{rem}

\begin{thm}\label{thm:get}
Suppose Assumptions~\ref{a:Gaussian}--\ref{a:feature} hold. Fix $\tau_1 \in [-\tau^\ast, \tau^\ast]$ and $\tau_2 \in [-1, 1]$. For every $\varepsilon \in (0, 1)$ and every finite constant $c$, we have
\begin{equation}\label{eq:get}
\begin{aligned}
    &\P(\abs{\Phi_{\mA}(\tau_1, \tau_2)/p - c} \ge 2\varepsilon)\\
    \le& \P(\abs{\Phi_{\mB}(\tau_1, \tau_2)/p - c} \ge \varepsilon) 
    + \frac{\polylog p}{\varepsilon \sqrt{p}}
\end{aligned}
\end{equation}
and
\begin{equation}\label{eq:get2}
\begin{aligned}
    &\P(\abs{\Phi_{\mB}(\tau_1, \tau_2)/p - c} \ge 2\varepsilon) \\
    \le& \P(\abs{\Phi_{\mA}(\tau_1, \tau_2)/p - c} \ge \varepsilon) + \frac{\polylog p}{\varepsilon \sqrt{p}},
\end{aligned}
\end{equation}
for $p \ge 1/\varepsilon^2$, where $\polylog p$ denotes some function that grows no faster than a polynomial of $\log p$. Consequently,
\begin{equation}\label{eq:get_cip}
\frac{\Phi_{\mA}(\tau_1, \tau_2)}{p} \cip c \quad \text{if and only if} \quad \frac{\Phi_{\mB}(\tau_1, \tau_2)}{p} \cip c,
\end{equation}
where $\cip$ denotes convergence in probability as $p \to \infty$.
\end{thm}

\begin{rem}
We prove this theorem in \sref{get}. A special case, with $\tau_1 = \tau_2 = 0$, implies that the training errors of the nonlinear feature model and its Gaussian surrogate must necessarily have the same asymptotic limit.
\end{rem}

The next result, whose proof can be found in \sref{proof_gen}, establishes the universality for the generalization error, under one additional assumption:
\begin{enumerate}[label={(A.\arabic*)}]
\setcounter{enumi}{\value{assumptions}}
\item \label{a:gen} There exists a limit function $q^\ast(\tau_1, \tau_2)$ such that $\frac{\Phi_{\mB}(\tau_1, \tau_2)}{n} \cip q^\ast(\tau_1, \tau_2)$ for all $\tau_1 \in [-\tau^\ast, \tau^\ast]$ and $\tau_2 \in [-1, 1]$. In addition, the partial derivatives of $q^\ast(\tau_1, \tau_2)$ exist at $\tau_1 = \tau_2 = 0$. Let them be denoted by  $\frac{\partial}{\partial \tau_1} q^\ast(0, 0) = \genvar^\ast$ and $\frac{\partial}{\partial \tau_2} q^\ast(0, 0) = \gencor^\ast$, respectively. We further assume that $\genvar^\ast \neq 0$.
\end{enumerate}

\begin{prop}\label{prop:gen}
Under Assumptions~\ref{a:Gaussian}--\ref{a:gen}, we have
\[
\gerrA \cip \gerrLim \quad \text{and} \quad \gerrB \cip \gerrLim,
\]
where
\[
\gerrLim \bydef \EE_{z_1, z_2}  \big[\fteacher(z_1) - {\fout}( \gencor^\ast z_1 + [\genvar^\ast - (\gencor^\ast)^2]^{1/2} z_2)\big]^2,
\]
and $z_1, z_2$ are two independent standard Gaussian random variables.

\end{prop}

\subsection{Related Work\label{subsec:relatedwork}}

The Gaussian equivalence phenomenon studied in this paper was stated in \cite{hastie2019surprises,mei2019generalization,montanari2019generalization,goldt2019modelling, Goldt2020Gaussian}, and explicitly exploited in \cite{montanari2019generalization,gerace2020generalisation,dhifallah2020precise,Dhifallah2021inherent} to derive the asymptotic limits of several learning problems. Related phenomena also appear in the context of random kernel matrices \cite{cheng2013spectrum,pennington2017nonlinear,louart2018random}, where it is shown that the impact of the nonlinear activation function [on the limiting singular value spectrum of the matrix $\mA$ in \eref{mat_AB}] can be captured by the three parameters in \eref{mu_constants}. However, these results on the asymptotic spectrum are not sufficient for our purpose. Except for the special case of ridge regression, the training and generalization errors of the learning problem in \eref{training} are not simple functions of the singular values/vectors of $\mA$. 

Recently, the authors of \cite{Goldt2020Gaussian} proved an interesting central limit theorem for the low-dimensional projections of $\va_t$ in \eref{a_vec} and $\vb_t$ in \eref{b_vec} onto generic low-dimensional subspaces. Specifically, for any $\vw \in \R^p$ with bounded $\ell_\infty$ norm and independent of $\va_t, \vb_t$, it is shown in \cite{Goldt2020Gaussian} that
\begin{equation}\label{eq:CLT_ind}
\big(\scp \va\tran_t \vw, \cvec\tran_t \sgl\big) \overset{\text{Law}}{\approx} \big(\scp\vb\tran_t \vw, \cvec\tran_t \sgl\big) \sim \mathcal{N}\Big(\mathbf{0}, \begin{bmatrix}\genvar & \pi\\ \gencor & 1 \end{bmatrix}\Big),
\end{equation}
where $\genvar= \vw\tran \mSig \vw / p$, with $\mSig$ defined in \eref{Sig}, and $\gencor = \mu_1 \sgl\tran \fmtx  \vw/{\sqrt{p}}$. This result is an important step towards a theoretical justification of the Gaussian equivalence, and indeed a quantitive version of \eref{CLT_ind} serves as a crucial ingredient of our proof. However, by itself the characterization in \eref{CLT_ind} does not imply the asymptotic equivalence stated in \eref{asymp_equiv}, as the training and generalization errors are all complicated functionals defined implicitly through the optimization problem \eref{training}. When calculating the generalization errors $\gerrA, \gerrB$ using \eref{gen_err}, for example, one will be dealing with two different weight vectors $\wt_{\mA}$ and $\wt_{\mB}$, respectively, as opposed to a single shared vector $\vw$ as in \eref{CLT_ind}. Showing that $\vw\tran_{\mA} \mSig \vw_{\mA} / p \approx \vw_{\mB}\tran \mSig \vw_{\mB} / p$ and $\sgl\tran \fmtx  \vw_{\mA}/{\sqrt{p}} \approx \sgl\tran \fmtx  \vw_{\mB}/{\sqrt{p}}$, which are the second-order statistics of the Gaussian distributions, is exactly among the technical challenges addressed in this work.

Our method for proving universality for the random feature model is based on the classical Lindeberg's principle \cite{lindeberg1922neue} and a leave-one-out analysis \cite{el2018impact} of the optimization problem in \eref{training}. Similar approaches have been used before to establish universality for various estimation problems \cite{oymak2018universality,korada2011applications,montanari2017universality,panahi2017universal,abbasi2019universality}. As a technical challenge in our problem, the entries of the regression vectors have a particular correlation structure, due to the presence of the random feature matrix $\fmtx$ in \eref{a_vec} and \eref{b_vec}. Thus, new techniques have to be developed to handle this correlation. Beyond the random feature model considered here, the Gaussian equivalence is a very general universality phenomenon that has been observed in many other models (see, \emph{e.g.}, \cite{Goldt2020Gaussian,seddik2020random,Dhifallah2021inherent,Loureiro2021capturing,gerace2022gaussian}).

\textcolor[rgb]{0.00,0.00,0.00}{After the initial release of this paper on arXiv, some of the results in this work have been used and adapted by other authors to rigorously establish the Gaussian equivalence phenomenon in several different settings.
Examples include minimum $\ell_1$ norm interpolated classification \cite{liang2020precise} and the feature learning in two-layer neural network \cite{ba2022high}.  
It will be interesting to extend the proof techniques in the current paper to handle some more general and challenging cases. Towards this direction, the recent paper \cite{montanari2022universality} by Montanari and Saeed studies the Gaussian equivalence of empirical risk minimization where the loss function and the regularizer do not need to be convex.}


\textcolor[rgb]{0.00,0.00,0.00}{Finally, it is worth mentioning that all the aforementioned works focus on the so-called linear asymptotics regime, \emph{i.e,} $n/d\to\alpha$ and $p/d\to\eta$, where $\alpha,\eta\in(0,\infty)$. Recently, the Gaussian equivalence in the more general polynomial asymptotic regime, where $n/d^\ell\to\alpha\in(0,\infty)$, with $\ell\in\mathbb{Z}^{+}$, has been studied.
For example, the papers \cite{lu2022equivalence,misiakiewicz2022spectrum,xiao2022precise} analyze the spectrum of random inner-product matrices in the polynomial asymptotics regime, where $n/d^\ell\to\alpha\in(0,\infty)$, with $\ell\in\mathbb{Z}^{+}$. Based on these results, the exact learning performance of kernel ridge regression with polynomial scalings was established in \cite{misiakiewicz2022spectrum,hu2022sharp,xiao2022precise}.}



\subsection{Paper Outline}

The rest of the paper is organized as follows.\ We prove Theorem~\ref{thm:get} and Proposition~\ref{prop:gen} in \sref{Proof-Sketch}. To emphasize readability, we only highlight the central ideas and key intermediate results there. In \sref{CLT}, we use Stein's method to provide an alternative proof of the central limit theorem for the nonlinear feature model. Heavier technical details are left to the appendix, where we compile all the auxiliary results. We conclude the paper in \sref{conclusion} with some additional remarks on how some of the technical assumptions in this work can be further relaxed.

%% file: proof_sketch.tex

\section{Proof of the Main Results}
\label{sec:Proof-Sketch}

\emph{Notation}: In our proof of Theorem~\ref{thm:get}, the parameters $\tau_1, \tau_2$ in \eref{gen_opt} are always kept fixed. Thus, to streamline the notation, we will write $\Phi_{\mA}(\tau_1, \tau_2)$ and $\Phi_{\mB}(\tau_1, \tau_2)$ simply as $\Phi_{\mA}$ and $\Phi_{\mB}$, when no confusion can arise. We will use $C$ and $c$ to denote generic constants that do not depend on the problem dimension $p$. To reduce the burden of bookkeeping, the exact values of $C$ and $c$ can change from one line to the next. In addition, $\polylog p$ stands for any function $B(p)$ that grows no faster than some polynomial of $\log p $, \ie,
\[
\abs{B(p)} \le C(1+\log^K\!p)
\]
for some finite $C > 0$ and $K \in \mathbb{Z}^+$. For a vector $\vx$, we use $\norm{\vx}$ to denote its 2-norm and $\norm{\vx}_\infty$ its $\ell_\infty$ norm. For a matrix $\mM$, its spectral and Frobenius norms are denoted by $\norm{\mM}$ and $\norm{\mM}_\text{F}$, respectively. Throughput the paper, we also adopt the following notational convention regarding conditional expectations. Given a family of independent random variables $X_1, X_2, \ldots, X_K$, we will write $\Ecnd{X_1}G(X_1, \ldots, X_K)$ for the conditional expectation of a function $G(\cdot)$ over $X_2, \ldots, X_K$, with $X_1$ kept fixed. A related notation is $\Eone{X_1}G(X_1, \ldots, X_K)$, where we take the expectation over $X_1$, conditional on all the other random variables. Finally, $\charfn_{\mathcal{A}}$ denotes the indicator function on a set $\mathcal{A}$, and $[n]$ stands for the set $\set{1, 2, \ldots, n}$.

\subsection{Test Functions}

We start by noting that, to prove the inequalities in \eref{get} and \eref{get2}, it suffices to show that
\begin{equation}\label{eq:test_func}
\begin{aligned}
        &\abs{\E \test(\tfrac{1}{p}{\Phi_{\mA}}) -  \E \test(\tfrac{1}{p}{\Phi_{\mB}})}  \\
    \le& \max\Big\{\norm{\test}_\infty, \norm{\test'}_\infty, \frac{\norm{\test''}_\infty}{\sqrt{p}}\Big\} \frac{\polylog p}{\sqrt{p}}
\end{aligned}
\end{equation}
for every bounded test function $\test(x)$ that also has bounded first and second derivatives. The precise connection between \eref{get}, \eref{get2} and \eref{test_func} will be made clear in \sref{get}, when we prove Theorem~\ref{thm:get}. For now, we focus on showing \eref{test_func}.

In our analysis, we first show a \emph{conditional} version of \eref{test_func}. Specifically, we will define a subset $\mathcal{A}$ of all $d \times p$ feature matrices, and show that
\begin{equation}\label{eq:cond_test}
\begin{aligned}
    &\sup_{\fmtx \in \mathcal{A}}\abs{\Ecnd{\fmtx}\test(\tfrac{1}{p}{\Phi_{\mA}})-  \Ecnd{\fmtx}\test(\tfrac{1}{p}{\Phi_{\mB}})} \\
    \le& \max\Big\{\norm{\test'}_\infty, \frac{\norm{\test''}_\infty}{\sqrt{p}}\Big\} \frac{\polylog p}{\sqrt{p}},
\end{aligned}
\end{equation}
where $\Ecnd{\fmtx}[\,\cdot\,]$ denotes the conditional expectation (over the input variables $\set{\cvec_i}$) for a fixed feature matrix $\fmtx$. We refer to $\mathcal{A}$ as the \emph{admissible set} of feature matrices, and its precise definition will be given in \sref{admissible}. To go from \eref{cond_test} to \eref{test_func}, we have
\begin{align}
&\abs{\EE \test(\tfrac{1}{p}{\Phi_{\mA}}) -  \EE \test(\tfrac{1}{p}{\Phi_{\mB}})} \nonumber \\
\le&  \EE\abs{\Ecnd{\fmtx}[\test(\tfrac{1}{p}{\Phi_{\mA}})] -  \Ecnd{\fmtx}[\test(\tfrac{1}{p}{\Phi_{\mB}})]}\nonumber\\
=&  \EE\abs{\Ecnd{\fmtx}[\test(\tfrac{1}{p}{\Phi_{\mA}})] -  \Ecnd{\fmtx}[\test(\tfrac{1}{p}{\Phi_{\mB}})]} (\charfn_{\mathcal{A}}(\fmtx) + \charfn_{\mathcal{A}^c}(\fmtx))\nonumber\\
\le& \sup_{\fmtx \in \mathcal{A}}\abs{\Ecnd{\fmtx}[\test(\tfrac{1}{p}{\Phi_{\mA}})] -  \Ecnd{\fmtx}[\test(\tfrac{1}{p}{\Phi_{\mB}})]} + 2\norm{\test}_\infty \P(\mathcal{A}^c).
\label{eq:test_conditional_decomp}
\end{align}

The remaining tasks are now clear: (1) Establish \eref{cond_test}; and (2) show $\P(\mathcal{A}^c)  = \mathcal{O}(\polylog p / \sqrt{p})$. But first, we need to define the admissible set $\mathcal{A}$.

\subsection{The Admissible Set of Feature Matrices}
\label{sec:admissible}

Recall that $\fmtx = [\vf_1, \vf_2, \ldots, \vf_p]$, where $\set{\vf_i}_{i\in[p]}$ are the feature vectors. For notational simplicity, we add one more vector by letting $\vf_0 \bydef \sgl$. The admissible set $\mathcal{A}$ is constructed as
\begin{equation}\label{eq:def_setA}
\mathcal{A} = \mathcal{A}_1 \cap \mathcal{A}_2 \cap \mathcal{A}_3,
\end{equation}
where
\begin{align}
\label{eq:def_setA1}
\mathcal{A}_1 &\bydef \Big\{\fmtx \in \R^{d \times p}: \max_{0 \le i \le j \le p} \abs{\vf_i\tran \vf_j - \delta_{ij}}\le \tfrac{(\log p)^2}{\sqrt{p}}\Big\}, \\
\intertext{with $\delta_{ij}$ denoting the Kronecker delta function, and}
\mathcal{A}_2 &\bydef \set{\fmtx \in \R^{d \times p}: \norm{\fmtx} \le 1+2\sqrt{\eta} },\label{eq:def_setA2}
\end{align}
where $\eta$ is the constant in Assumption~\ref{a:dimensions}. Before defining $\mathcal{A}_3$, which requires some additional notation, we first note that $\mathcal{A}_1 \text{ and } \mathcal{A}_2$ are all high-probability events under Assumption~\ref{a:feature}. Specifically, standard concentration inequalities for sub-Gaussian random vectors give us
\begin{equation}\label{eq:A1A2_p}
\P(\mathcal{A}_1) \ge 1 - ce^{-\left(\log p\right)^{2}/c}
\end{equation}
for some $c > 0$. (See Lemma~\ref{lem:A1_high_prob} in Appendix~\ref{appendix:Concentration-of-GaussianVec} for a proof.) Similarly, applying matrix concentration inequalities [\eqref{eq:spectral_norm_Wishart} in Appendix~\ref{appendix:matrix}], we can conclude that
\begin{equation}\label{eq:A3_p}%
\P(\mathcal{A}_2) \ge 1 - 2 e^{-cp}
\end{equation}
for some constant $c>0$.

The definition of the last set $\mathcal{A}_3$ in \eref{def_setA} is a bit technical. Consider a family of optimization problems
\begin{align}
\Phi_{k}\bydef & \min_{\vw\in\R^{p}}\Big\{\textstyle\sum_{t=1}^{k}\loss(\tfrac{1}{\sqrt{p}}\vb_{t}\tran\vw;y_{t})+\sum_{t=k+1}^{n}\loss(\tfrac{1}{\sqrt{p}}\va_{t}\tran\vw;y_{t})\nonumber \\
&\hspace{2.8em}+\sum_{j=1}^{p}\regu(w_{j})+Q(\vw)\Big\},\label{eq:interpolation_k}\\
\wt_{k}\bydef & \argmin{\vw\in\R^{p}}\Big\{\textstyle\sum_{t=1}^{k}\loss(\tfrac{1}{\sqrt{p}}\vb_{t}\tran\vw;y_{t})+\sum_{t=k+1}^{n}\loss(\tfrac{1}{\sqrt{p}}\va_{t}\tran\vw;y_{t}) \nonumber\\
&\hspace{3.9em}+\sum_{j=1}^{p}\regu(w_{j})+Q(\vw)\Big\},\label{eq:interpolation_k_wt}
\end{align}
for $0 \le k \le n$, where $\set{\va_\didx}$ and $\set{\vb_\didx}$ are the regressors in \eref{a_vec} and \eref{b_vec}, respectively, and
\begin{align}
\label{eq:Q_def}
Q(\vw) \bydef {\tau_1} \vw\tran \mSig \vw + \tau_2\mu_1\sqrt{p} \sgl\tran \fmtx \vw.
\end{align}
The reason for considering this sequence of problems will become clear in \sref{Lindeberg}. For now, just note that our quantities of interest, namely $\Phi_{\mA}$ and $\Phi_{\mB}$, are just the starting and end point of this sequence, \ie, $\Phi_{0}=\Phi_{\mA}$ and $\Phi_{n}=\Phi_{\mB}$. We then have
\begin{align}
\label{eq:def_setA3}
\mathcal{A}_3 \bydef  \Big\{\fmtx \in \R^{d \times p}: \Big[\max_{0 \le k \le n} \Ecnd{\fmtx}\norm{\wt_k}_\infty^2\Big] \le \left(\log p\right)^{7+4K_1}\Big\},
\end{align}
where $K_1$ is the constant in Assumption~\ref{a:loss}.
\begin{prop}\label{prop:wt_bnd}
Under Assumptions~\ref{a:Gaussian}--\ref{a:feature}, there exists some $c>0$ such that
\begin{equation}\label{eq:A4_p}
\P({\cal A}_{3})\geq 1-c e^{-\left(\log p\right)^{2}/c}.
\end{equation}
\end{prop}
This result, whose proof can be found in Appendix~\ref{appendix:linfty_boundedness}, shows that $\mathcal{A}_3$ is still a high-probability event. In light of \eref{A1A2_p}, \eref{A3_p} and \eref{A4_p}, there exists $c>0$ such that
\begin{align}
\label{eq:A_small_prob}
\P(\mathcal{A}^c) \le \P(\mathcal{A}_1^c) + \P(\mathcal{A}_2^c) + \P(\mathcal{A}_3^c) \le  ce^{-\left(\log p\right)^{2}/c}.
\end{align}

\subsection{The Lindeberg Method}
\label{sec:Lindeberg}


In what follows, we prove \eref{cond_test} by using Lindeberg's method \cite{lindeberg1922neue,korada2011applications,panahi2017universal}.
The idea is simple: The sequence shown in \eqref{eq:interpolation_k} serves as an interpolation path that allows us to go from $\Phi_{\mA}$ to $\Phi_{\mB}$. To prove \eref{cond_test}, it suffices to show that the difference between any two neighboring
points on the interpolation path is small. Indeed, as there are only $n = \mathcal{O}(p)$ such pairwise comparisons, we just need to show that
\[
\abs{\Ecnd{\fmtx} \big[\test\big(\tfrac{1}{p}\Phi_{k}\big)\big]-\Ecnd{\fmtx}\big[ \test\big(\tfrac{1}{p}\Phi_{k-1}\big)\big]} = \mathcal{O}\Big(\frac{\polylog p}{p^{3/2}}\Big),
\]
uniformly over $\fmtx \in \mathcal{A}$ and $1 \le k \le n$.

By construction, the optimization problems associated with $\Phi_k$ and $\Phi_{k-1}$ differ only in their choice of the $k$th regressor. The former uses $\vb_k$, whereas the latter uses $\va_k$. Consequently, both $\Phi_k$ and $\Phi_{k-1}$ can be seen as a perturbation of a common ``leave-one-out'' problem:
\begin{align}
\label{eq:LOO_objctive_function}
\Phi_{\bs k}\bydef & \min_{\vw\in\R^p} \big\{\textstyle\sum_{t=1}^{k-1}\loss(\tfrac{1}{\sqrt{p}}\vb_{t}\tran\vw;y_{t})+\sum_{t=k+1}^{n}\loss(\tfrac{1}{\sqrt{p}}\va_{t}\tran\vw;y_{t}) \nonumber \\
&\hspace{3em}+\sum_{j=1}^{p}\regu(w_{j})+Q(\vw)\big\}.
\end{align}
As $\Phi_k \approx \Phi_{\bs k}$, it is natural to apply Taylor's expansion around $\Phi_{\bs k}$, which gives us
\begin{equation}\label{eq:phik_expansion}
\begin{aligned}
    \test(\tfrac{1}{p}\Phi_{k}) =& \test(\tfrac{1}{p}\Phi_{\bs k}) + \tfrac{1}{p} \test'(\tfrac{1}{p}\Phi_{\bs k}) ({\Phi_{k}} - {\Phi_{\bs k}}) \nonumber \\
    &+ \tfrac{1}{2p^2} \test''(\theta) ({\Phi_{k}} - {\Phi_{\bs k}})^2,
\end{aligned}
\end{equation}
with $\theta$ denoting some value that lies between $\tfrac{1}{p}\Phi_{k}$ and $\tfrac{1}{p}\Phi_{\bs k}$. Writing an analogous expansion for $\btest(\Phi_{k-1})$ around $\Phi_{\bs k}$, and then subtracting it from \eref{phik_expansion}, we can get
\begin{equation}\label{eq:testfunc_interpolation}
\begin{aligned}
&\left|\Ecnd{\fmtx}\big[ \test\big(\tfrac{1}{p}\Phi_{k}\big)\big]-\Ecnd{\fmtx}\big[ \test\big(\tfrac{1}{p}\Phi_{k-1}\big)\big]\right| \\
\le&  \tfrac{\left\Vert \test'(x)\right\Vert _{\infty}}{p}\Ecnd{\fmtx}\left|\E_{k}\left(\Phi_{k}-\Phi_{k-1}\right)\right|\\
 &+\tfrac{\left\Vert \test''(x)\right\Vert _{\infty}}{2p^{2}}\left[\Ecnd{\fmtx}\left(\Phi_{k}-\Phi_{\bs k}\right)^{2}+\Ecnd{\fmtx}\left(\Phi_{k-1}-\Phi_{\bs k}\right)^{2}\right],
\end{aligned}
\end{equation}
where $\E_k$ denotes the conditional expectation over the random vectors $\{\va_k, \vb_k\}$ associated with the $k$th training sample, while keeping everything else, \ie, $\{\va_t, \vb_t\}_{t\neq k}$ and $\fmtx$, fixed.

To make further progress, we need to introduce a surrogate optimization problem:
\begin{equation}\label{eq:quadratic_approx_objctive_function}
\begin{aligned}
    \Psi_{k}(\vr)\bydef&
    \Phi_{\bs k}+\min_{\vw\in\R^p}\Big\{\frac{1}{2}(\vw-\loowt{\bs k})\tran\mH_{\bs k}(\vw-\loowt{\bs k}) \\
    &\hspace{6em}+\loss(\tfrac{1}{\sqrt{p}}\vr\tran\vw;y_{k})\Big\},
\end{aligned}
\end{equation}
where $\loowt{\bs k}$ is the leave-one-out optimal solution of \eqref{eq:LOO_objctive_function}, and
\begin{equation}\label{eq:looH}
\begin{aligned}
\mH_{\bs k} \bydef& \frac{1}{p}\textstyle\sum_{t=1}^{k-1}\loss''(\tfrac{1}{\sqrt{p}}\vb_{t}\tran\loowt{\bs k};y_{t})\vb_{t}\vb_{t}\tran \\
&+\displaystyle\frac{1}{p}\textstyle\sum_{t=k+1}^{n}\loss''(\tfrac{1}{\sqrt{p}}\va_{t}\tran\loowt{\bs k};y_{t})\va_{t}\va_{t}\tran\\
&+\diag\Big\{ h''\big(w_{\bs k,i}^{*}\big)\Big\} +\nabla^{2}Q(\loowt{\bs k})
\end{aligned}
\end{equation}
is the Hessian matrix of the objective function in \eref{LOO_objctive_function} evaluated at $\loowt{\bs k}$. We note that $\Psi_k(\vr)$ has a simple interpretation: By setting $\vr = \vb_k$, we can see that the optimization problem associated with $\Psi_k(\vb_k)$ is simply a quadratic approximation of the one associated with $\Phi_k$ in \eref{interpolation_k}. Similarly, $\Psi_k(\va_k)$ is a quadratic approximation of $\Phi_{k-1}$. The following lemma, whose proof can be found in Appendix~\ref{appendix:quad_approx}, quantifies the accuracy of such approximation.
\begin{lem}
\label{lem:quad_approx}We have
\begin{equation}\label{eq:quad_approx_psi_phi_k}
\begin{aligned}
   &\max\{\Ecnd{\fmtx}\left(\Psi_{k}(\vb_k) - \Phi_{\bs k}\right)^2,\Ecnd{\fmtx}\left(\Psi_{k}(\va_k) - \Phi_{\bs k}\right)^2\} \\
   \leq& \polylog p, 
\end{aligned}
\end{equation}
and
\begin{equation}\label{eq:quad_approx_phi_psi}
\begin{aligned}
    &\max\set{\Ecnd{\fmtx}\left(\Psi_{k}(\vb_k)-\Phi_{k}\right)^2,\Ecnd{\fmtx}\left(\Psi_{k}(\va_k)-\Phi_{k-1}\right)^2} \\
    \leq& \frac{\polylog p}{p},
\end{aligned}
\end{equation}
both of which hold uniformly over $\fmtx\in \mathcal{A}$ and $k\in[n]$.
\end{lem}




Using this lemma, we can now bound the terms on the right-hand side of \eqref{eq:testfunc_interpolation} as follows:
\begin{align}
\label{eq:quad_decomp_1}
&\Ecnd{\fmtx}\left|\E_{k}\left(\Phi_{k}-\Phi_{k-1}\right)\right| \nonumber \\
\le& \Ecnd{\fmtx}\absb{\E_{k}[\Psi_{k}(\vb_k)-\Psi_{k}(\va_k)]} + \Ecnd{\fmtx}\left|\Psi_{k}(\vb_k)-\Phi_{k}\right| \nonumber \\
&+\Ecnd{\fmtx}\left|\Psi_{k}(\va_k)-\Phi_{k-1}\right|\nonumber\\
\le&  \Ecnd{\fmtx}\absb{\E_{k}[\Psi_{k}(\vb_k)-\Psi_{k}(\va_k)]} + \polylog p / \sqrt{p},
\end{align}
where to reach the last step we have used H\"{o}lder's inequality and \eref{quad_approx_phi_psi}. Meanwhile, combining \eref{quad_approx_phi_psi} and \eref{quad_approx_psi_phi_k} gives us
\begin{equation}\label{eq:quad_decomp_2}
\begin{aligned}
    &\Ecnd{\fmtx}\left(\Phi_{k}-\Phi_{\bs k}\right)^{2} \\ \leq& 2\Ecnd{\fmtx}\left(\Phi_{k}-\Psi_{k}(\vb_k)\right)^{2} + 2\Ecnd{\fmtx}\left(\Psi_{k}(\vb_k)-\Phi_{\bs k}\right)^{2} \\
    \le& \polylog p,
\end{aligned}
\end{equation}
and similarly,
\begin{equation}\label{eq:quad_decomp_3}
\Ecnd{\fmtx}\left(\Phi_{k-1}-\Phi_{\bs k}\right)^{2} \le \polylog p.
\end{equation}
In light of \eref{quad_decomp_1}, \eref{quad_decomp_2}, and \eref{quad_decomp_3}, we just need to show that
\[
\Ecnd{\fmtx}\absb{\E_{k}[\Psi_{k}(\vb_k)-\Psi_{k}(\va_k)]} = o(1)
\]
to get a useful bound for the left-hand side of \eref{testfunc_interpolation}.

We are now in a position to show why we introduce and work with $\Psi_k(\vr)$. Let $\calM_{k}(x;\gamma)$ denote the \emph{Moreau envelope} of the loss function
$\loss\left(x;y_{k}\right)$, \ie,
\begin{equation}\label{eq:Moreau_envelope}
\calM_{k}(x;\gamma)\bydef\min_{z}\Big\{\loss\left(z;y_{k}\right)+\frac{(x-z)^{2}}{2\gamma}\Big\},
\end{equation}
where $\gamma > 0$ is some fixed parameter. It is straightforward to show (see Lemma~\ref{lem:quadapprox_optsol} in Appendix~\ref{appendix:opt_properties}) that
\begin{equation}\label{eq:Rk_cr_decomp_general}
\Psi_{k}(\vr)=\Phi_{\bs k}+\calM_{k}\big(\scp{\vr\tran\loowt{\bs k}};\gamma_{k}(\vr)\big),
\end{equation}
where
\begin{equation}\label{eq:gamma_k}
\gamma_{k}(\vr)\bydef({\vr\tran\mH_{\bs k}^{-1}\vr})/{p}.
\end{equation}
It then follows that
\begin{equation}\label{eq:Moreau_diff}
\begin{aligned}
    \Psi_k(\vb_k) - \Psi_k(\va_k) =& \calM_{k}\big(\scp{\vb_k\tran\loowt{\bs k}};\gamma_{k}(\vb_k)\big) \nonumber \\
    &- \calM_{k}\big(\scp{\va_k\tran\loowt{\bs k}};\gamma_{k}(\va_k)\big).
\end{aligned}
\end{equation}
By construction, both $\va_k$ and $\vb_k$ are independent of the leave-one-out solution $\loowt{\bs k}$ and the Hessian matrix $\mH_{\bs k}$. It is this independent structure that significantly simplifies our analysis.

As $p \to \infty$, the scalars $\gamma_{k}(\vb_k)$ and $\gamma_{k}(\va_k)$ in \eref{Moreau_diff} concentrate around a common value $\gamma_k \bydef  \E_{k}\gamma_{k}(\vb_{k})$. This then prompts us to write the following decomposition
\begin{equation}\label{eq:Phi_k_decomp_2}
\begin{aligned}
 &\Ecnd{\fmtx}\left|\E_{k}(\Psi_{k}(\vb_k)-\Psi_{k}(\va_k))\right|\\
  \le& \Ecnd{\fmtx}\underbrace{\absb{\E_{k}\calM_{k}\big(\tfrac{1}{\sqrt{p}}\vb_{k}\tran\loowt{\bs k};\gamma_{k}\big)-\E_{k}\calM_{k}\big(\tfrac{1}{\sqrt{p}}\va_{k}\tran\loowt{\bs k};\gamma_{k}\big)}}_{\Delta_{\text{CLT}}} \\
  &+ \Delta_1 + \Delta_2,
 \end{aligned}
\end{equation}
where
\begin{equation}\label{eq:Phi_decomp_D1}
\begin{aligned}
    &\Delta_1  
    \bydef \Ecnd{\fmtx}\absb{\E_{k}\calM_{k}\big(\tfrac{1}{\sqrt{p}}\vb_{k}\tran\loowt{\bs k};\gamma_{k}(\vb_{k})\big) \\
    &\hspace{11em}-\E_{k}\calM_{k}\big(\tfrac{1}{\sqrt{p}}\vb_{k}\tran\loowt{\bs k};\gamma_{k}\big)}
\end{aligned}
\end{equation}
and
\begin{equation}\label{eq:Phi_decomp_D2}
\begin{aligned}
    &\Delta_2 
    \bydef\Ecnd{\fmtx}\absb{\E_{k}\calM_{k}\big(\tfrac{1}{\sqrt{p}}\va_{k}\tran\loowt{\bs k};\gamma_{k}(\va_{k})\big) \\
    &\hspace{11em}-\E_{k}\calM_{k}\big(\tfrac{1}{\sqrt{p}}\va_{k}\tran\loowt{\bs k};\gamma_k\big)}.
\end{aligned}
\end{equation}
These last two terms are easy to control, due to the concentrations of $\gamma_{k}(\vb_k)$ and $\gamma_{k}(\va_k)$ around $\gamma_k$. As shown in Lemma~\ref{lem:pathdiff_D1D2_bd} in Appendix~\ref{appendix:Thm2_satisfy}, we have
\begin{equation}\label{eq:D1D2_max}
\max\set{\Delta_1, \Delta_2} \le \frac{\polylog p}{\sqrt{p}},
\end{equation}
uniformly over $\fmtx \in \mathcal{A}$ and $k\in [n]$.






It is more challenging to bound the term $\Delta_\text{CLT}$, whose subscript alludes to the fact that we will be using a version of the central limit theorem. To see that, we first recall from \eref{Moreau_envelope} that the Moreau envelope $\calM_{k}(x;\gamma_k)$ depends on the training label $y_k$. The latter is generated by the model in \eref{teacher}, with a teacher function $\fteacher(x)$. Introducing a two-dimensional test function
\begin{equation}\label{eq:test_M}
\btest(x; s) \bydef\min_{z}\loss\left(z;\fteacher(s)\right)+\frac{(x-z)^{2}}{2\gamma_k},
\end{equation}
we can then write
\[
\Delta_\text{CLT} = \absb{\E_{k}\btest\big(\tfrac{1}{\sqrt{p}}\va_{k}\tran\loowt{\bs k};\cvec_k\tran \sgl\big)-\E_{k}\btest\big(\tfrac{1}{\sqrt{p}}\vb_{k}\tran\loowt{\bs k};\cvec_k\tran \sgl\big)}.
\]
That $\Delta_\text{CLT} = o(1)$ is due to the following fact: When conditioned on $\fmtx$ and $\loowt{\bs k}$, we have
\begin{equation}
\label{eq:clt_informal}
\big(\scp{\va\tran_k \loowt{\bs k}}, \cvec_k\tran \sgl\big) \overset{\text{Law}}{\approx} \big(\scp{\vb_k\tran \loowt{\bs k}}, \cvec_k\tran \sgl\big) \sim \text{jointly Gaussian}.
\end{equation}
Making \eqref{eq:clt_informal} precise is the focus of Theorem~\ref{thm:CLT_general} in \sref{CLT}. It is easy to verify that the test function defined in \eref{test_M} indeed satisfies the assumptions of Theorem~\ref{thm:CLT_general}. (See Lemma~\ref{lem:Thm2_satisfy} in Appendix~\ref{appendix:Thm2_satisfy}.) Consequently, for every $\fmtx \in \mathcal{A}$, Theorem~\ref{thm:CLT_general} gives us
\begin{align}
&\Ecnd{\fmtx} [\Delta_\text{CLT}] \nonumber\\
\overset{(a)}{\le}&  \Ecnd{\fmtx} \big[(1 + \norm{\loowt{\bs k}}_\infty [1+\BPf])(1 + (\scp\norm{\loowt{\bs k}})^{K})\big]\frac{\polylog p}{\sqrt{p}}\nonumber\\
\le&  \Ecnd{\fmtx} \big[1 + (1+\BPf)^2 \norm{\loowt{\bs k}}_\infty^2 + (\scp\norm{\loowt{\bs k}})^{2K}\big]\frac{\polylog p}{\sqrt{p}}\nonumber\\
\overset{(b)}{\le}& \frac{\polylog p}{\sqrt{p}}.\label{eq:D_clt}
\end{align}
In (a), $\BP$ is the bound in \eref{BP}, and $K$ is some positive constant. To reach (b), we have used the fact that $\fmtx \in \mathcal{A}_2$, which then implies that $\BP \le \polylog p$, and $\fmtx \in \mathcal{A}_3$, which guarantees the boundedness of $\Ecnd{\fmtx} \norm{\loowt{\bs k}}_\infty^2$. Finally, the boundedness of $\Ecnd{\fmtx}(\scp\norm{\loowt{\bs k}})^{2K}$ is verified in Lemma \ref{lem:norm_optwt_Expec_bd_A2}.

We can now retrace our steps to reach our goal of proving \eref{cond_test}. Specifically, substituting \eref{D_clt} and \eref{D1D2_max} into \eref{Phi_k_decomp_2} gives us $\Ecnd{\fmtx}\absb{\E_{k}(\Psi_{k}(\vb_k)-\Psi_{k}(\va_k))} \le \polylog p / \sqrt{p}$, which, together with \eref{quad_decomp_1}, \eref{quad_decomp_2}, \eref{quad_decomp_3}, and \eref{testfunc_interpolation}, leads to
\begin{equation}\label{eq:cond_test_k}
\begin{aligned}
    &\absb{\Ecnd{\fmtx}\big[ \test\big(\tfrac{1}{p}\Phi_{k}\big))\big]-\Ecnd{\fmtx}\big[ \test\big(\tfrac{1}{p}\Phi_{k-1}\big)\big]} \\
&\hspace{2.8em} \le  \max\Big\{\left\Vert \test'(x)\right\Vert _{\infty}, \frac{\left\Vert \test''(x)\right\Vert _{\infty}}{\sqrt{p}}\Big\}\frac{\polylog p}{p^{3/2}}.
\end{aligned}
\end{equation}
Note that the upper bound is uniform over all $\fmtx \in \mathcal{A}$ and all $k \in [n]$. Now let us recall the construction of the interpolation sequence in \eref{interpolation_k}. Since $\Phi_{0}=\Phi_{\mA}$ and $\Phi_{n}=\Phi_{\mB}$, we obtain \eref{cond_test} from \eref{cond_test_k} via triangle inequality. Finally, given the decomposition in \eref{test_conditional_decomp} and the probability bound in \eref{A_small_prob}, we establish the inequality in \eref{test_func}.

\begin{figure*}[t]
\begin{centering}
\subfigure[$\lambda=1$]{\begin{centering}
\includegraphics[scale=0.32]{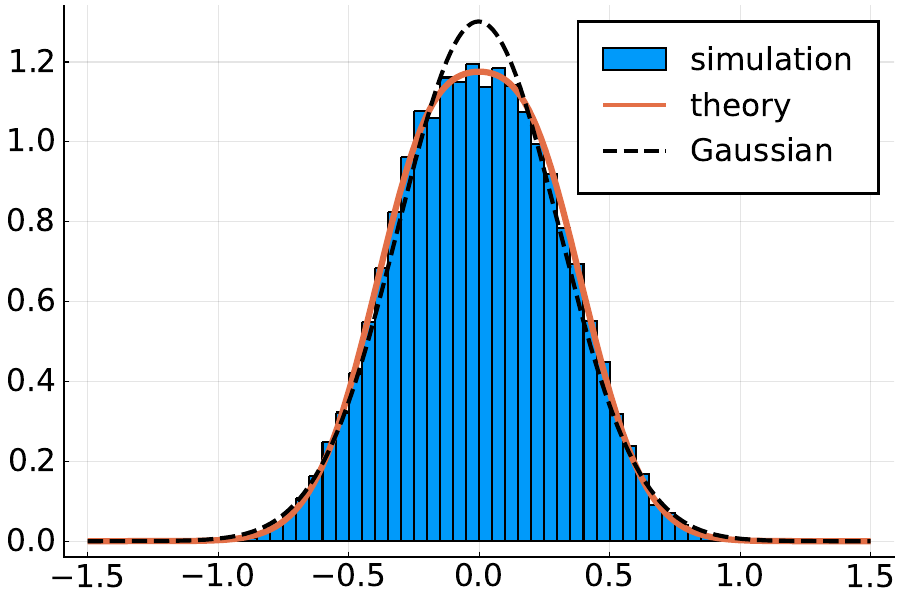}
\par\end{centering}

}\hphantom{}\subfigure[$\lambda=10^{-2}$]{\begin{centering}
\includegraphics[scale=0.32]{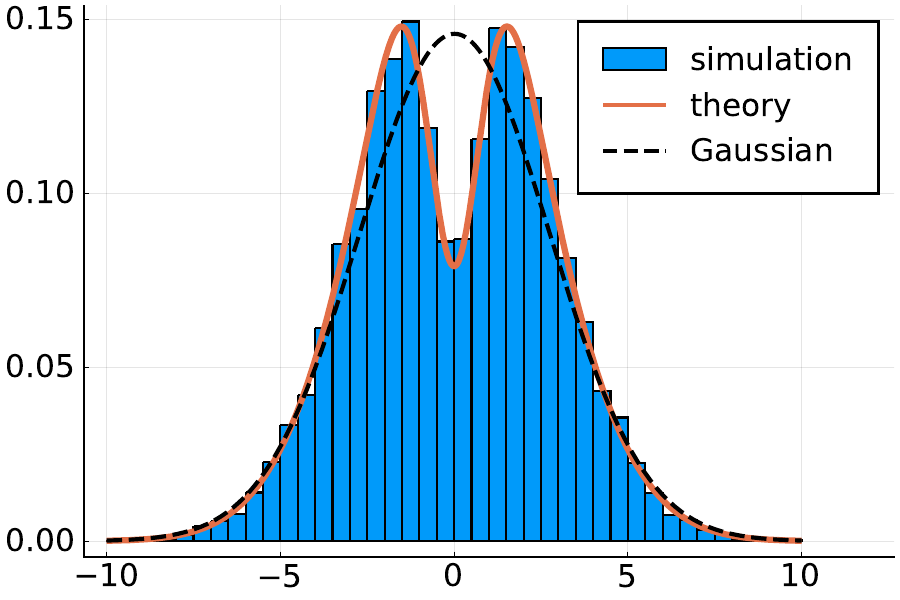}
\par\end{centering}

}\hphantom{}\subfigure[$\lambda=10^{-3}$]{\begin{centering}
\includegraphics[scale=0.32]{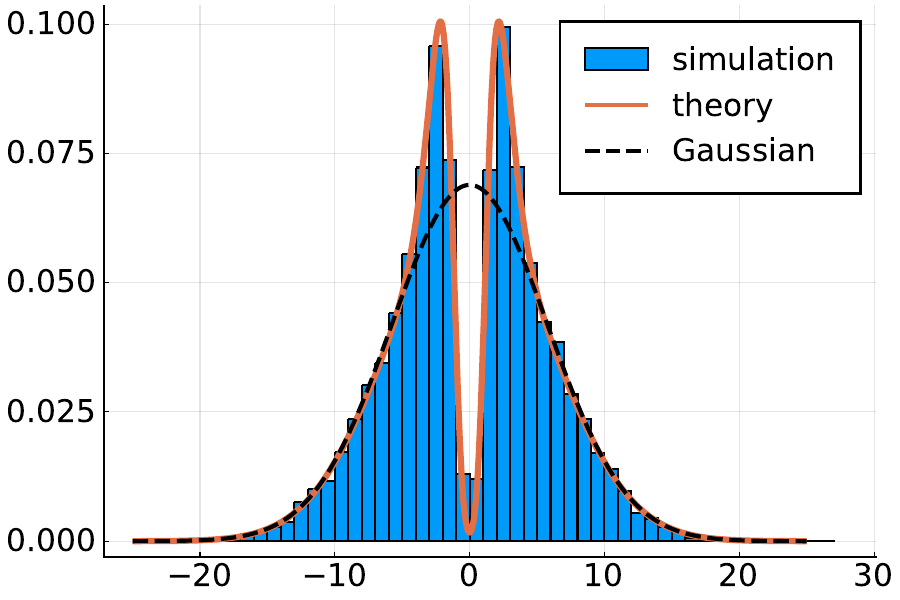}
\par\end{centering}
}
\par\end{centering}
\begin{centering}
\subfigure[$\lambda=1$]{\begin{centering}
\includegraphics[scale=0.32]{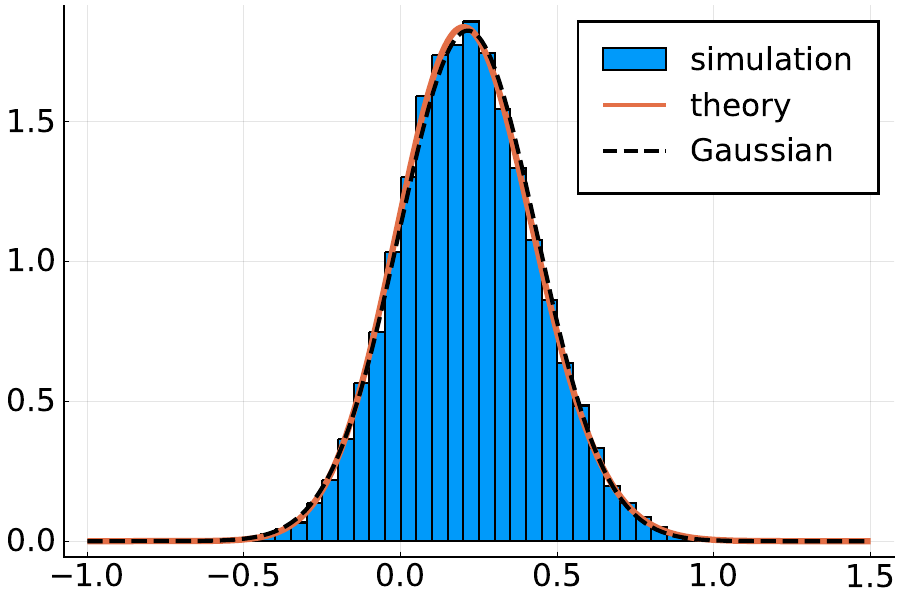}
\par\end{centering}
}\hphantom{}\subfigure[$\lambda=10^{-2}$]{\begin{centering}
\includegraphics[scale=0.32]{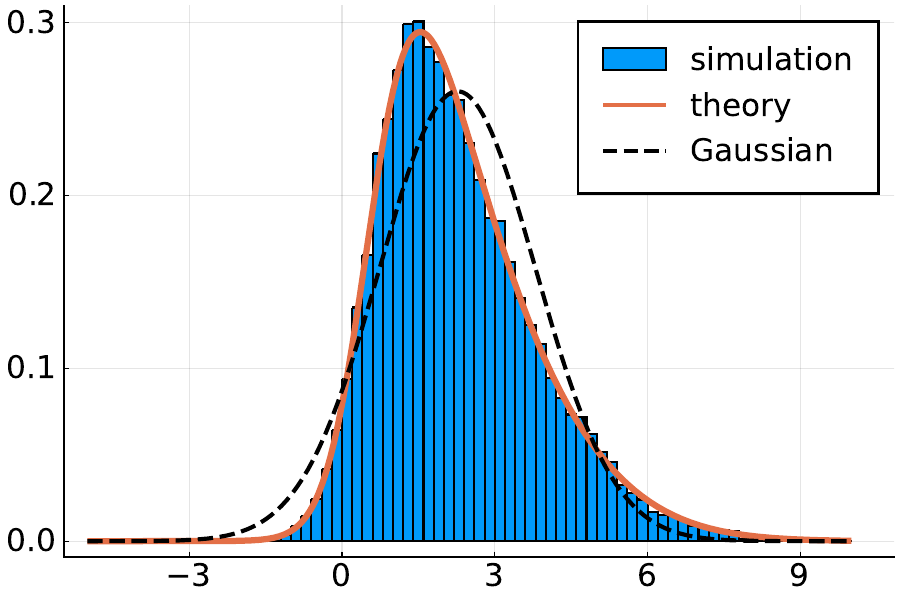}
\par\end{centering}
}\hphantom{}\subfigure[$\lambda=10^{-3}$]{\begin{centering}
\includegraphics[scale=0.32]{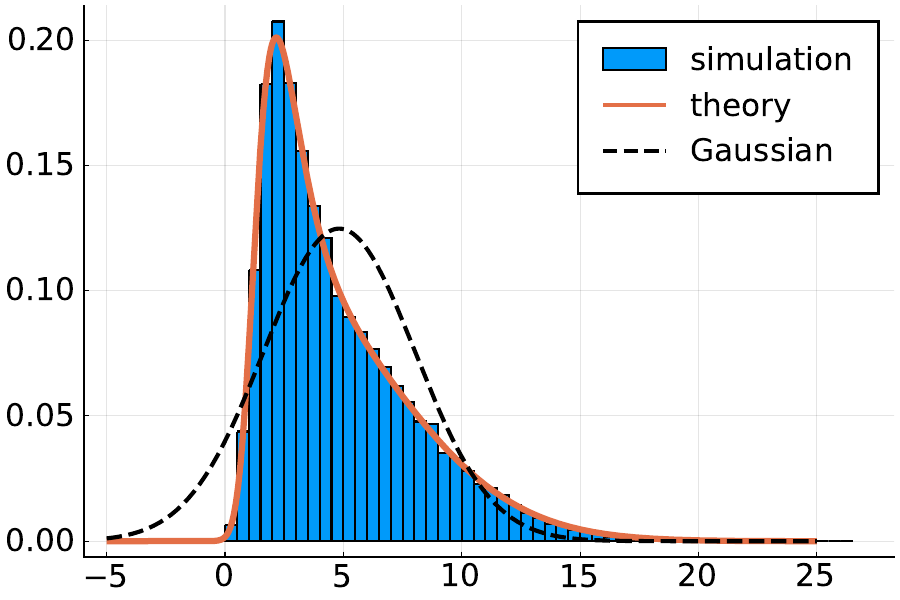}
\par\end{centering}
}
\par\end{centering}
\caption{Empirical distributions of $\scp{\va\tran_k \wt_{k}}$,
{[}(a)-(c){]} and $\scp{y_k\va\tran_k \wt_{k}}$,
{[}(d)-(f){]}. Here $\kernel(x) = \tanh(x)$, $\ell(x; y)$ is the logistic loss, $h(x)=\tfrac{\lambda}{2}x^2$, and $\fteacher(x) = \fout(x) = \sign(x)$. We fix $d=600$, $p=900$ and $n=1800$, while considering three different values of $\lambda$. The histograms are plotted on the values
of $\{\scp{\va\tran_k \wt_{k}}\}_{k\in[n]}$ and $\{\scp{y_k\va\tran_k \wt_{k}}\}_{k\in[n]}$
from 10 independent runs.  The dashed lines show Gaussian PDFs with the same empirical means and variances of the histograms. Observe that the empirical distributions of $\{\scp{\va\tran_k \wt_{k}}\}_{k\in[n]}$ and $\{\scp{y_k\va\tran_k \wt_{k}}\}_{k\in[n]}$ are not Gaussian, and the difference becomes increasing noticeable as $\lambda$ becomes smaller. The correct limit distributions are obtained by using Lemma \ref{lem:quadapprox_optsol} and Lemma \ref{lem:qwt_wt_deterministic_bd}.\label{fig:Empirical-distributions-of}}
\end{figure*}

Before proceeding to the proof of Theorem~\ref{thm:get}, we pause and point out a subtle issue regarding the central limit theorem stated informally in \eref{clt_informal}. It is important that the weight vector in \eref{clt_informal} is the leave-one-out solution $\loowt{\bs k}$, which is independent of both $\va_k$ and $\vb_k$. The situation will be very different if we use the original optimal solution $\wt_{k}$ instead.
In this case, the asymptotic distribution of $\scp{\va\tran_k \wt_{k}}$ is not Gaussian (\emph{i.e.}, the central limit theorem is no longer valid), due to the weak yet non-negligible correlation between $\wt_{k}$ and $\va_k$.We illustrate this fact in Fig. \ref{fig:Empirical-distributions-of}. The theoretical prediction of the limit distributions shown in the figure can be found by using Lemma \ref{lem:quadapprox_optsol} and Lemma \ref{lem:qwt_wt_deterministic_bd} in Appendix \ref{appendix:opt_properties}.

\subsection{Proof of Theorem~\ref{thm:get}}
\label{sec:get}

Equipped with \eref{test_func}, we just need to construct a suitable test function in order to complete the proof. For any fixed $\veps>0$ and $c$, let
\begin{equation}
\test_{\veps}(x)=(\charfn_{|x|\geq3\veps/2}*\mol_{\veps/2})(x-c),\label{eq:smoothed_test_fn}
\end{equation}
where $\mol_{\veps/2}(x)$ is a scaled mollifier defined in \eref{sc_mol} in Appendix \ref{appendix:truncation}. By properties of $\mol_{\veps/2}(x)$, it is easy to check that $\norm{\test_{\veps}'}_{\infty}<{C}/{\veps}$ and $\norm{\test_{\veps}''}_{\infty}<{C}/{\veps^{2}}$. Moreover,
\begin{equation}\label{eq:smoothed_test_fn_bd}
\charfn_{|x-c|\geq 2\veps}\leq\test_{\veps}(x)\leq\charfn_{|x-c|\geq\veps}.
\end{equation}
Letting $x = \Phi_{\mA}/p$ and taking expectation over the functions in \eref{smoothed_test_fn_bd}, we have
\[
P(\abs{\Phi_{\mA}/p-c}\ge2\varepsilon)\le\E\test_{\veps}(\Phi_{\mA}/p).
\]
Changing $x$ to $\Phi_{\mB}/p$ yields
\[
\E\test_{\veps}(\Phi_{\mB}/p) \le P(\abs{\Phi_{\mB}/p-c}\ge \varepsilon).
\]
Applying \eref{test_func}, we then have
\begin{align*}
    P(\abs{\Phi_{\mA}/p-c}\ge2\varepsilon)\le P(\abs{\Phi_{\mB}/p-c}\ge \varepsilon) &\\
&\hspace{-8em}+ \max\Big\{\veps, 1, \frac{1}{\veps\sqrt{p}}\Big\} \frac{\polylog p}{\veps\sqrt{p}},
\end{align*}
which leads to \eref{get} for $\veps \in (0, 1)$ and $p\geq \tfrac{1}{\veps^2}$. The proof of \eref{get2} is analogous, as the above procedure is completely symmetric with respect to $\Phi_{\mA}$ and $\Phi_{\mB}$.

\subsection{Proof of Proposition~\ref{prop:gen}}
\label{sec:proof_gen}

Let $\cvec_\text{new} \sim \mathcal{N}(0, \mI_d)$ be a Gaussian vector independent of the existing training samples and the feature matrix. Substituting \eref{teacher} into \eref{gen_err}, we can then write the generalization errors as
\[
\gerrA = \Eone{\cvec_\text{new}}[\fteacher(\cvec_\text{new}\tran \sgl) - \fout(\scp \va_\text{new}\tran \wt_{\mA})]^2
\]
and
\[
\gerrB = \Eone{\cvec_\text{new}}[\fteacher(\cvec_\text{new}\tran \sgl) - \fout(\scp \vb_\text{new}\tran \wt_{\mB})]^2,
\]
respectively. Here, $\va_\text{new} = \kernel(\fmtx\tran \cvec_\text{new})$ and $\vb_\text{new} = \mu_1 \fmtx\tran \cvec_\text{new} + \mu_2 \vz_\text{new}$, where $\vz_\text{new} \sim \mathcal{N}(0, \mI_p)$ is an independent Gaussian vector. Note that $(\cvec_\text{new}\tran \sgl, \scp \vb_\text{new}\tran \wt_{\mB})$ are jointly Gaussian, and thus their distributions are completely determined by their covariance matrix. As $\norm{\sgl} = 1$, we have $\E (\cvec_\text{new}\tran \sgl)^2 = 1$. Let $\genvar_B \bydef \E (\scp \vb_\text{new}\tran \wt_{\mB})^2$ and $\gencor_B \bydef \E (\cvec_\text{new}\tran \sgl) (\scp \vb_\text{new}\tran \wt_{\mB})$. Clearly,
\begin{equation}\label{eq:genvar}
\genvar_B = \frac{[\wt_{\mB}]\tran \mSig \wt_{\mB}}{p} \quad \text{and} \quad \gencor_B = \frac{\mu_1 \sgl\tran \fmtx  \wt_{\mB}}{\sqrt{p}},
\end{equation}
where $\mSig$ is the matrix in \eref{Sig}. It is also easy to check that $\gerrB = G(\genvar_B, \gencor_B)$, where
\begin{equation}\label{eq:G_func}
G(\genvar, \gencor) \bydef \E_{z_1, z_2} [\fteacher(z_1) - \fout(\gencor z_1 + [{\genvar - \gencor^2}]^{1/2} z_2)]^2,
\end{equation}
with $z_1, z_2 \overset{\text{i.i.d.}}{\sim}{\mathcal{N}(0, 1)}$.

The rest of the proof falls naturally into three parts: (a) We will first show that $\genvar_B \to \genvar^\ast = \frac{\partial}{\partial \tau_1} q^\ast(0, 0)$ and $\gencor_B \to \gencor^\ast = \frac{\partial}{\partial \tau_2} q^\ast(0, 0)$, where $q^\ast(\tau_1, \tau_2)$ is the limit function in Assumption~\ref{a:gen}; (b) By replacing $\wt_{\mB}$ in \eref{genvar} with $\wt_{\mA}$, we introduce the analogous quantities $\genvar_A$ and $\gencor_A$. We will show that $\genvar_A, \gencor_A$ have the same limits as $\genvar_B, \gencor_B$; (c) Finally, we will show that $\gerrA \approx G(\genvar_A, \gencor_A)$ with high probability, where $G(\cdot, \cdot)$ is the function in \eref{G_func}.

We start with part (a). By the definition of the optimization problem in \eref{gen_opt}, we have
\[
\Phi_{\mB}(\tau_1, \tau_2) \le \Phi_{\mB}(0, 0) + \tau_1 ([\wt_{\mB}]\tran \mSig \wt_{\mB}) + \tau_2 (\sqrt{p} \mu_1 \sgl\tran \fmtx \wt_{\mB})
\]
for any $\tau_1, \tau_2$. It follows that, for any $\tau > 0$,
\begin{equation}\label{eq:gen_bnds}
\frac{\Phi_{\mB}(\tau, 0) - \Phi_{\mB}(0, 0)}{p\tau} \le \genvar_B \le \frac{\Phi_{\mB}(-\tau, 0) - \Phi_{\mB}(0, 0)}{-p\tau}.
\end{equation}
Fix $\varepsilon > 0$. By Assumption~\ref{a:gen}, the limit function $q^\ast(\tau_1, \tau_2)$ is differentiable at the origin. Thus, there is some $\delta > 0$ such that
\[
\Big\vert\frac{q^\ast(\delta, 0) - q^\ast(0, 0)}{\delta} - \genvar^\ast\Big\vert \le \varepsilon / 3.
\]
The first inequality in \eref{gen_bnds}, with $\tau$ substituted by $\delta$, then gives us
\begin{align}
&\P(\genvar_B  - \genvar^\ast < -\varepsilon) \le \P\Big(\frac{\Phi_{\mB}(\delta, 0) - \Phi_{\mB}(0, 0)}{p\delta} - \genvar^\ast < -\varepsilon\Big)\nonumber\\
&\qquad\le \P(\absb{\Phi_{\mB}(\delta, 0)/p - q^\ast(\delta, 0)} > \delta \varepsilon / 3) \nonumber\\
&\hspace{6em}+ \P(\absb{\Phi_{\mB}(0, 0)/p - q^\ast(0, 0)} > \delta \varepsilon / 3).\label{eq:genvar_bnd}
\end{align}
By our assumption, $\Phi_{\mB}(\delta, 0)/p \cip q^\ast(\delta, 0)$ and $\Phi_{\mB}(0, 0)/p \cip q^\ast(0, 0)$. It then follows from \eref{genvar_bnd} that $\lim_{p \to \infty} \P(\genvar_B  - \genvar^\ast < -\varepsilon) = 0$. The same reasoning, applied to the second inequality in \eref{gen_bnds}, will give us $\lim_{p \to \infty} \P(\genvar_B  - \genvar^\ast > \varepsilon) = 0$, and thus $\genvar_B \cip \genvar^\ast$. The proof that $\gencor_B \cip \gencor^\ast$ is completely analogous and it is omitted.

Next, we move on to part (b) and establish the limits for $\genvar_A$ and $\gencor_A$. This is easy, in light of the universality laws given by Theorem~\ref{thm:get}. Specifically, \eref{get_cip} gives us $\Phi_{\mA}(\tau_1, \tau_2) / p \cip q^\ast(\tau_1, \tau_2)$. Replicating the same steps in part (a), with $B$ replaced by $A$, allows us to conclude that
\begin{equation}\label{eq:genA_sig_lim}
\genvar_A \cip \genvar^\ast \quad \text{and} \quad \gencor_A \cip \gencor^\ast.
\end{equation}
We can also show the function $G(\genvar, \gencor)$ is continuous at any point $(\genvar, \gencor)$ satisfying $\genvar \geq \gencor^2$ and $\rho\neq 0$. Let $z_1,z_2\iid \mathcal{N}(0,1)$ and $\{(\genvar_k,\gencor_k)\}_{k\geq 1}$ be a sequence converging to $(\genvar, \gencor)$, with $\genvar_k \geq \gencor_k^2$. Correspondingly, define $X_k:=[\fteacher(z_1) - \fout(\gencor_k z_1 + [{\genvar_k - \gencor^2_k}]^{1/2} z_2)]^2$ and $X:=[\fteacher(z_1) - \fout(\gencor z_1 + [{\genvar - \gencor^2}]^{1/2} z_2)]^2$. By Assumption \ref{a:fout}, $\fout$ is continuous almost everywhere, so if $\genvar \geq \gencor^2$ and $\rho\neq 0$, we can get $X_k\asconv X$, where $\asconv$ denotes almost sure convergence. On the other hand, since there exist some constants $C>0$ and $K_2\in\mathbb{Z}^+$ such that $\max\set{\abs{\fteacher(x)}, \abs{\fout(x)}} \le C(1 + \abs{x}^{K_2})$ by Assumption \ref{a:fout}, we have $|X_k|\leq C'(1+|z_1|^{2K_2}+|z_2|^{2K_2})$ for any $k\geq 1$, where $C'>0$ is a constant. Then by dominated convergence theorem, $G(\genvar_k, \gencor_k)=\E X_k \to \E X = G(\genvar, \gencor)$. This verifies the continuity of $G(\genvar, \gencor)$.
As a result,
\begin{equation}\label{eq:genAB_lim}
\gerrB \cip G(\genvar^\ast, \gencor^\ast) \quad \text{and} \quad G(\genvar_A, \gencor_A) \cip G(\genvar^\ast, \gencor^\ast).
\end{equation}

To complete the proof, we just need to establish part (c), namely, $\gerrA \approx G(\genvar_A, \gencor_A)$. To that end, we first write $\gerrA = \E_{\cvec_\text{new}}\btest(\scp \va_\text{new}\tran \wt_{\mA}, \cvec_\text{new}\tran \sgl)$, where
\[
\btest(x; s) \bydef (\fteacher(s) - \fout(x))^2.
\]
By Assumption~\ref{a:fout}, $\btest(x; s)$ is differentiable with respect to $x$ except at a finite number of points. Moreover, it is easy to check that
\[
\max\set{\abs{\btest(x; s)}, \abs{\btest'(x; s)}} \le C (1+\abs{s}^{2K_2})(1 + \abs{x}^{2K_2}),
\]
where $C > 0$ and $K_2\in \mathbb{Z}^+$ are the constants in Assumption~\ref{a:fout}. Our goal is to apply Proposition~\ref{prop:CLT_piecewise}, but we first need to put forth some additional restrictions. Let
\[
\mathcal{B} = \set{\norm{\wvv}_{\infty} \le \left(\log p\right)^{3+2K_{1}}},
\]
where $K_1$ is the constant in Assumption~\ref{a:loss} and
\[
\mathcal{C} = \set{\genvar_A = [\wt_{\mA}]\tran \mSig \wt_{\mA}/ p \ge \genvar^\ast/ 2}.
\]
Also recall the admissible set $\mathcal{A}$ defined in \sref{admissible}. We can verify that the assumptions of Proposition~\ref{prop:CLT_piecewise} (as stated and shown in \sref{CLT_disc}) hold for any $\fmtx \in \mathcal{A}$ and $\wvv = \wt_{\mA} \in \mathcal{B} \cap \mathcal{C}$. Thus, conditioned on $\mathcal{A} \cap \mathcal{B} \cap \mathcal{C}$, we can apply Proposition~\ref{prop:CLT_piecewise} to get
\begin{equation}\label{eq:gen_CLT}
\begin{aligned}
   &\abs{\E_{\cvec_\text{new}}\btest(\scp \va_\text{new}\tran \wt_{\mA}, \cvec_\text{new}\tran \sgl) - \E_{\cvec_\text{new}}\btest(\scp \vb_\text{new}\tran \wt_{\mA}, \cvec_\text{new}\tran \sgl)} \\
   \le& \frac{\polylog p}{p^{1/8}}. 
\end{aligned}
\end{equation}

Observe that $\E_{\cvec_\text{new}}\btest(\scp \vb_\text{new}\tran \wt_{\mA}, \cvec_\text{new}\tran \sgl) = G(\genvar_A, \gencor_A)$. Fix $\varepsilon > 0$.  For all sufficiently large $p$, we have ${\polylog p}/({p^{1/8}}) \le \varepsilon$. It then follows from \eref{gen_CLT} that
\begin{align}
&\P(\absb{\gerrA - G(\genvar_A, \gencor_A)} > \varepsilon) \nonumber \\
\le& \P(\mathcal{A}^c) + \P(\mathcal{B}^c) + \P(\mathcal{C}^c)\nonumber\\
\le& C e^{-C (\log p)^2} + \P(\absb{\genvar_A - \genvar^\ast} \ge \genvar^\ast / 2),\label{eq:genA_lim}
\end{align}
where to reach the last inequality we have used the probability estimates in \eref{A_small_prob} [for $\P(\mathcal{A}^c)$] and Lemma~\ref{lem:loowt_l_infty_bd} in Appendix~\ref{appendix:linfty_boundedness} [for $\P(\mathcal{B}^c)$]. Combining \eref{genA_lim}, \eref{genAB_lim}, and \eref{genA_sig_lim}, we complete the proof.

%% file: clt_new.tex
\section{A Central Limit Theorem for the Feature Model}
\label{sec:CLT}

In this section, we prove a central limit theorem (CLT) related to the nonlinear feature model. Let
\begin{equation}\label{eq:ab_vec}
\va = \kernel(\fmtx\tran \cvec) \quad \text{and} \quad \vb = \mu_1 \fmtx\tran \cvec+ \mu_2 \vz,
\end{equation}
where $\cvec \sim \mathcal{N}(0, \mI_d)$ and $\vz \sim \mathcal{N}(0, \mI_p)$ are two independent Gaussian vectors, $\fmtx = [\vf_1, \ldots, \vf_p]$ is a collection feature vectors in $\R^d$, and $\mu_1, \mu_2$ are constants as defined in \eref{mu_constants}. Given the teacher vector $\sgl$ in \eref{teacher} and a second vector $\wvv \in \R^{p}$, we show that
\begin{equation}\label{eq:CLT_approx}
\big(\scp{\va\tran \wvv}, \cvec\tran \sgl\big) \overset{\text{Law}}{\approx} \big(\scp{\vb\tran \wvv}, \cvec\tran \sgl\big)
\end{equation}
as $p \to \infty$. Here, we consider the setting where $\wvv, \sgl$ and the feature vectors are all \emph{deterministic}, and the only sources of randomness come from $\cvec$ and $\vz$. Thus, the right-hand side of \eref{CLT_approx} are just two jointly Gaussian random variables. 
\textcolor[rgb]{0.00,0.00,0.00}{
CLT in the form of \eref{CLT_approx} was first studied and proved in \cite{Goldt2020Gaussian} (see our discussions in Section \ref{subsec:relatedwork} and Remark \ref{rem:CLT} below).}
It will be useful in bounding the term $\Delta_\text{CLT}$ in \eqref{eq:Phi_k_decomp_2}, a critical step in our application of the Lindeberg method. It also plays an important role in our proof of Proposition~\ref{prop:gen}, where we establish the universality of the generalization error.

To state the theorem, we first need to put some restrictions on the feature vectors and the teacher vector $\sgl$. Let $\fvec_0 \bydef \sgl$, and let $\delta_{ij}$ denote the Kronecker delta function. We assume that
\begin{equation}\label{eq:BP}
\max_{0 \le i, j \le p} \abs{\vf_i\tran \vf_j - \delta_{ij}} \le \frac{\BP}{\sqrt{p}}
\end{equation}
for some $\BP = \mathcal{O}(p^{1/8 - \gamma})$ and $\gamma > 0$. Moreover,
\begin{equation}\label{eq:fnorm}
\norm{\fmtx} \le \polylog p.
\end{equation}
Note that, for the random feature vectors considered in this paper [see Assumption~\ref{a:feature} and the admissible condition in \eref{def_setA1}], the upper bound $\BP$ can actually be as small as $\polylog p$, and the spectral norm $\norm{\fmtx}$ can be set to be of $\mathcal{O}(1)$. However, since we believe that the central limit theorem could be of independent interest in other problems beyond this paper, we are going to prove it under the more relaxed assumption in \eref{BP}.

\begin{thm}\label{thm:CLT_general}
Suppose that the feature vectors satisfy \eref{BP} and \eref{fnorm}, and the activation function $\sigma(x)$ satisfies the conditions in Assumption~\ref{a:kernel}. Let $\set{\btest_p(x; s)}$ be a sequence of two-dimensional test functions that are differentiable with respect to $x$. Moreover, for each $p$,
\begin{equation}\label{eq:btest_condition}
\max\set{ \abs{\btest_p(x, s)}, \abs{\btest'_p(x, s)}} \le B_p(s)(1 + \abs{x}^K)
\end{equation}
for some constant $K \ge 1$ and some function $B_p(s)$. For any fixed vectors $\wvv \in \R^p$ and $\sgl \in \R^d$ with $\norm{\sgl} = 1$, it holds that
\begin{equation}\label{eq:CLT_general}
\begin{aligned}
    &\Big|\E \btest_p\big(\tfrac{1}{\sqrt{p}}\va\tran \wvv; \cvec\tran \sgl\big)-\E \btest_p\big(\tfrac{1}{\sqrt{p}}\vb\tran \wvv; \cvec\tran \sgl\big)\Big| \\
    &\hspace{6.5em}\le\frac{[\E B_p^4(z)]^{1/4}P({\wvv}, \BP)\polylog p}{\sqrt{p}},
\end{aligned}
\end{equation}
where $z\sim\mathcal{N}(0,1)$ and $P({\wvv}, \BP) = [1 + \norm{\wvv}_\infty (1+\BPf)][1 + (\scp\norm{\wvv})^{2K+1}]/\mu_2^2$.
\end{thm}
\begin{rem}
\label{rem:CLT}
We prove this theorem in \sref{CLT_general}, after first establishing two lemmas in \sref{CLT_1D} and \sref{CLT_cond}. 
\textcolor[rgb]{0.00,0.00,0.00}{As mentioned in Section \ref{subsec:relatedwork}},
a CLT in the form of \eref{CLT_approx} was first proved in \cite{Goldt2020Gaussian}. In principle, we could have adapted the proof there. However, as the CLT needs to be integrated with other components of our proof in \sref{Proof-Sketch}, we find it more convenient to derive an alternative proof, with a bound in \eref{CLT_general} that brings forth the explicit dependence of the approximation error on the $\ell_\infty$ norm of $\wvv$. The emphasis on $\norm{\wvv}_\infty$ is an important point. Later, when the CLT is applied [see \eref{Phi_k_decomp_2}], the vector $\wvv$ in \eref{CLT_general} will be $\loowt{\bs k}$, \emph{i.e.}, the leave-one-out optimal solution of \eqref{eq:LOO_objctive_function}. Showing that $\norm{\loowt{\bs k}}_\infty$ is bounded with high probability turns out to be a nontrivial challenge (see Lemma~\ref{lem:loowt_l_infty_bd} and Proposition~\ref{prop:wt_bnd}). 

The settings of the CLT shown in \cite{Goldt2020Gaussian} are also somewhat different from ours. On the one hand, the one in \cite{Goldt2020Gaussian} is more general in that it does not require the nonlinear activation function $\kernel(x)$ to be an odd function. On the other hand, Theorem~\ref{thm:CLT_general} is more relaxed in terms of the test function $\btest(x; s)$, which only needs to be differentiable with respect to the first variable $x$. In addition, we further relax this restriction in \sref{CLT_disc}, where a characterization similar to \eref{CLT_general} is given for piecewise differentiable test functions, at the cost of a slower decay rate than the right-hand side of \eref{CLT_general}. This extension will be needed when we study the universality of the generalization error in \eref{gen_err}. Finally, the new proof technique here, based on Stein's method \cite{stein1972bound,barbour2005introduction}, might be of interest in its own right.
\end{rem}


\subsection{A Reduced Form of Theorem~\ref{thm:CLT_general}}
\label{sec:CLT_1D}

\begin{lem}\label{lem:CLT_1D}
Consider a sequence of activation functions $\set{\kernel_p(x)}$ and differentiable test functions $\set{\btest_p(x)}$ such that, for every $p$,
\begin{enumerate}
\item $\kernel_p(x)$ is an odd function;
\item $\max\set{\norm{\sigma'_p(x)}_\infty, \norm{\sigma''_p(x)}_\infty, \norm{\sigma'''_p(x)}_\infty} \le \polylog p$;
\item $\kernel_p(x)$ is compactly supported. Specifically, there is some threshold $\tau_p \le \polylog p$ such that $\kernel_p(x) = 0$ for all $\abs{x} \ge \tau_p$;
\item $\max\set{\norm{\btest_p(x)}_\infty, \norm{\btest'_p(x)}_\infty} \le B_p$ for some $B_p < \infty$.
\end{enumerate}
For any fixed vector $\wvv \in \R^p$, it holds that
\begin{equation}\label{eq:CLT_1D}
\Big|\E \btest_p\Big(\frac{\va\tran \wvv}{\sqrt{p}}\Big)-\E \btest_p\Big(\frac{\vb\tran \wvv}{\sqrt{p}}\Big)\Big|\le \frac{B_p(1 + \BPf) \norm{\wvv}_\infty \polylog p }{\mu_{2,p}^2\sqrt{p}}.
\end{equation}
Here, $\va = \kernel(\fmtx\tran \cvec)$ and $\vb = \mu_{1,p} \fmtx\tran \cvec+ \mu_{2,p} \vz$, where $\cvec \sim \mathcal{N}(0, \mI_d)$ and $\vz \sim \mathcal{N}(0, \mI_p)$ are two independent Gaussian vectors, $\fmtx = [\vf_1, \vf_2, \ldots, \vf_p]$ is a collection of feature vectors satisfying \eref{BP} and \eref{fnorm}, and
\begin{equation}\label{eq:mu_12p}
\mu_{1, p} = \EE[z\sigma_p(z)], \qquad  \quad \mu_{2, p} = \sqrt{\EE \sigma^2_p(z) - \mu_{1, p}^2},
\end{equation}
with $z \sim \mathcal{N}(0, 1)$.
\end{lem}
\begin{rem}
Lemma~\ref{lem:CLT_1D} is essentially a reduced form of Theorem~\ref{thm:CLT_general}. The characterization in \eref{CLT_1D} guarantees that $\frac{\va\tran \wvv}{\sqrt{p}}$ has an asymptotical Gaussian law, whereas \eref{CLT_general} needs to consider the joint distribution of $\frac{\va\tran \wvv}{\sqrt{p}}$ and $\cvec\tran \sgl$. Moreover, Lemma~\ref{lem:CLT_1D} puts some further constraints on $\sigma_p(x)$ and $\btest_p(x)$, requiring the former to have compact supports and the latter to be bounded and to have bounded derivatives.
\end{rem}

\begin{proof}
To lighten the notation in the proof, we will omit the subscript $p$ in $\sigma_p(x)$ and $\btest_p(x)$. Also note that, if $\norm{\wvv} = 0$, the left-hand side of \eref{CLT_1D} is $0$; if $\mu_{2, p} = 0$, the right-hand side is $\infty$. In either case, \eref{CLT_1D} holds trivially. Therefore, we assume $\norm{\wvv} > 0$ and $\mu_{2, p} > 0$ in what follows.

Our proof is based on Stein's method \cite{stein1972bound,Chen2011normal}. We start by observing that $\frac{\vb\tran \wvv}{\sqrt{p}}$ is a Gaussian random variable with zero mean and variance
\begin{equation}\label{eq:nu}
\nu^2\bydef \wvv\tran \mSig_b \wvv/p \quad \text{where} \quad \mSig_b \bydef \mu_{1, p}^2 \fmtx\tran \fmtx + \mu_{2, p}^2 \mI.
\end{equation}
It follows that we can rewrite the left-hand side of \eref{CLT_1D} as
\begin{equation}\label{eq:CLT_1Da}
\Big|\E \btest\Big(\frac{\va\tran \wvv}{\sqrt{p}}\Big)-\E \btest\Big(\frac{\vb\tran \wvv}{\sqrt{p}}\Big)\Big| =
\Big|\E \btest\Big(\nu\frac{\va\tran \wvv}{\nu\sqrt{p}}\Big)-\E \btest(\nu z)\Big|
\end{equation}
for $z \sim \mathcal{N}(0, 1)$. Next, we introduce the following ``Stein transform'':
\[
\bts(x)\bydef e^{\frac{x^{2}}{2}}\int_{-\infty}^{x}e^{-\frac{y^{2}}{2}}\left[\btest\left(\nu y\right)-\E \btest\left(\nu z\right)\right]dy.
\]
Key to Stein's method is the following identity
\begin{equation}
\bts'(x)-x\bts(x)=\btest\left(\nu x\right)-\E \btest\left(\nu z\right),\label{eq:stein_identity}
\end{equation}
which can be directly verified from the definition of $\bts(x)$. Moreover, since $\norm{\btest'(x)}_\infty \le B_p$, we have from \cite[Lemma 2.4]{Chen2011normal} that
\begin{equation}
\max\set{\norm{\bts(x)}_\infty, \norm{\bts'(x)}_\infty, \norm{\bts''(x)}_\infty} \le 2\nu B_p.\label{eq:phi_deri_bd}
\end{equation}

In light of (\ref{eq:stein_identity}) and \eref{CLT_1Da}, showing \eref{CLT_1D} boils down to bounding $\left|\E\bts'\left(\frac{\va\tran \wvv}{\nu\sqrt{p}}\right)-\E\frac{\va\tran \wvv}{\nu\sqrt{p}}\bts\left(\frac{\va\tran \wvv}{\nu\sqrt{p}}\right)\right|$. To proceed, we define for every $(i, j)$,
\begin{equation}\label{eq:rhoij}
\rho_{ij}\bydef\frac{\fvec_{i}\tran\fvec_{j}}{\|\fvec_{i}\|^{2}}
\end{equation}
and
\[
\ajmi\bydef\kernel(\cvec\tran \fvec_j - \rho_{ij} \cvec\tran \fvec_i) = \kernel\left(\cvec\tran\left(\mI-\mP_i\right)\fvec_{j}\right),
\]
where $\mP_i = \frac{\fvec_{i}\fvec_{i}\tran}{\|\fvec_{i}\|^{2}}$ denotes the orthogonal projection onto the 1-D space spanned by $\fvec_i$. It is easy to
check that $a_{i}=\kernel(\cvec\tran \fvec_i)$ is independent of $\ajmi$ for all $j\neq i$. It follows that
\begin{equation}\label{eq:ai_aij}
\EE a_i \bts\Big(\tfrac{1}{\nu\sqrt{p}}\sum_{j \neq i} \ajmi\wv_j\Big) = \EE a_i \,\EE \bts\Big(\tfrac{1}{\nu\sqrt{p}}\sum_{j \neq i} \ajmi\wv_j\Big) = 0,
\end{equation}
where the last equality uses the assumption that $\E a_i = 0$ due to $\sigma(x)$ being an odd function. Applying \eref{ai_aij} and after some manipulations, we can verify the following decomposition:
\begin{equation}\label{eq:stein_identity_decomp}
\begin{aligned}
 &\E\Big[\frac{\va\tran \wvv}{\nu\sqrt{p}}\bts\Big(\frac{\va\tran \wvv}{\nu\sqrt{p}}\Big)\Big]-\E\bts'\Big(\frac{\va\tran \wvv}{\nu\sqrt{p}}\Big) \\
=&  \underbrace{\E\Big[\Big(\frac{1}{\nu\sqrt{p}}\sum_{i=1}^{p}\wv_{i}a_{i}\delta_{i}-1\Big)\bts'\Big(\frac{\va\tran \wvv}{\nu\sqrt{p}}\Big)\Big]}_{\text{(a)}} + \\
 & \underbrace{\E\Big\{ \sum_{i=1}^{p}\frac{\wv_{i}a_{i}}{\nu\sqrt{p}}\Big[\bts\Big(\frac{\va\tran \wvv}{\nu\sqrt{p}}\Big)-\bts\Big(\frac{\va\tran \wvv}{\nu\sqrt{p}}-\delta_{i}\Big)-\bts'\Big(\frac{\va\tran \wvv}{\nu\sqrt{p}}\Big)\delta_{i}\Big]\Big\} }_{\text{(b)}},
\end{aligned}
\end{equation}
where
\begin{equation}
\delta_{i}=\frac{\va\tran \wvv}{\nu\sqrt{p}}-\frac{\sum_{j\neq i}\ajmi \wv_{j}}{\nu\sqrt{p}}.\label{eq:delta_i_def}
\end{equation}
\textcolor[rgb]{0.00,0.00,0.00}{By Stein's identity, when  $\frac{\va\tran \wvv}{\nu\sqrt{p}}$ follows the standard Gaussian distribution, the left-hand side of \eqref{eq:stein_identity_decomp} exactly equals to zero. Intuitively, this quantity should be approximately equal to zero when $\frac{\va\tran \wvv}{\nu\sqrt{p}}$ is approximately standard Gaussian. This is what we are going to prove next.}
In what follows, we derive bounds for the two parts on the right-hand side of (\ref{eq:stein_identity_decomp}), separately.

We start with part (a). To simplify the notation, we let $\chi = \frac{1}{\nu\sqrt{p}}\sum_{i=1}^{p}\wv_ia_{i}\delta_{i}$. Applying the bound on $\left\Vert \bts'(x)\right\Vert _{\infty}$ in \eref{phi_deri_bd} gives us
\begin{align}
\abs{\text{part (a)}} &\le (2\nu B_p) \E\abs{\chi - 1}\nonumber\\
&\le (2\nu B_p) (\E\abs{\chi- \E\chi} + \abs{\E \chi -1})\nonumber\\
&\le (2\nu B_p) (\sqrt{\text{var}(\chi)} +\abs{\E \chi -1}),\label{eq:parta}
\end{align}
where the last step is due to H\"{o}lder's inequality. It is now clear what to do: to show $\text{part (a)} \to 0$, we just need to verify that $\E \chi \to 1$ and $\text{var}(\chi) \to 0$.

Calculating $\E \chi$ is easy. Applying the independence property \eref{ai_aij}, we have
\[
\E \chi = \frac{1}{\nu^2 p} \E \Big[\sum_{i\le p} \wv_i a_i (\va\tran \wvv)\Big] = \frac{1}{\nu^2 p} \wvv\tran \mSig_a \wvv,
\]
where $\mSig_a = \EE \va \va\tran$. One can show that $\mSig_a \approx \mSig_b$, where the latter is defined in \eref{nu}. Specifically, Lemma~\ref{lem:sigma_ab} in Appendix~\ref{appendix:covariance} gives us
\[
\begin{aligned}
    \norm{\mSig_a - \mSig_b} \le& \frac{(1 + \BPf + \norm{\mF}^4)  \polylog p}{\sqrt{p}} \\
    \le& \frac{(1 + \BPf)  \polylog p}{\sqrt{p}},
\end{aligned}
\]
with the second inequality due to \eref{fnorm}. Recall the definition of $\nu$ in \eref{nu}. We then have
\begin{align}
\abs{\E \chi - 1} &= \frac{\abs{\wvv\tran (\mSig_a - \mSig_b) \wvv}}{\nu^2 p} \le \frac{(1 + \BPf ) \polylog p}{\sqrt{p}} \Big(\frac{\norm{\wvv}^2}{\nu^2 p}\Big)\nonumber\\
&\le \frac{(1 + \BPf ) \norm{\wvv}_\infty \norm{\wvv} \polylog p}{\nu^2 p},\label{eq:Echi}
\end{align}
where in the last step we use a simple inequality ($\norm{\wvv}^2 \le \norm{\wvv}_\infty \norm{\wvv} \sqrt{p}$) to bring the final bound to a convenient form.

Next, we consider the variance term in \eref{parta}. Introducing the shorthand notation $u_k = \cvec\tran \fvec_k, 1 \le k \le p$, we rewrite $\delta_i$ in \eref{delta_i_def} as
\begin{align}
\delta_i =& \frac{\wv_i \kernel(u_i)}{\nu \sqrt{p}}  + \frac{1}{\nu \sqrt{p}} \sum_{j \neq i} \wv_j [\kernel(u_j) - \kernel(u_j - \rho_{ij} u_i)]
\nonumber\\
=& \frac{\wv_i \kernel(u_i)}{\nu \sqrt{p}} + \frac{1}{\nu \sqrt{p}} \sum_{j \neq i} \wv_j [\kernel'(u_j)u_i \rho_{ij} - \tfrac{1}{2}\kernel''(u_j)(u_i\rho_{ij})^2 \nonumber\\
&+ \tfrac{1}{6}\kernel'''(\theta_{ij})(u_i\rho_{ij})^3 )],\label{eq:delta_series}
\end{align}
where to reach the second equality we have used Taylor's expansion, with $\theta_{ij}$ denoting a point between $u_j - \rho_{ij} u_i$ and $u_j$. Substituting \eref{delta_series} into the expression for $\chi$ leads to
\begin{equation}\label{eq:X_decomp}
\begin{aligned}
\chi &= \Gamma + \Delta,
\end{aligned}
\end{equation}
where
\begin{equation}
\begin{aligned}
    \Gamma =& \frac{1}{\nu^{2}p}\sum_{i=1}^{p}\left[\wv_i \kernel(u_i)\right]^{2}+\frac{1}{\nu^{2}p}\sum_{i \neq j}\wv_i \wv_j \kernel(u_{i})\big[\kernel'(u_{j})u_{i}\rho_{ij} \\
    &-\frac{1}{2}\kernel''(u_{j})\left(u_{i}\rho_{ij}\right)^{2}\big]
\end{aligned}    
\end{equation}
and
\begin{equation}
    \Delta = \frac{1}{6\nu^{2}p}\sum_{i\neq j}\wv_i \wv_j \kernel(u_{i})\kernel'''(\theta_{ij})\left(u_{i}\rho_{ij}\right)^{3}.
\end{equation}

This then allows us to write
\begin{equation}\label{eq:chi_gamma_delta}
\begin{aligned}
    \sqrt{\text{var}(\chi)} &= \sqrt{\text{var}(\Gamma + \Delta)} \le \sqrt{2 \text{var}(\Gamma) + 2 \E [\Delta^2]} \\
    &\le \sqrt{2 \text{var}(\Gamma)} + \sqrt{2 \E[\Delta^2]}.
\end{aligned}
\end{equation}

The term involving $\Delta$ on the right-hand side of \eref{chi_gamma_delta} is easy to bound, even deterministically. Using our assumptions about the function $\sigma_p(x)$ stated in the lemma, namely it has a compact support and bounded third derivatives, we have $\abs{\kernel(u_i) u_i^3} \le \polylog p$ and $\abs{\kernel'''(\theta_{ij})} \le \polylog p$. In addition, since the feature vectors satisfy \eref{BP}, we can verify from the definition \eref{rhoij} that $\max_{i \neq j} \abs{\rho_{ij}} \le \frac{c \BP}{\sqrt{p}}$ for some constant $c$. It follows that
\begin{equation}\label{eq:Delta_bd}
\left|\Delta\right|  \leq\frac{\BPt \polylog p}{\nu^2p^{5/2}}\sum_{i \neq j} \abs{\wv_i \wv_j} \le \frac{\BPt \norm{\wvv}_\infty \norm{\wvv}\polylog p}{{\nu^2}p},\end{equation}
where the second inequality is due to the simple bound that $\sum_{i \neq j} \abs{\wv_i \wv_j} \le p \norm{\wvv}_\infty  \sum_i \abs{\wv_i} \le p^{3/2} \norm{\wvv}_\infty \norm{\wvv}$.

Now we tackle the more challenging task of bounding $\text{var}(\Gamma)$ in \eref{chi_gamma_delta}. We first note that, since $u_i = \cvec\tran \fvec_i$ and $u_j = \cvec\tran \fvec_j$, we can view $\Gamma$ as a differentiable function of $\cvec$, denoted by $\Gamma(\cvec)$, with $\cvec \sim \mathcal{N}(0, \mI_d)$. The Gaussian Poincar\'{e} inequality (see, \eg, \cite[Theorem 3.20]{boucheron2013concentration}) then gives us
\begin{equation}\label{eq:GP}
\text{var}(\Gamma(\cvec)) \le \E \norm{\nabla \Gamma(\cvec)}^2,
\end{equation}
where the gradient $\nabla \Gamma(\cvec)$ can be computed, with some diligence, as
\begin{align}
 &\nabla\Gamma(\cvec) \nonumber\\
 =& \frac{1}{\nu^{2}p}\Big(\sum_{i\le p}\wv_i^{2}q_1(u_i)\fvec_{i} + \sum_{i \neq j}[\wv_i q'_2(u_i)] [\wv_j q_3(u_j)] \rho_{ij}\fvec_i \nonumber \\ &\hspace{6em} + \sum_{i \neq j}[\wv_i q_2(u_i)] [\wv_j q'_3(u_j)] \rho_{ij}\fvec_j \nonumber \\
 &\hspace{6em} +\sum_{i \neq j}[\wv_i q'_4(u_i)] [\wv_j q_5(u_j)] \rho^2_{ij}\fvec_i \nonumber \\
 &\hspace{6em} + \sum_{i \neq j}[\wv_i q_4(u_i)] [\wv_j q'_5(u_j)] \rho^2_{ij}\fvec_j\Big),\label{eq:grad_gamma}
\end{align}
where $q_1(u) = 2 \kernel(u) \kernel'(u)$, $q_2(u) = \kernel(u) u$, $q_3(u) = \kernel'(u)$, $q_4(u) = -\tfrac{1}{2}\kernel(u) u^2$, and $q_5(u) = \kernel''(u)$.
In light of \eref{GP}, we just need to show that $\norm{\nabla\Gamma(\cvec)}$ is properly bounded. We do so by controlling the norm of each term on the right-hand side of \eref{grad_gamma}.

Note that our assumptions about the function $\sigma_p(x)$ implies that $\norm{q_i(u)}_\infty \le \polylog p$ and $\norm{q'_i(u)}_\infty \le \polylog p$ for $1 \le i \le 5$. Moreover, $\norm{\fmtx} \le \polylog p$ by assumption. Thus, the first term on the right-hand side of \eref{grad_gamma} can be bounded as
\begin{equation}\label{eq:grad_t1}
\Big\Vert{\frac{1}{\nu^2 p}\sum_{i\le p}\wv_i^{2}q_1(u_i)\fvec_{i}}\Big\Vert \le \frac{\norm{\wvv}_\infty \norm{\wvv}\polylog p}{\nu^2 p}.
\end{equation}
For the second term, we first rewrite it in the form of a matrix-vector multiplication as
\[
\frac{1}{\nu^{2}p} \sum_{i \neq j}[\wv_i q'_2(u_i)] [\wv_j q_3(u_j)] \rho_{ij}\fvec_i = \frac{1}{\nu^2 p} \fmtx\mD_1 \mM \mD_2 \wvv,
\]
where $\mD_1 = \diag\left\{ \wv_i q'_2(u_{i})\right\}$, $\mM=\diag\left\{ \|\fvec_{i}\|^{-2}\right\} \fmtx\tran\fmtx-\mI$, and $\mD_2 = \diag\left\{q_3(u_{j})\right\}$.  Clearly, $\norm{\mD_1} \le \norm{\wvv}_\infty \polylog p$ and $\norm{\mD_2} \le \polylog p$. We can also verify that
\begin{equation}\label{eq:rho_mtx}
\norm{\mM} \le c (\norm{\mF}^2 + 1) \le \polylog p.
\end{equation}
It follows that
\begin{equation}\label{eq:grad_t2}
\Big\Vert \frac{1}{\nu^2 p} \sum_{i \neq j}[\wv_i q'_2(u_i)] [\wv_j q_3(u_j)] \rho_{ij}\fvec_i \Big \Vert \le \frac{ \norm{\wvv}_\infty \norm{\wvv}\polylog p}{\nu^2 p}.
\end{equation}
Similarly, the fourth term on the right-hand side of \eref{grad_gamma} can be rewritten as $\frac{1}{\nu^2 p} \fmtx\widetilde{\mD}_1 \widetilde{\mM} \widetilde{\mD}_2 \wvv$, where $\widetilde{\mD}_1 = \diag\left\{ \wv_i q'_4(u_{i})\right\}$, $\widetilde{\mD}_2 = \diag\left\{q_5(u_{j})\right\}$, and $\widetilde{\mM} = \mM \circ \mM$, with $\circ$ denoting the Hadamard product of two matrices. The spectral norm of $\widetilde{\mM}$ can be bounded as
\begin{equation}\label{eq:rho2_mtx}
\norm{\widetilde{\mM}} \le \norm{\widetilde{\mM}}_\text{F} = \big[\textstyle\sum_{i \neq j} \rho_{ij}^4\big]^{1/2} \le c\BPs,
\end{equation}
for some constant $c$, where the last inequality is due to \eref{BP}. This then allows us to bound the norm of the fourth term of the gradient expression as
\begin{equation}\label{eq:grad_t3}
\begin{aligned}
    &\Big\Vert \frac{1}{\nu^2 \sqrt{p}} \sum_{i \neq j}[\wv_i q'_4(u_i)] [\wv_j q_5(u_j)] \rho^2_{ij}\fvec_i\Big\Vert \\
    \le& \frac{ \norm{\wvv}_\infty \norm{\wvv}\BPs\polylog p}{\nu^2 p}.
\end{aligned}
\end{equation}
The situations for the third and fifth term on the right-hand side of \eref{grad_gamma} are completely analogous, and thus we avoid the repetitions. With the bounds in \eref{grad_t1}, \eref{grad_t2} and \eref{grad_t3}, we can now apply \eref{GP} to get
\[
\sqrt{\text{var}(\Gamma)} \le \frac{(1+ \BPs) \norm{\wvv}_\infty \norm{\wvv}\polylog p}{\nu^2 p}.
\]
Combining this bound with those in \eref{Delta_bd}, \eref{chi_gamma_delta}, \eref{Echi}, we can retrace our steps back to \eref{parta} and conclude
\begin{equation}\label{eq:parta_bnd}
\abs{\text{part (a)}} \le   \frac{B_p(1 + \BPf )\norm{\wvv}_\infty\polylog p}{\mu_{2, p} \sqrt{p}},
\end{equation}
where the last inequality also uses the fact that $\mSig_b \succeq \mu_{2, p}^2 \mI$ and thus
\begin{equation}\label{eq:nu_bnd}
\nu \ge \mu_{2,p} \norm{\wvv}/\sqrt{p}.
\end{equation}

Now the remaining task is to bound the part (b) in \eref{stein_identity_decomp} before we can complete the proof. Using Taylor's expansion, we have
\[
\left|\text{part (b)}\right|  =\Big|\frac{1}{2\nu\sqrt{p}}\E\sum_{i=1}^{p}\wv_{i}a_{i}{\bts''(\theta_{i})}\delta_{i}^{2}\Big|
\]
where $\theta_i$ is some point between $\frac{\va\tran \wvv}{\nu\sqrt{p}} - \delta_i$ and $\frac{\va\tran \wvv}{\nu\sqrt{p}}$. By assumption, the function $\kernel(x)$ considered in this lemma is supported on $[-\tau_p, \tau_p]$ for some $\tau_p \le \polylog p$. We can then write $a_i = \kernel(u_i) = \kernel(u_i) \charfn_{[-\tau_p, \tau_p]}(u_i)$. This step of introducing an indicator function is not strictly necessary, but it helps to simplify some of our later arguments. We now have
\begin{align}
\left|\text{part (b)}\right| & =\Big|\frac{1}{2\nu\sqrt{p}}\E\sum_{i=1}^{p}\wv_{i}a_{i}{\bts''(\theta_{i})}\delta_{i}^{2}\charfn_{[-\tau_p, \tau_p]}(u_i)\Big|\nonumber \\
 & \leq\frac{B_p \polylog p}{\sqrt{p}}\linf{\wvv}\sum_{i=1}^{p}\E[\delta_{i}^{2}\charfn_{[-\tau_p, \tau_p]}(u_i)],\label{eq:part_b_bd}
\end{align}
where to reach the last inequality we have also used (\ref{eq:phi_deri_bd}) and the boundedness of $a_i = \kernel(u_i)$. Using a similar Taylor's expansion as in \eref{delta_series} but only to the second order, we have
\[
\begin{aligned}
    &\sum_{i=1}^{p}\E[\delta_{i}^{2}\charfn_{[-\tau_p, \tau_p]}(u_i)] \\
    =& \frac{1}{\nu^2 p}\sum_{i=1}^{p}\E\Big[\wv_{i}a_{i}+\sum_{j\neq i}\big(\kernel'(u_{j})\tilde{u}_{i}\rho_{ij}-\frac{1}{2}\kernel''(\theta_{ij})\left(\tilde{u}_{i}\rho_{ij}\right)^{2}\big)\wv_{j}\Big]^{2},
\end{aligned}
\]
where $\tilde{u}_i \bydef u_i \charfn_{[-\tau_p, \tau_p]}(u_i)$. Expanding the right-hand side of this expression then gives us
\begin{align}
&\abs{\text{part (b)}} \nonumber \\
\leq& \frac{B_p \norm{\wvv}_\infty \polylog p}{\nu^2 p^{3/2}}\Big(\norm{\wvv}^2 +\sum_{i=1}^{p}\E\big[\sum_{j\neq i}\kernel'(u_{j})\rho_{ij}\wv_{j}\big]^{2} \nonumber \\
&\hspace{9em}+\sum_{i=1}^{p}\E\big[\sum_{j\neq i}\kernel''(\theta_{ij})\rho^2_{ij}\wv_{j}\big]^{2}\Big)\nonumber\\
&\le \frac{B_p \norm{\wvv}_\infty \polylog p}{\nu^2 p^{3/2}} (\norm{\wvv}^2 + \E \norm{\mM \diag\set{\kernel'(u_j)} \wvv}^2 \nonumber \\
&\hspace{10em}+\E \norm{\widetilde{\mM} \widetilde{\wvv}}^2),
\end{align}
where $\mM, \widetilde{\mM}$ are the matrices considered in \eref{rho_mtx} and \eref{rho2_mtx}, respectively, and $\widetilde{\wvv} = [\abs{\wv_1}, \ldots, \abs{\wv_p}]\tran$. Using the spectral bounds given in \eref{rho_mtx} and \eref{rho2_mtx}, and the inequality \eref{nu_bnd}, we get
\[
\abs{\text{part (b)}} \le \frac{B_p(1 + \BPf) \norm{\wvv}_\infty \polylog p }{\mu_{2,p}^2\sqrt{p}}.
\]
Substituting this inequality and \eref{parta_bnd} into \eref{stein_identity_decomp}, and using the fact that $\mu_{2,p} \le \polylog p$, we are done.
\end{proof}

\subsection{Joint Distributions}
\label{sec:CLT_cond}

Lemma~\ref{lem:CLT_1D} shows that $\frac{\va\tran \wvv}{\sqrt{p}}$ has an asymptotically Gaussian distribution. Using this result, we can easily show that the asymptotic distribution of $\frac{\va\tran \wvv}{\sqrt{p}}$ and $\cvec\tran \sgl$ is jointly Gaussian, via a conditioning technique.

\begin{lem}\label{lem:CLT_2D}
Consider a sequence of activation functions $\set{\kernel_p(x)}$ and two-dimensional test functions $\set{\btest_p(x; s)}$ such that, for every $p$,
\begin{enumerate}
\item $\kernel_p(x)$ is an odd function;
\item $\max\set{\norm{\sigma'_p(x)}_\infty, \norm{\sigma''_p(x)}_\infty, \norm{\sigma'''_p(x)}_\infty} \le \polylog p$;
\item $\kernel_p(x)$ is compactly supported. Specifically, there is some threshold $\tau_p \le \polylog p$ such that $\kernel_p(x) = 0$ for all $\abs{x} \ge \tau_p$;
\item $\btest_p(x; s)$ is differentiable with respect to $x$. Moreover, there is a function $B_p(s)$ such that
\begin{equation}\label{eq:bps_bnd}
\max\set{\norm{\btest_p(x; s)}_\infty, \norm{\btest'_p(x; s)}_\infty} \le B_p(s).
\end{equation}
\end{enumerate}
For any fixed vectors $\wvv \in \R^p$ and $\sgl \in \R^d$ with $\norm{\sgl} = 1$, it holds that
\begin{equation}\label{eq:CLT_2D}
\begin{aligned}
    &\Big|\EE \btest_p\Big(\frac{\va\tran \wvv}{\sqrt{p}}; \cvec\tran \sgl\Big)-\EE \btest_p\Big(\frac{\vb\tran \wvv}{\sqrt{p}}; \cvec\tran \sgl\Big)\Big| \nonumber \\
    &\hspace{6em}\le \frac{[{\E B_p^2(z)}]^{1/2}(1 + \BPf ) \norm{\wvv}_\infty \polylog p }{\mu_{2,p}^2\sqrt{p}}.
\end{aligned}
\end{equation}
Here, $\vz \sim \mathcal{N}(0, 1)$, $\va, \vb$ are defined the same way as in Lemma~\ref{lem:CLT_1D}, and $\fmtx = [\vf_1, \vf_2, \ldots, \vf_p]$ is a collection of feature vectors satisfying \eref{BP} and \eref{fnorm}.
\end{lem}

\begin{proof}
To lighten the notation, we will omit the subscript $p$ in $\sigma_p(x)$ and $\btest_p(x; s)$ in the proof. The key idea in our proof is to rewrite the jointly Gaussian random variables $\cvec\tran \fmtx$ and $\cvec\tran \sgl$ via an equivalent representation through conditioning. It is easy to check that
\[
(\cvec\tran \fmtx, \cvec\tran \sgl) \overset{\text{Law}}{=} (s \sgl\tran \fmtx + \widetilde{\cvec}\tran (\mI - \sgl \sgl\tran) \fmtx, s),
\]
where $s \sim \mathcal{N}(0, 1)$ and $\widetilde{g} \sim \mathcal{N}(0, \mI_d)$ are two independent sets of Gaussian random variables. Let
\begin{equation}\label{eq:rho_f_tilde}
\rho_i \bydef \sgl\tran \fvec_i, \quad \widetilde{\vf}_i \bydef (\mI - \sgl \sgl\tran) \fvec_i, \quad \text{and} \quad \widetilde{u}_i = \widetilde{\cvec}\tran \widetilde{\fvec}_i.
\end{equation}
We can then redefine the entries of $\va$ and $\vb$ as
\begin{equation}\label{eq:ab_c}
a_i = \kernel(s\rho_i  + \widetilde{u}_i) \quad \text{and} \quad b_i = \mu_{1, p} (s\rho_i  + \widetilde{u}_i) + \mu_{2, p} z_i,
\end{equation}
without changing their probability distributions. The reason we do such decomposition is that $\widetilde{\cvec}\tran \widetilde{\fvec}_i$ is independent of $s$. This convenient independence structure allows us to calculate the expectations in \eref{CLT_2D} by first conditioning on $s$.

Applying Taylor's expansion to the expression for $a_i$ in \eref{ab_c}, we get
\begin{align}
a_{i} & =\kernel(\widetilde{u}_i) + \kernel'(\widetilde{u}_i) s \rho_i + \frac{1}{2}\kernel''\left(\theta_{i}\right)\left(s\rho_{i}\right)^{2}\nonumber \\
 & =\kernel(\widetilde{u}_i) + \mu_{1, p} s \rho_i + [\kernel'(\widetilde{u}_i) -\E\kernel'(\widetilde{u}_i)] (s \rho_i) \nonumber \\
 &~~~+ [\E\kernel'(\widetilde{u}_i) -\mu_{1,p}] (s \rho_i)+ \frac{1}{2}\kernel''\left(\theta_{i}\right)\left(s\rho_{i}\right)^{2},\label{eq:ai_sgl}
\end{align}
where $\theta_i$ is some point between $\widetilde{u}_i$ and $\widetilde{u}_i + s \rho_i$. This expansion then leads to
\[
\E\btest\Big(\frac{\va\tran \wvv}{\sqrt{p}}; \cvec\tran \sgl\Big) = \E\btest\Big(\frac{\sum_i \wv_i \kernel(\widetilde{u}_i)}{\sqrt{p}} +  \frac{s \mu_{1, p}\sum_i \wv_i \rho_i}{\sqrt{p}} + \Delta_1 + \Delta_2; s\Big),
\]
where
\begin{align*}
    \Delta_1 =  \frac{s \sum_i \wv_i \rho_i[\kernel'(\widetilde{u}_i) -\E\kernel'(\widetilde{u}_i)]}{\sqrt{p}}
\end{align*}
and
\begin{align*}
    \quad \Delta_2 = \frac{s \sum_i \wv_i \rho_i[\E\kernel'(\widetilde{u}_i) -\mu_{1,p} + \tfrac{1}{2} \kernel''(\theta_i) s\rho_i]}{\sqrt{p}}.
\end{align*}
Using the bounded derivative assumption in \eref{bps_bnd}, we have
\begin{equation}\label{eq:clt_D1D2}
\begin{aligned}
&\abs{\E\btest\Big(\frac{\va\tran \wvv}{\sqrt{p}}; \cvec\tran \sgl\Big) - \E\btest\Big(\frac{\sum_i \wv_i \kernel(\widetilde{u}_i)}{\sqrt{p}} +  \frac{s \mu_{1, p}\sum_i \wv_i \rho_i}{\sqrt{p}}; s\Big)}\\
&\qquad\qquad\qquad \le \E [B_p(s) (\abs{\Delta_1} + \abs{\Delta_2})] \nonumber\\
&\qquad\qquad\qquad \le [{\E B_p^2(s)}]^{1/2} ([\E \Delta_1^2]^{1/2} +[\E \Delta_2^2]^{1/2}).
\end{aligned}
\end{equation}
Next, we show that the terms involving $\Delta_{1}$ and $\Delta_{2}$ in \eref{clt_D1D2} are small.

The quantity $\E \Delta_1^2$ is small due to concentration. To see that, let $\Gamma_1(\widetilde{\cvec}) \bydef \big[\sum_i \wv_i \rho_i\kernel'(\widetilde{\cvec}\tran \widetilde{\fvec}_i)\big]/\sqrt{p}$. Clearly, $\Delta_1 = s [\Gamma_1(\widetilde{\cvec}) - \E \Gamma_1(\widetilde{\cvec})]$. From the independence of $s$ and $\widetilde{\cvec}$,
\begin{equation}\label{eq:clt_D1_GP}
\E \Delta_1^2 = \text{var}(\Gamma_1(\widetilde{\cvec})) \le \E \norm{\nabla \Gamma_1(\widetilde{\cvec})}^2,
\end{equation}
with the last step being the Gaussian Poincar\'{e} inequality. Recall the definition of $\rho_i$ and $\widetilde{\fvec}_i$ in \eref{rho_f_tilde}. One can verify that
\begin{align}
\norm{\nabla  \Gamma_1(\widetilde{\cvec})} &= \norm{(\mI-\sgl \sgl\tran) \fmtx \diag\set{\rho_i \kernel''(\widetilde{u}_i)} \wvv/\sqrt{p}}\nonumber\\
&\le {\BP (\norm{\wvv}/\sqrt{p}) \polylog p  }/{\sqrt{p}}\label{eq:clt_D1_bnd},
\end{align}
where to reach \eref{clt_D1_bnd} we have used the bound $\max_i \abs{\rho_i} \le \BP/\sqrt{p}$ due to \eref{BP}. Substituting \eref{clt_D1_bnd} into \eref{clt_D1_GP} then gives us
\begin{equation}\label{eq:D1_sq_bnd}
[\E {\Delta_1}^2]^{1/2} \le {\BP  (\norm{\wvv}/\sqrt{p}) \polylog p  }/{\sqrt{p}}.
\end{equation}

To bound $\E \Delta_2^2$, we first note that $\E\kernel'(\widetilde{u}_i) \approx \mu_{1,p}$. More precisely, a simple bound \eref{cov_bnd1} in Appendix~\ref{appendix:covariance} yields
\[
\begin{aligned}
\abs{\E\kernel'(\widetilde{u}_i) - \mu_{1,p}} &\le \polylog p \,\absb{\norm{\widetilde{\fvec}_i}^2 - 1}\\
&= \polylog p \abs{\norm{\fvec_i}^2 - 1 - (\sgl\tran \fvec_i)^2}\\
&\le \BP \polylog p / \sqrt{p}.
\end{aligned}
\]
This then gives us  
$$
\abs{\Delta_2} \le (s^2+\abs{s}) \BPs (\sum_i \abs{\wv_i} / p) \polylog p / \sqrt{p},
$$ and thus
\begin{equation}\label{eq:D2_sq_bnd}
[\E {\Delta_2}^2]^{1/2} \le \BPs (\norm{\wvv}/\sqrt{p}) \polylog p / \sqrt{p}.
\end{equation}
In light of \eref{D1_sq_bnd} and \eref{D2_sq_bnd}, the left-hand side of \eref{clt_D1D2} is well under control.

Using the equivalent representation for $\vb$ in \eref{ab_c}, we have
\[
\EE\btest\Big(\frac{\vb\tran \wvv}{\sqrt{p}}; \cvec\tran \sgl\Big) = \EE{\btest}_\text{shift}\Big(\frac{\sum_i \wv_i (\mu_{1,p} \widetilde{u}_i + \mu_{2, p}z_i)}{\sqrt{p}}; s\Big),
\]
where $\btest_\text{shift}(x; s) \bydef \btest(x + \frac{s \mu_{1, p}\sum_i \wv_i \rho_i}{\sqrt{p}}; s)$ is simply a shifted version of $\btest(x; s)$.
Combining this with \eref{clt_D1D2}, \eref{D1_sq_bnd} and \eref{D2_sq_bnd}, we can now bound the left-hand side (LHS) of \eref{CLT_2D} as
\begin{align}
&\text{LHS of } \eref{CLT_2D} \nonumber \\
\le& \Big|\EE \btest_\text{shift}\Big(\frac{\sum_i \wv_i \widetilde{a}_i}{\sqrt{p}}; s\Big)-\EE \btest_\text{shift}\Big(\frac{\sum_i \wv_i \widetilde{b}_i}{\sqrt{p}}; s\Big)\Big| + \Delta_3\nonumber\\
\le& \E \Big|\EE\big[\btest_\text{shift}(\textstyle\scp\sum_i \wv_i \widetilde{a}_i; s) \mid s\big] \nonumber \\
&\hspace{8em}-\EE\big[\btest_\text{shift}(\scp\sum_i \wv_i \widetilde{b}_i; s) \mid s\big]\Big| + \Delta_3,\label{eq:CLT_2D_cond}
\end{align}
where $\widetilde{a}_i = \kernel(\widetilde{\cvec}\tran \widetilde{\fvec}_i)$, $\widetilde{b}_i = \mu_{1, p} \widetilde{\cvec}\tran \widetilde{\fvec}_i + \mu_{2, p} z_i$, $\E[\cdot \vert s]$ denotes conditional expectation given $s$, and the ``remainder'' term is
\begin{align}
\Delta_3 &=  [{\E B_p^2(s)}]^{1/2}(\BPs + 1) (\norm{\wvv}/\sqrt{p}) \polylog p / \sqrt{p}\nonumber\\
&\le [{\E B_p^2(s)}]^{1/2}(\BPs + 1) \norm{\wvv}_\infty \polylog p / \sqrt{p}\label{eq:remainder}
\end{align}
Note that, for any fixed $s$, we can use Lemma~\ref{lem:CLT_1D} to control the conditional expectation in the first term on the right-hand side of \eref{CLT_2D_cond}. Indeed, with $s$ fixed, $\btest_\text{shift}(x; s)$ can be viewed as a one-dimensional test function and it satisfies all the assumptions stated in Lemma~\ref{lem:CLT_1D}. The only thing that is different here is that we are now using $\{\widetilde{\fvec}_i\}$ as the feature vectors. Thus, to apply Lemma~\ref{lem:CLT_1D}, we need to check that this modified set of feature vectors still satisfy the condition in \eref{BP}. But this is easy to do. Recall that  $\widetilde{\fvec}_i = (\mI - \sgl \sgl\tran) \vf_i$, with $\{\fvec_i\}$ satisfying \eref{BP} for some $\BP = \mathcal{O}(p^{1/8})$. Thus, for all $i, j$,
\[
\begin{aligned}
    \abs{\widetilde{\fvec}_i\tran \widetilde{\fvec}_j-\delta_{ij}} =& \abs{{\fvec_i\tran (\mI - \sgl \sgl\tran) \fvec_j}-\delta_{ij}} \\
    \le& \abs{\fvec_i\tran \fvec_j-\delta_{ij}} + \abs{\sgl\tran \fvec_i} \abs{\sgl\tran \fvec_j} \\
    \le& \frac{\BP}{\sqrt{p}} + \frac{\BPs}{p} \le \frac{c\BP}{\sqrt{p}}
\end{aligned}
\]
for some positive constant $c$. Finally, by substituting the bounds \eref{CLT_1D} [with $B_p$ there replaced by $B_p(s)$] and \eref{remainder} into \eref{CLT_2D_cond}, we reach the target inequality in \eref{CLT_2D}.
\end{proof}

\subsection{Proof of Theorem~\ref{thm:CLT_general}}
\label{sec:CLT_general}

To go from Lemma~\ref{lem:CLT_2D} to Theorem~\ref{thm:CLT_general}, we just need to remove the following two restrictions in the assumptions of Lemma~\ref{lem:CLT_2D}: (1) $\kernel(x)$ is compactly supported on $[-\tau_p, \tau_p]$ for some $\tau_p = \polylog p$; and (2) $\btest(x; s)$ and its derivatives are bounded [see \eref{bps_bnd}]. The main ingredient of our proof is to show, via a standard truncation technique, that the central limit theorem characterization still holds even if we relax these two assumptions.

Let $\btest(x; s)$ be a test function satisfying \eref{btest_condition}. We can construct a smoothly truncated version of this function via
\[
\widehat{\btest}_p(x; s) \bydef \btest(x; s) \strunc_{T_p, 1}(x),
\]
where $\strunc_{T_p, 1}(x)$ is the smooth window function defined in \eref{smooth_ind} in Appendix~\ref{appendix:truncation} and
\begin{equation}\label{eq:Tp}
T_p = (\norm{\fmtx} + 1) (\norm{\wvv}/\sqrt{p}) \sqrt{C_T\log p}
\end{equation}
for some positive constant $C_T$. The threshold $T_p$ in \eref{Tp} is chosen strategically. 
With this choice, we can show
\begin{equation}\label{eq:trunc_test_a_bnd}
\begin{aligned}
    &\E \big|\btest\big(\tfrac{1}{\sqrt{p}}\va\tran \wvv; \cvec\tran \sgl\big)- \widehat{\btest}_p\big(\tfrac{1}{\sqrt{p}}\va\tran \wvv; \cvec\tran \sgl\big)\big| \\
    \le& [\E B_p^4(z)]^{1/4}(1 + (\norm{\wvv}/\sqrt{p})^{K})\polylog p/\sqrt{p}.
\end{aligned}
\end{equation}
and
\begin{equation}\label{eq:trunc_test_b_bnd}
\begin{aligned}
    &\E \big|\btest\big(\tfrac{1}{\sqrt{p}}\vb\tran \wvv; \cvec\tran \sgl\big)- \widehat{\btest}_p\big(\tfrac{1}{\sqrt{p}}\vb\tran \wvv; \cvec\tran \sgl\big)\big| \\
    \le& [\E B_p^4(z)]^{1/4}(1 + (\norm{\wvv}/\sqrt{p})^{K})\polylog p/\sqrt{p}.
\end{aligned}
\end{equation}
The detailed proof of \eref{trunc_test_a_bnd} and \eref{trunc_test_b_bnd} are provided in Appendix \ref{subsec:theorem2proof_auxiliary}
 Together, \eref{trunc_test_a_bnd} and \eref{trunc_test_b_bnd} show that replacing the original test function $\btest(x; s)$ with its smoothly truncated approximation $\widehat{\btest}_p(x; s)$ only incurs a small price of $\mathcal{O}(\polylog p/\sqrt{p})$.

Next, we consider the activation function $\kernel(x)$. Using the smooth window function in \eref{smooth_ind} again, we can build a truncated approximation
\begin{equation}\label{eq:kernel_trunc}
\widehat{\kernel}_p(x) \bydef \kernel(x) \strunc_{\tau_p, 1}(x), \quad \text{where }\tau_p = \sqrt{2C_\tau \log p}
\end{equation}
for some positive constant $C_\tau$. It is easy to verify that $\widehat{\kernel}_p(x)$ satisfies all the assumptions stated in Lemma~\ref{lem:CLT_2D} concerning the activation functions. With this truncated activation function, define
\begin{equation}\label{eq:abh_vec}
\widehat{\va} \bydef \widehat{\kernel}_p(\cvec\tran \fmtx) \quad \text{and} \quad \widehat{\vb} \bydef \mu_{1, p} \cvec\tran \fmtx + \mu_{2, p} \vz
\end{equation}
as the counterparts of $\va$ and $\vb$ in \eref{ab_vec}. Here, $\mu_{1, p}, \mu_{2, p}$ are the constants defined in \eref{mu_12p}.
Our goal is to show that $\scp \va\tran \wvv \approx \scp \widehat{\va}\tran \wvv$ and $\scp \vb\tran \wvv \approx \scp \widehat{\vb}\tran \wvv$. 
Specifically, we can get (details are relegated to Appendix \ref{subsec:theorem2proof_auxiliary})
\begin{align}
&\E \big|\widehat{\btest}_p\big(\tfrac{1}{\sqrt{p}}\va\tran \wvv; \cvec\tran \sgl\big)- \widehat{\btest}_p\big(\tfrac{1}{\sqrt{p}}\widehat{\va}\tran \wvv; \cvec\tran \sgl\big)\big| \nonumber\\ 
\le&  [\E B_p^4(z)]^{1/4}[1 + (\scp\norm{\wvv})^{2K+1}]  {\polylog p}/{\sqrt{p}}.\label{eq:trunc_kernel_a}
\end{align}
and
\begin{align}
&\E \big|\widehat{\btest}_p\big(\tfrac{1}{\sqrt{p}}\vb\tran \wvv; \cvec\tran \sgl\big)- \widehat{\btest}_p\big(\tfrac{1}{\sqrt{p}}\widehat{\vb}\tran \wvv; \cvec\tran \sgl\big)\big| \nonumber \\
\le& [\E B_p^4(z)]^{1/4}[1 + (\scp\norm{\wvv})^{2K+1}]\frac{\polylog p}{\sqrt{p}}.\label{eq:trunc_kernel_b}
\end{align}
Given the inequalities in \eref{trunc_test_a_bnd}, \eref{trunc_test_b_bnd}, \eref{trunc_kernel_a} and \eref{trunc_kernel_b}, we have
\[
\begin{aligned}
&\Big|\E \btest\big(\tfrac{1}{\sqrt{p}}\va\tran \wvv; \cvec\tran \sgl\big)-\E \btest\big(\tfrac{1}{\sqrt{p}}\vb\tran \wvv; \cvec\tran \sgl\big)\Big| \nonumber\\ 
\le& [\E B_p^4(z)]^{1/4}[1 + (\scp\norm{\wvv})^{2K+1}]\frac{\polylog p}{\sqrt{p}}\\
&\qquad + \Big|\E \widehat{\btest}_p\big(\tfrac{1}{\sqrt{p}}\widehat{\va}\tran \wvv; \cvec\tran \sgl\big)-\E \widehat{\btest}_p\big(\tfrac{1}{\sqrt{p}}\widehat{\vb}\tran \wvv; \cvec\tran \sgl\big)\Big|.
\end{aligned}
\]
We can use Lemma~\ref{lem:CLT_2D} to bound the second term on the right-hand side, since its test function $\widehat{\btest}_p(x; s)$ and the activation function $\widehat{\kernel}_p(x)$ satisfy the assumptions stated in that lemma. Using \eref{CLT_2D} and the property that $\abs{\mu_2 - \mu_{2,p}} \le \polylog p / \sqrt{p}$, we reach the main result \eref{CLT_general} of the theorem.

\subsection{Extension to Piecewise Smooth Test Functions}
\label{sec:CLT_disc}

In what follows, we generalize Theorem~\ref{thm:CLT_general} to test functions that are only piecewise differentiable. This auxiliary result will be needed in our proof of Proposition~\ref{prop:gen} for the case where the ``output function'' $\fout(y)$ in the generalization error \eref{gen_err} lacks smoothness [\eg, $\fout(y) = \sign(y)]$.

\begin{prop}\label{prop:CLT_piecewise}
Consider the same assumptions of Theorem~\ref{thm:CLT_general} with ``$\btest_p(x; s)$ is differentiable with respect to $x$'' replaced by ``$\btest_p(x; s)$ differentiable with respect to $x$ except at a finite number of points $\set{x_1, x_2, \ldots, x_L}$''. Additionally, we also assume that
\begin{enumerate}
\item The upper bound $\BP \le \polylog p$ in \eref{BP}.

\item $\norm{\wvv}_\infty \le \polylog p$.

\item Let $\nu^2 = \wvv\tran \mSig \wvv / p$, where $\mSig$ is the covariance matrix in \eref{Sig}. Then $\nu^2 \ge c > 0$ for some constant $c$.
\end{enumerate}
It then holds that
\begin{equation}\label{eq:CLT_piecewise}
\begin{aligned}
    &\Big|\E \btest_p\big(\tfrac{1}{\sqrt{p}}\va\tran \wvv; \cvec\tran \sgl\big)-\E \btest_p\big(\tfrac{1}{\sqrt{p}}\vb\tran \wvv; \cvec\tran \sgl\big)\Big| \\
    \le& \frac{[\E B^4_p(z)]^{1/4}\polylog p}{p^{1/8}},
\end{aligned}
\end{equation}
where $z\sim\mathcal{N}(0,1)$.
\end{prop}
\begin{rem}
It is possible to improve the convergence rate on the right-hand side of \eref{CLT_piecewise} from $\mathcal{O}(p^{-1/8} \polylog p)$ to $\mathcal{O}(p^{-1/4} \polylog p)$, by requiring higher moments of $B_p(z)$ to be bounded. We do not pursue this optimization as the current form is sufficient for our proof of Proposition~\ref{prop:gen}.
\end{rem}
\begin{proof}
See Appendix \ref{subsec:Proposition3proof}
\end{proof}

%% file: conclusion.tex

\section{Conclusion and Final Remarks}
\label{sec:conclusion}

In this paper, we have proved the asymptotic equivalence of a nonlinear random feature model and a surrogate linear Gaussian models in terms of their training and generalization errors. As a consequence of this universality theorem, the learning performance of high-dimensional random feature models can be precisely characterized by studying their linear Gaussian counterparts, which are much more amenable to theoretical analysis. Our proof, which builds on the classical Lindeberg approach, makes several technical assumptions on the loss function, the nonlinear activation function, and the feature matrix. We close the paper by discussing how some of these assumptions can be further relaxed.

\emph{Non-differentiable loss functions.} In Assumption~\ref{a:loss}, we require the loss function $\loss(x; y)$ to have bounded third partial derivatives with respect to $x$. Many loss functions used in practice are not differentiable everywhere. A notable example is the hinge loss for binary classification, where $\loss(x; y) = \ell_\text{hinge}(yx)$ with $\ell_\text{hinge}(x) = \max(0, 1-x)$. One way to extend our current analysis to such non-differentiable functions is to consider a smoothed approximation via convolution. In the case of the hinge loss, let
\[
\ell_{\text{hinge},\delta}(x) =  \int_{\R} \ell_\text{hinge}(x-z)\mol_\delta(z) dz,
\]
where $\mol_\delta(z)$ is a scaled mollifier defined in \eref{sc_mol}. It is easy to verify that, for every $\delta > 0$, $\ell_{\text{hinge},\delta}(x)$ is convex, $\norm{\ell'''_{\text{hinge},\delta}(x)}_\infty \le C / \delta^3$, and
\begin{equation}\label{eq:hinge_smooth_bnd}
\norm{\ell_\text{hinge}(x) - \ell_{\text{hinge},\delta}(x)}_\infty \le C\delta,
\end{equation}
for some $C > 0$. Recall that $\Phi_{\mA}$ and $\Phi_{\mB}$ denote the training errors [\emph{i.e.},\ the minimum value of the optimization problem in \eref{gen_opt}] of the nonlinear feature model and the linear Gaussian model, respectively. We now use $\Phi_{\mA}^\delta$ and $\Phi_{\mB}^\delta$ to denote the corresponding quantities if we replace the hinge loss in \eref{gen_opt} by its smooth version $\ell_{\text{hinge},\delta}(x)$. It follows from \eref{hinge_smooth_bnd} that $\abs{\tfrac{1}{p}\Phi_{\mA} - \tfrac{1}{p}\Phi_{\mA}^\delta} \le C(n/p)\delta$ and $\abs{\tfrac{1}{p}\Phi_{\mB} - \tfrac{1}{p}\Phi_{\mB}^\delta} \le C(n/p)\delta$. The left-hand side of \eref{test_func} can now be bounded as
\begin{equation}\label{eq:test_hinge_smooth}
\begin{aligned}
    &\abs{\E \test(\tfrac{1}{p}{\Phi_{\mA}}) -  \E \test(\tfrac{1}{p}{\Phi_{\mB}})} \\
    &\quad\le 2C(n/p) \norm{\test'}_\infty \delta + \abs{\E \test(\tfrac{1}{p}{\Phi_{\mA}^\delta}) -  \E \test(\tfrac{1}{p}{\Phi_{\mB}^\delta})}.
\end{aligned}
\end{equation}
Since $\ell_{\text{hinge},\delta}(x)$ satisfies Assumption~\ref{a:loss}, we can apply our current analysis to bound the second term on the right-hand side of \eref{test_hinge_smooth}. We have the freedom in choosing the parameter $\delta$. Clearly, $\delta$ must go to $0$ as $p \to \infty$, but it cannot be too small. This is because $\norm{\ell'''_{\text{hinge},\delta}(x)}_\infty \le C / \delta^3$, and this bound on the third derivative is hidden in our estimates in Lemma~\ref{lem:qwt_wt_deterministic_bd}, Lemma~\ref{lem:optwt_qwt_prob_bd} and Lemma~\ref{lem:wk_qwtk_diff_moments_bd}. By choosing an optimal rate of decay for $\delta$, we can show that the left-hand side of \eref{test_hinge_smooth} tends to $0$ as $p$ grows, albeit with a slower convergence rate than that given in \eref{test_func}. Note that similar smoothing techniques can also be used to extend our analysis to non-differentiable activation functions and regularizers.

\emph{More general activation functions.} As a main limitation of our current work, we have assumed that the activation function $\kernel(x)$ is odd. Under this assumption, the regression vectors $\set{\va_t}$ in \eref{a_vec} and $\set{\vb_t}$ in \eref{b_vec} have zero mean and this simplifies our proof. As shown in \fref{quadratic}, the universality phenomenon holds under more general activation functions, including \emph{e.g.}, $\kernel(x) = \max(x, 0)$. One possible way to extend our work to such cases is the following. Let $\widetilde{\va}_t = \va_t - \mu_0 \vone$ and $\widetilde{\vb}_t = \vb_t - \mu_0 \vone$, where $\mu_0$ is the constant in \eref{mu_constants}. Then $\widetilde{\va}_t$ and $\widetilde{\vb}_t$ have (approximately) zero mean. We also rewrite the optimization problem in \eref{gen_opt} as an equivalent two stage process: $\Phi_{\mA}(\tau_1, \tau_2) = \min_{c \in \R} \Phi_{\mA}(c, \tau_1, \tau_2)$, where
\[
\begin{aligned}
    &\Phi_{\mA}(c, \tau_1, \tau_2) \\
    =& \inf_{\vone\tran \vw / \sqrt{p} = c} \big\{\textstyle\sum_{\didx=1}^{n} \loss(\tfrac{1}{\sqrt{p}}\widetilde{\va}_\didx\tran \vw + \mu_0 c; y_\didx) + \\
    &\qquad\qquad\quad \sum_{j = 1}^p \regu(w_j) + {\tau_1} (\vw\tran \mSig \vw) + \tau_2 (\sqrt{p} \mu_1\sgl\tran \fmtx \vw)\big\}.
\end{aligned}
\]
We can define $\Phi_{\mB}(c, \tau_1, \tau_2)$ in a similar way. Since $\E[\widetilde{\va}_t] \approx \mathbf{0}$ and $\E[\widetilde{\vb}_t] = \mathbf{0}$, it is not difficult to extend our current analysis to show that $\Phi_{\mA}(c, \tau_1, \tau_2)/p \approx \Phi_{\mB}(c, \tau_1, \tau_2) / p$. The remaining challenge is to show that this approximate equivalence holds uniformly over $c$, potentially by exploiting the convexity of the functions $\Phi_{\mA}(c, \tau_1, \tau_2)$ and $\Phi_{\mA}(c, \tau_1, \tau_2)$ with respect to $c$.

\emph{Deterministic feature matrices.} Yet another limitation of our work is that we have only considered cases where the columns of the feature matrix $\fmtx$ are independent Gaussian vectors. In fact, most of our technical results (such as those stated in \sref{Lindeberg}) have been obtained when we condition on a \emph{fixed} $\fmtx$ that satisfies \eref{BP} and \eref{fnorm}. The only place where we use the randomness of $\fmtx$ is in Lemma~\ref{lem:loowt_l_infty_bd} and Proposition~\ref{prop:wt_bnd}, where we show that the $\ell_\infty$-norm of the optimal weight vector $\wt_k$ is bounded by $\polylog p$ with high probability. This bound on the $\ell_\infty$-norm is needed in the central limit theorem stated in Theorem~\ref{thm:CLT_general}. [See, in particular, \eref{CLT_general}.] Thus, an important open problem is to check if $\norm{\wt_k}_\infty \le \polylog p$ with high probability for deterministic feature matrices that satisfy \eref{BP} and \eref{fnorm}.

%% file: appendix_truncation.tex

\subsection{Smoothing and Truncation}
\label{appendix:truncation}

In our proofs, we often need to apply smoothing and truncation to certain functions. This appendix collects the background and auxiliary results associated with such operations. First, we recall the construction of a standard mollifier
\[
\mol(x) = \begin{cases}
c \,e^{-1/(1-x^2)}, \quad &\text{if } \abs{x} < 1\\
0, \quad&\text{if } \abs{x} \ge 1,
\end{cases}
\]
where the constant $c$ ensures that $\int_\R \mol(x) dx = 1$. By construction, $\mol(x)$ is compactedly supported and nonnegative. It is also easy to show that $\mol(x)$ is infinitely differentiable and that
\begin{equation}\label{eq:mollifier_bnd}
\max\set{\norm{\mol(x)}_\infty, \norm{\mol'(x)}_\infty, \norm{\mol''(x)}_\infty} \le C
\end{equation}
for some numerical constant $C$. For each $\delta > 0$, we can rescale the mollifier as
\begin{equation}\label{eq:sc_mol}
\mol_\delta(x) = \delta^{-1} \mol(x/\delta)
\end{equation}
so that the resulting function is supported on $[-\delta, \delta]$. For any piecewise-smooth function $\genfn(x)$, we can obtain a smooth approximation by convolving it with a mollifier, \ie,
\[
\genfn_\delta(x) \bydef (\genfn \ast \mol_\delta)(x) = \int_{\R} \mol_\delta(x - y) \genfn(y) dy.
\]
A special case, frequently used in our proofs, is when $\genfn(x)$ is the indicator function defined on certain intervals. In particular, for $T > 0, \delta > 0$, we define
\begin{equation}\label{eq:smooth_ind}
\strunc_{T, \delta}(x) =  (\charfn_{[-T-\delta/2, T+\delta/2]} \ast \mol_{\delta/2})(x)
\end{equation}
as a smooth ``window function''. It is easy to check that $\strunc_{T, \delta}(x) = 1$ for $\abs{x} \le T$, $\strunc_{T, \delta}(x) = 0$ for $\abs{x} \ge T + \delta$, and $0 \le \strunc_{T, \delta}(x) \le 1$ for $x$ in the smooth ``transition bands''. Moreover, it follows from \eref{mollifier_bnd} that $\norm{\strunc'_{T, \delta}(x)}_\infty \le C/\delta$.

\begin{lem}\label{lem:piecewise_bnd}
Let $\genfn(x)$ be a function that is differentiable everywhere except at a finite number of points $\set{x_1, x_2, \ldots, x_L}$. If there is a function $B(x)$ such that
\begin{equation}\label{eq:smooth_bnd}
\abs{\genfn'(x)} \le B(x), \text{ for } x \not\in \set{x_1, x_2, \ldots, x_k} \text{ and } \abs{\genfn(x)} \le B(x)
\end{equation}
then for every $\delta > 0$,
\begin{equation}\label{eq:smooth_approx}
\abs{\genfn(x) - \genfn_\delta(x)} \le B_\delta(x) \delta + 2 B_\delta(x) \sum_{i = 1}^L \strunc_{2\delta, \delta}(x-x_i),
\end{equation}
where $B_\delta(x) \bydef \sup_{\abs{c} \le \delta}B(x+c)$ and $\strunc_{2\delta, \delta}(\cdot)$ is a smoothed window function as defined in \eref{smooth_ind}. Moreover,
\begin{equation}\label{eq:smooth_mol_bnd}
\abs{\genfn_\delta(x)} \le B_\delta(x) \quad \text{and} \quad \abs{\genfn'_\delta(x)} \le \frac{C B_\delta(x)}{\delta}
\end{equation}
for some numerical constant $C$.
\end{lem}
\begin{proof}
Let $\mathcal{D} = \cup_{1 \le i \le L} [x_i - 2\delta, x_i+2\delta]$. For any $x \not\in \mathcal{D}$, the function $\genfn(x)$ is differentiable on the interval $[x-\delta, x+\delta]$. For such $x$, we have
\begin{align}
\abs{\genfn(x) - \genfn_\delta(x)} &\overset{(a)}{=}  \abs{\int_{\abs{y-x} \le \delta} [\genfn(x) - \genfn(y)]  \mol_\delta(x-y) dy}\nonumber\\
&\le \int_{\abs{y-x} \le \delta} \abs{\genfn(x) - \genfn(y)}  \mol_\delta(x-y) dy\nonumber\\
&\overset{(b)}{\le} B_\delta(x)\delta,\label{eq:smooth_bnd2}
\end{align}
where $(a)$ uses the property that $\int_\R  \mol_\delta(x-y) dy = 1$, and $(b)$ is due to the intermediate value theorem and \eref{smooth_bnd}. For any $x \in \mathcal{D}$, we directly use the bound on $\genfn(x)$ to get
\begin{equation}
\abs{\genfn(x) - \genfn_\delta(x)} \le \int_{\abs{y-x} \le \delta} \abs{\genfn(x) - \genfn(y)}  \mol_\delta(x-y) dy \le 2 B_\delta(x).\label{eq:disc_bnd2}
\end{equation}

Combining \eref{smooth_bnd2} and \eref{disc_bnd2} gives us
\[
\begin{aligned}
\abs{\genfn(x) - \genfn_\delta(x)} &= \abs{\genfn(x) - \genfn_\delta(x)} \charfn_{\mathcal{D}^c}(x) +  \abs{\genfn(x) - \genfn_\delta(x)} \charfn_{\mathcal{D}}(x)\\
&\le B_\delta(x)\delta + 2B_\delta(x) \sum_{i=1}^L \charfn_{[x_i-2\delta, x_i+2\delta]}(x).
\end{aligned}
\]
The desired inequality in \eref{smooth_approx} then follows from the simple observation that $\charfn_{[x_i-2\delta, x_i+2\delta]}(x) \le \strunc_{2\delta, \delta}(x-x_i)$, which can be easily verified from the definition in \eref{smooth_ind}.

The first inequality in \eref{smooth_mol_bnd} is obvious. To get the second inequality, we have
\[
\genfn'_\delta(x) = \int_{\abs{y-x} \le \delta} \genfn(y) \frac{1}{\delta^2} \mol'\big(\frac{x-y}{\delta}\big) dy \le  \frac{B_\delta(x)}{\delta} \norm{\mol'(x)}_\infty,
\]
and this completes the proof.
\end{proof} 
\subsection{Auxiliary Results for the Proof of Theorem~\ref{thm:CLT_general}\label{subsec:theorem2proof_auxiliary}}
\textbf{1. Proof of \eref{trunc_test_a_bnd} and \eref{trunc_test_b_bnd}.}


\textcolor[rgb]{0.00,0.00,0.00}{Let $\mathcal{B}$ be the event that $\set{\abs{\scp \va\tran \wvv} \le T_p}$. Applying Lemma~\ref{lem:linear_concentrate} in Appendix~\ref{appendix:Concentration_Lipschitz}, we get $\P(\mathcal{B}) \ge 1- 2 p^{-C_T / C}$, where $C > 0$ is some fixed numerical constant. Thus, by using a sufficiently large $C_T$, we have
\begin{equation}\label{eq:PB_bnd}
\P(\mathcal{B}) \ge 1- 2 /p.
\end{equation}
The standard trick in a truncation method is to introduce two indicator functions defined on $\mathcal{B}$ and $\mathcal{B}^c$, respectively. Since $\big|\btest\big(\tfrac{1}{\sqrt{p}}\va\tran \wvv; \cvec\tran \sgl\big) - \widehat{\btest}_p\big(\tfrac{1}{\sqrt{p}}\va\tran \wvv; \cvec\tran \sgl\big)\big| \charfn_{\mathcal{B}} \equiv 0$,
\begin{align}
&\E \big|\btest\big(\tfrac{1}{\sqrt{p}}\va\tran \wvv; \cvec\tran \sgl\big)- \widehat{\btest}_p\big(\tfrac{1}{\sqrt{p}}\va\tran \wvv; \cvec\tran \sgl\big)\big| \nonumber\\
=&\E \Big[\big|\btest\big(\tfrac{1}{\sqrt{p}}\va\tran \wvv; \cvec\tran \sgl\big)- \widehat{\btest}_p\big(\tfrac{1}{\sqrt{p}}\va\tran \wvv; \cvec\tran \sgl\big)\big| \charfn_\mathcal{B}^c\Big]\nonumber\\
\le& 2\big[\E \btest^2\big(\tfrac{1}{\sqrt{p}}\va\tran \wvv; \cvec\tran \sgl\big)\big]^{1/2} \big[1-\P(\mathcal{B})\big]^{1/2},\label{eq:trunc_test_a}
\end{align}
where to reach \eref{trunc_test_a} we have used H\"{o}lder's inequality and the fact that $\abs{\btest(x; s)} \ge \abs{\widehat{\btest}_p(x; s)}$. To bound the first term on the right-hand side of \eref{trunc_test_a}, we can use \eref{btest_condition} and get
\begin{align}
\E \btest^2\big(\tfrac{1}{\sqrt{p}}\va\tran \wvv; \cvec\tran \sgl\big) &\le \EE B^2(\cvec\tran \sgl) (1 + \big|\tfrac{1}{\sqrt{p}}\va\tran \wvv\big|^K)^2\nonumber\\
&\hspace{-2em}\le 2\sqrt{2} \, [\E B_p^4(\cvec\tran \sgl)]^{1/2} [1+ \E (\tfrac{1}{\sqrt{p}}\va\tran \wvv)^{4K}]^{1/2}\nonumber\\
&\hspace{-2em}\le [\E B_p^4(z)]^{1/2}(1 + (\norm{\wvv}/\sqrt{p})^{2K})\polylog p,\label{eq:trunc_test_a_bnd0}
\end{align}
where the last inequality is obtained by using the moment estimate \eref{akw_bd_moments} in Lemma~\ref{lem:linear_concentrate}. Substituting \eref{PB_bnd} and \eref{trunc_test_a_bnd0} into \eref{trunc_test_a}, we can get \eref{trunc_test_a_bnd}.
The steps leading to \eref{trunc_test_b_bnd} are completely analogous to what we did to reach \eref{trunc_test_a_bnd}, so we omit the details here.}

\noindent\textbf{2. Proof of \eqref{eq:trunc_kernel_a} and \eqref{eq:trunc_kernel_b}.}

\textcolor[rgb]{0.00,0.00,0.00}{First, we prove \eqref{eq:trunc_kernel_a}.
Let
\begin{equation}\label{eq:D_event}
\mathcal{D} \bydef \big\{\max_{i \le p} \abs{\cvec\tran \fvec_i} \le \tau_p\big\}.
\end{equation}
By construction, $\va = \widehat{\va}$ when the event $\mathcal{D}$ holds. Next, we show that $\mathcal{D}$ is indeed a high-probability event. Recall that $\cvec\tran \fvec_i \overset{\text{Law}}{=} \norm{\fvec_i} z$ for $z \sim \mathcal{N}(0,1)$. Moreover, the condition in \eref{BP} implies that $\max_i \norm{\fvec_i}^2 \le C$ for some fixed constant $C$. A standard Gaussian tail bound $\P(\abs{z} \ge t) \le 2 e^{-t^2/2}$ then gives us
\begin{equation}\label{eq:D_bnd}
\begin{aligned}
    \P(\mathcal{D}^c) &\le \sum_{i\le p} \P\Big(\abs{z} \ge \frac{\tau_p}{\norm{\fvec_i}}\Big) \\
    &\le 2p e^{-\tau_p^2 / (2C)} \le 2 p^{-(C_\tau / C - 1)} \le 2/p
\end{aligned}
\end{equation}
for all sufficiently large $C_\tau$. [Without loss of generality, we should also assume that $C_\tau \ge 2$, as this is needed in the proof of an auxiliary result in Appendix~\ref{appendix:covariance}.] 
On the other hand, by the construction of $\strunc_{T_p, 1}(x)$ and the assumption in \eref{btest_condition}, we can easily verify that
\begin{equation}\label{eq:btest_trunc_unif_bnd}
\begin{aligned}
    &\max\set{\norm{\widehat{\btest}_p(x; s)}_\infty, \norm{\widehat{\btest}'_p(x; s)}_\infty} \\
    \le& \widehat{B}_p(s) \bydef \big[1 + (\scp\norm{\wvv})^{2K}\big] B_p(s) \polylog p,
\end{aligned}
\end{equation}
where $K$ is the constant in \eref{btest_condition}.
Then using the boundedness of $\widehat{\btest}'_p(x; s)$ given in \eref{btest_trunc_unif_bnd} and defining $\charfn_{\mathcal{D}^c}$ as the indicator function supported on $\mathcal{D}^c$, we have
\begin{align}
&\E \big|\widehat{\btest}_p\big(\tfrac{1}{\sqrt{p}}\va\tran \wvv; \cvec\tran \sgl\big)- \widehat{\btest}_p\big(\tfrac{1}{\sqrt{p}}\widehat{\va}\tran \wvv; \cvec\tran \sgl\big)\big| \nonumber \\
\le& \E \Big[\widehat{B}_p(\cvec\tran \sgl) \abs{\scp \va\tran \wvv - \scp \widehat{\va}\tran \wvv} \charfn_{\mathcal{D}^c}\Big]\nonumber\\
\overset{(a)}{\le}& 8^{1/4} [\E \widehat{B}_p^4(\cvec\tran \sgl)]^{1/4}\Big([\EE (\scp \va\tran \wvv)^4]^{1/4} \nonumber\\
&\hspace{10em}+ [\EE (\scp \widehat{\va}\tran \wvv)^4]^{1/4}\Big) \sqrt{\P(\mathcal{D}^c)}\nonumber\\
\overset{(b)}{\le}&  [\E B_p^4(z)]^{1/4}[1 + (\scp\norm{\wvv})^{2K}] (\scp\norm{\wvv}) {\polylog p}/{\sqrt{p}}\nonumber\\
\le&  [\E B_p^4(z)]^{1/4}[1 + (\scp\norm{\wvv})^{2K+1}]  {\polylog p}/{\sqrt{p}},
\end{align}
which is \eqref{eq:trunc_kernel_a}.
Here, (a) is based on a generalized H\"{o}lder's inequality: $\EE \abs{XYZ} \le (\EE X^4 \EE Y^4)^{1/4} (\EE Z^2)^{1/2}$. To reach (b), we use \eref{btest_trunc_unif_bnd} and the moment bound \eref{akw_bd_moments} in Lemma~\ref{lem:linear_concentrate}.}

\textcolor[rgb]{0.00,0.00,0.00}{Next we prove \eqref{eq:trunc_kernel_b}. It follows from the definition in \eref{abh_vec} that
\[
\begin{aligned}
\abs{\scp \vb\tran \wvv - \scp \widehat{\vb}\tran \wvv} &\le \absb{\mu_1 - \mu_{1, p}} \absb{\scp \cvec\tran \fmtx\wvv} \\
&\hspace{6em}+ \absb{\mu_2 - \mu_{2, p}} \absb{\scp \vz\tran \wvv}\\
&\le \frac{\polylog p}{\sqrt{p}}\big( \absb{\scp \cvec\tran \fmtx\wvv} + \absb{\scp \vz\tran \wvv}\big),
\end{aligned}
\]
where the last inequality uses the estimate given in Lemma~\ref{lem:mu_p_diff} in Appendix~\ref{appendix:covariance}. We now have
\begin{align}
&\E \big|\widehat{\btest}_p\big(\tfrac{1}{\sqrt{p}}\vb\tran \wvv; \cvec\tran \sgl\big)- \widehat{\btest}_p\big(\tfrac{1}{\sqrt{p}}\widehat{\vb}\tran \wvv; \cvec\tran \sgl\big)\big| \nonumber\\
\le&  \E\big[\widehat{B}_p(\cvec\tran \sgl)\big( \absb{\scp \cvec\tran \fmtx\wvv} + \absb{\scp \vz\tran \wvv}\big)\big]\frac{\polylog p}{\sqrt{p}}\nonumber\\
\le& [\E B_p^4(z)]^{1/4}[1 + (\scp\norm{\wvv})^{2K}]\Big(\sqrt{\E (\scp \cvec\tran \fmtx \wvv)^2} \nonumber \\
&\hspace{12em}+ \sqrt{\E (\scp \vz\tran \wvv)^2}\Big)\frac{\polylog p}{\sqrt{p}}\nonumber\\
\le& [\E B_p^4(z)]^{1/4}[1 + (\scp\norm{\wvv})^{2K+1}]\frac{\polylog p}{\sqrt{p}},
\end{align}
which is \eqref{eq:trunc_kernel_b}.
}

\subsection{Proof of Proposition~\ref{prop:CLT_piecewise}\label{subsec:Proposition3proof}}
\textcolor[rgb]{0.00,0.00,0.00}{For any ${\delta_p} \in (0, 1)$, let
\[
\btest_{\delta_p}(x; s) = \int \btest_p(y; s) \mol_{\delta_p}(x-y) dy
\]
be a smoothed version of the test function, where $\mol_{\delta_p}(x)$ is the mollifier introduced in Appendix~\ref{appendix:truncation}. The main idea of the proof is choosing a diminishing sequence of ${\delta_p}$ so that the left-hand side of \eref{CLT_piecewise} is well-approximated by a similar term involving the smooth function $\btest_{\delta_p}(x; s)$. To shorten notation, in what follows, we abbreviate $\btest_p\big(\tfrac{1}{\sqrt{p}}\va\tran \wvv; \cvec\tran \sgl\big)$ and $\btest_{\delta_p}\big(\tfrac{1}{\sqrt{p}}\va\tran \wvv; \cvec\tran \sgl\big)$ to $\btest(\va)$ and $\btest_{\delta_p}(\va)$, respectively. The meaning of the notation $\btest(\vb)$ and $\btest_{\delta_p}(\vb)$ should also be clear. Since
\begin{equation}\label{eq:piecewise_decomp}
\begin{aligned}
    \abs{\E \btest(\va)-\E \btest(\vb)} &\le \abs{\E \btest_{\delta_p}(\va)-\E \btest_{\delta_p}(\vb)} \\
    &+ \E \abs{\btest(\va)-\btest_{\delta_p}(\va)}+ \E \abs{\btest(\vb)-\btest_{\delta_p}(\vb)},
\end{aligned}
\end{equation}
we just need to bound the three terms on the right-hand side.}

\textcolor[rgb]{0.00,0.00,0.00}{The first term can be controlled by Theorem~\ref{thm:CLT_general}, as $\btest_{\delta_p}(x; s)$ is differentiable. By assumption, $\abs{\btest(x; s)} \le B_p(s)(1+\abs{x}^K)$ for some $K \ge 1$. Using the simple bound \eref{smooth_mol_bnd} in Lemma~\ref{lem:piecewise_bnd} (see Appendix~\ref{appendix:truncation}), we can check that, for any ${\delta_p} < 1$,
\[
\begin{aligned}
    &\max\set{ \abs{\btest_{\delta_p}(x, s)}, \absb{\btest'_{\delta_p}(x, s)}} \\
    \le& \frac{C B_p(s)[1 + (\abs{x}+{\delta_p})^K]}{{\delta_p}} \le \frac{C' B_p(s)[1 + \abs{x}^K]}{{\delta_p}},
\end{aligned}
\]
where $C$ is some numerical constant and $C' = (2^{K-1} +1)C$. Theorem~\ref{thm:CLT_general} then gives us
\begin{equation}\label{eq:test_ab_delta}
\abs{\E \btest_{\delta_p}(\va)-\E \btest_{\delta_p}(\vb)} \le \frac{[\E B_p^4(z)]^{1/4}\polylog p}{{\delta_p}\sqrt{p}},
\end{equation}
where we have simplified the term $P(\wvv, \BP)$ in \eref{CLT_general} by using the additional assumption that $\BP \le \polylog p$ and $\norm{\wvv}_\infty \le \polylog p$.}

\textcolor[rgb]{0.00,0.00,0.00}{To control the second term on the right-hand side of \eref{piecewise_decomp}, we apply Lemma~\ref{lem:piecewise_bnd} again. Using a shorthand notation $\widehat{B}_p(\va) = C'B_p(s)(1 + \absb{\scp \va\tran \wvv}^K)$, we have, from \eref{smooth_approx},
\begin{align}
&\E \abs{\btest(\va)-\btest_{\delta_p}(\va)} \nonumber \\ \le& {\delta_p}\, \E \widehat{B}_p(\va) + 2 \sum_{i \le L} \E[\widehat{B}_p(\va) \, \strunc_{2{\delta_p}, {\delta_p}}( \scp\va\tran \wvv - x_i)]\nonumber\\
\le& \sqrt{\E \widehat{B}^2_p(\va)}\Big[{\delta_p} + 2 \sum_{i\le L}  \sqrt{\E \strunc_{2{\delta_p}, {\delta_p}}^2( \scp\va\tran \wvv - x_i)}\,\Big]\nonumber\\
\le& [\E B_p^4(z)]^{1/4} \polylog p \nonumber \\
&\hspace{3em} \times \Big[{\delta_p} + 2 \sum_{i\le L}  \sqrt{\E \strunc_{2{\delta_p}, {\delta_p}}^2( \scp\va\tran \wvv - x_i)}\,\Big],
\label{eq:piecewise_a}
\end{align}
where in reaching the last step we have used the moment bound obtained in \eref{trunc_test_a_bnd0}. The same reasoning also yields
\begin{equation}\label{eq:piecewise_b}
\begin{aligned}
    &~~~\E \abs{\btest(\vb)-\btest_{\delta_p}(\vb)}\\
    &\le [\E B_p^4(z)]^{1/4} \polylog p \\
    &\hspace{3em}\times\Big[{\delta_p} + 2 \sum_{i\le L}  \sqrt{\E \strunc_{2{\delta_p}, {\delta_p}}^2( \scp\vb\tran \wvv - x_i)}\,\Big].
\end{aligned}
\end{equation}}

\textcolor[rgb]{0.00,0.00,0.00}{Note that $ \scp\vb\tran \wvv$ is a Gaussian random variable with zero mean and variance $\nu^2$. (Recall the definition of $\nu^2$ in the statement of the proposition.) As the function $\strunc_{2{\delta_p}, {\delta_p}}^2(x-x_i) \le 1$ with a compact support of width $6{\delta_p}$, we have
\begin{equation}\label{eq:b_density_bnd}
\E \strunc_{2{\delta_p}, {\delta_p}}^2( \scp\vb\tran \wvv - x_i) \le \frac{6{\delta_p}}{\sqrt{2\pi\nu^2}} \le C \delta_p,
\end{equation}
where the second inequality is by the assumption that $\nu^2 \ge c > 0$ for some fixed $c$. This bound can also be leveraged to control $\E \strunc_{2{\delta_p}, {\delta_p}}^2( \scp\va\tran \wvv - x_i)$. Indeed, $\strunc_{2{\delta_p}, {\delta_p}}^2(x-x_i)$ is a smooth and bounded test function whose derivative is bounded by $C/{\delta_p}$. By Theorem~\ref{thm:CLT_general},
\[
\absb{\E \strunc_{2{\delta_p}, {\delta_p}}^2( \scp\va\tran \wvv - x_i) - \E \strunc_{2{\delta_p}, {\delta_p}}^2( \scp\vb\tran \wvv - x_i)} \le \frac{\polylog p}{{\delta_p}\sqrt{p}},
\]
and thus
\begin{equation}\label{eq:a_density_bnd}
\E \strunc_{2{\delta_p}, {\delta_p}}^2( \scp\va\tran \wvv - x_i) \le \Big[\delta_p +  \frac{1}{{\delta_p}\sqrt{p}}\Big] \polylog p.
\end{equation}
Substituting \eref{a_density_bnd}, \eref{b_density_bnd}, \eref{piecewise_a}, \eref{piecewise_b}, \eref{test_ab_delta} into \eref{piecewise_decomp}, and after some simplifications, we get
\[
\begin{aligned}
    \abs{\E \btest(\va)-\E \btest(\vb)} &\le [\E B^4_p(z)]^{1/4} \polylog p \\
    &\hspace{1em} \times \Big[\delta_p + (\delta_p \sqrt{p})^{-1}+ \sqrt{\delta_p + (\delta_p \sqrt{p})^{-1}}\,\Big].
\end{aligned}
\]
The convergence rate of the right-hand side can be optimized by setting $\delta_p = p^{-1/4}$. This then leads to the claim in \eref{CLT_piecewise}.}

%% file: appendix_covariance.tex

\subsection{Asymptotic Equivalence of the Covariance Matrices}
\label{appendix:covariance}

Consider a sequence of activation functions $\set{\kernel_p(x)}$ such that, for every $p$, $\kernel_p(x)$ is an odd function and
\[
\max\set{\norm{\sigma'_p(x)}_\infty, \norm{\sigma''_p(x)}_\infty, \norm{\sigma'''_p(x)}_\infty} \le \polylog p.
\]
Given a set of feature vectors $\fmtx = [\fvec_1, \fvec_2, \ldots, \fvec_p] \in \R^{d \times p}$, we define
\[
\va \bydef \kernel(\fmtx\tran \cvec) \quad \text{and} \quad \vb = \mu_{1,p} \fmtx\tran \cvec+ \mu_{2,p} \vz,
\]
where $\cvec \sim \mathcal{N}(0, \mI_d)$ and $\vz \sim \mathcal{N}(0, \mI_p)$ are two independent Gaussian vectors, and $\mu_{1, p} = \EE[z\sigma_p(z)]$, $\mu_{2, p} = \sqrt{\EE \sigma^2_p(z) - \mu_{1, p}^2}$, with $z \sim \mathcal{N}(0, 1)$, are two constants. The primary goal of this appendix is to quantify the difference between the covariance matrices
\[
\mSig_a = \EE \va\va\tran \quad \text{and}\quad \mSig_b = \EE \vb \vb\tran = \mu_{1,p}^2 \fmtx\tran \fmtx + \mu_{2, p}^2 \mI_p.
\]

We start by noting that $\mu_{1,p} = \E \kernel'_p(z)$ and thus
\begin{align}
\abs{\E \kernel'_p(\cvec\tran \fvec_i) - \mu_{1,p}} &\le \E \abs{\kernel'_p(\norm{\fvec_i} z) - \kernel'_p(z)}\nonumber\\
&\le \norm{\sigma''_p(x)}_\infty (\E \abs{z}) \absb{\norm{\fvec_i} - 1}\nonumber\\
&\le (\polylog p) \absb{\norm{\fvec_i}^2 - 1}.\label{eq:cov_bnd1}
\end{align}

\begin{lem}\label{lem:sigma_ab}
Suppose that the feature vectors satisfy \eref{BP} with some $\BP$. We have
\begin{equation}\label{eq:sigma_ab}
\norm{\mSig_a - \mSig_b} \le \frac{(1 + \BPt+ \norm{\mF}^4)\polylog p}{\sqrt{p}}.
\end{equation}
\end{lem}
\begin{proof}
The $(i,j)$th entry of $\mSig_a$ is $\E[\kernel_p(\cvec\tran \fvec_i)\kernel_p(\cvec\tran \fvec_j)]$. Since $(\cvec\tran \fvec_i, \cvec\tran \fvec_j)$ are jointly Gaussian, we can rewrite their joint distribution as that of $(z_i, \rho_{ij}  z_i + \sqrt{1 - \rho_{ij} \rho_{ji}}  z_j)$, where $z_i \sim \mathcal{N}(0, \norm{\fvec_i}^2), z_j\sim \mathcal{N}(0, \norm{\fvec_j}^2)$ are two independent Gaussian random variables and $\rho_{ij} \bydef \fvec_i\tran \fvec_j / \norm{\fvec_i}^2$. Note that the definition of $\rho_{ij}$ is not symmetric: $\rho_{ij} \neq \rho_{ji}$ unless $\norm{\fvec_i} = \norm{\fvec_j}$. With this new representation, we have, for $i \neq j$,
\begin{align}
&\mSig_a(i, j) \nonumber \\
=& \E[ \kernel_p( z_i) \kernel_p(\rho_{ij}  z_i + \sqrt{1 - \rho_{ij} \rho_{ji}} z_j)]\nonumber\\
\overset{(a)}{=}& \E [\kernel_p(z_i) \kernel_p(\sqrt{1 - \rho_{ij} \rho_{ji}}  z_j)] \nonumber \\
 &\hspace{5em}+ \rho_{ij}\E[\kernel_p(z_i)z_i \kernel'_p(\sqrt{1 - \rho_{ij} \rho_{ji}}  z_j)]\nonumber\\
 &\hspace{5em}+ \tfrac{1}{2}\rho_{ij}^2 \E[\kernel_p(z_i)z_i^2 \kernel_p''(\sqrt{1 - \rho_{ij} \rho_{ji}}  z_j)] \nonumber \\
 &\hspace{5em}+ \tfrac{1}{6}\rho_{ij}^3 \E[\kernel_p(z_i)z_i^3 \kernel_p'''(\theta_{ij})]\nonumber\\
\overset{(b)}{=}& (\fvec_i\tran \fvec_j) \E\kernel_p'(z_i) \E\kernel_p'(\sqrt{1 - \rho_{ij} \rho_{ji}}  z_j) \nonumber \\
&\hspace{6em}+ \tfrac{1}{6}\rho_{ij}^3 \E[\kernel_p(z_i)z_i^3 \kernel'''_p(\theta_{ij})]\nonumber\\
\overset{(c)}{=}& (\fvec_i\tran \fvec_j) \E\kernel'_p(z_i) \E\kernel'_p( z_j) + R_{ij}.\label{eq:Siga_ij}
\end{align}
Here, (a) comes from Taylor's series expansion, with $\theta_{ij}$ being some point between $\sqrt{1 - \rho_{ij} \rho_{ji}} z_j$ and $\rho_{ij}  z_i + \sqrt{1 - \rho_{ij} \rho_{ji}} z_j$. To reach (b), we have used the independence between $z_i$ and $z_j$, and the following identities: $\E \kernel_p(z_i) = \E[\kernel_p(z_i)z_i^2] = 0$ (due to $\kernel_p(x)$ being an odd function) and $\E[\kernel_p(z_i)z_i] = \norm{\fvec_i}^2 \E[\kernel_p'(z_i)]$. In (c), $R_{ij}$ is the remainder term, defined as
\begin{equation}\label{eq:remainder_R}
\begin{aligned}
    R_{ij} &= (\fvec_i\tran \fvec_j) \E\kernel'_p(z_i)(\E\kernel'_p(\sqrt{1 - \rho_{ij} \rho_{ji}}  z_j) - \E\kernel'_p( z_j)) \\
    &\hspace{3em}+ \tfrac{1}{6}\rho_{ij}^3 \E[\kernel_p(z_i)z_i^3 \kernel'''_p(\theta_{ij})].
\end{aligned}
\end{equation}
For the case of $i = j$, we define $R_{ii} = 0$.

Using \eref{Siga_ij}, we can verify the following decomposition of $\mSig_a$:
\[
\mSig_a = (\mu_{1,p}\mI + \mD_1) \fmtx\tran \fmtx (\mu_{1, p} \mI + \mD_1) + \mu_{2,p}^2 \mI +\mD_2 + \mD_3 + \mR
\]
where $\mD_1 = \diag\set{\E\kernel'_p(z_i) - \mu_{1,p}}$,
\[
\mD_2 = \diag\set{\mu_{1,p}^2-\norm{\fvec_i}^2[\E \kernel'_p(z_i)]^2},
\]
and
\[
\mD_3 = \diag\set{\E \kernel_p^2(z_i) - \mu_{1,p}^2 - \mu_{2,p}^2}.
\]
Since $\mSig_b = \mu_{1,p}^2 \fmtx\tran \fmtx +\mu_{2,p}^2 \mI$, we must have
\begin{equation}\label{eq:SaSb_d}
\begin{aligned}
\norm{\mSig_a - \mSig_b} &\le (2\mu_{1,p} + \norm{\mD_1}) \norm{\fmtx}^2 \norm{\mD_1} \\
&\hspace{5em }+ \norm{\mD_2} + \norm{\mD_3} + \norm{\mR}.
\end{aligned}
\end{equation}
Recall the assumptions about the feature vectors in \eref{BP}. It then follows from \eref{cov_bnd1} that $\norm{\mD_1} \le \BP \polylog p / \sqrt{p}$. Similarly, we also have $\norm{\mD_2} \le \BP \polylog p / \sqrt{p}$. Controlling $\norm{\mD_3}$ requires a few more steps. Let $z \sim \mathcal{N}(0, 1)$ and $T = \sqrt{2 \log p}$.
\begin{align}
&\abs{\E\kernel_p^2(z_i) - \mu_{1,p}^2 - \mu_{2,p}^2} \nonumber\\
=& \abs{\E \kernel_p^2(\norm{\fvec_i} z) - \E \kernel_p^2(z)}\nonumber\\
\le& \E \big[\abs{\kernel_p^2(\norm{\fvec_i} z) - \kernel_p^2(z)} (\charfn_{\abs{z} > T} + \charfn_{\abs{z} \le T})\big]\nonumber\\
\overset{(a)}{\le}& \sqrt{2} [\E \kernel_p^4(\norm{\fvec_i} z) + \E \kernel_p^4(z)]^{1/2} \sqrt{\P(\abs{z} > T)} \nonumber \\
&\hspace{10em}+ \polylog p \absb{\norm{\fvec_i}^2 - 1}\nonumber\\
\overset{(b)}{\le}& \frac{(\BP + 1) \polylog p}{\sqrt{p}}.\label{eq:D3_diag_bnd}
\end{align}
Here, (a) uses Holder's inequality and the fact that the derivative of $\kernel_p^2(x)$ is bounded by $\polylog p$ within the interval $\abs{x} \le \max\set{\norm{\fvec_1}, 1} T$; (b) applies the standard tail bound $\P(z > T) \le 2e^{-T^2/2}$. As \eref{D3_diag_bnd} holds for all $i \le p$, we have $\norm{\mD_3} \le {(\BP + 1) \polylog p}/{\sqrt{p}}$. The last term to consider is the remainder matrix $\mR$. From its definition in \eref{remainder_R}, we can easily verify that
\[
\max_{1\le i,j \le p} \abs{R_{ij}} \le \frac{\BPt \polylog p}{p^{3/2}}.
\]
It follows that $\norm{\mR} \le \norm{\mR}_\text{F} = \sqrt{\sum_{i,j} R^2_{ij}} \le \BPt \polylog p/\sqrt{p}$. Substituting our bounds for $\norm{\mD_1}, \norm{\mD_2},\norm{\mD_3}$ and $\norm{\mR}$ into \eref{SaSb_d}, we then reach the bound \eref{sigma_ab} in the statement of the lemma.
\end{proof}

Next, we prove an auxiliary result that will be used in the proof of Theorem~\ref{thm:CLT_general}. Here, we consider a particular sequence of activation functions $\set{\widehat{\kernel}_p(x)}$ as defined in \eref{kernel_trunc}. They form a family of smoothly truncated versions of a fixed activation function $\kernel(x)$.

\begin{lem}\label{lem:mu_p_diff}
Let $\mu_1, \mu_2$ and $\mu_{1, p}, \mu_{2, p}$ be the constants associated with $\kernel(x)$ and $\widehat{\kernel}_p(x)$, respectively. If the threshold $\tau_p = \sqrt{2 C_\tau \log p}$ in \eref{kernel_trunc} is chosen with a constant $C_\tau \ge 2$, then
\begin{equation}\label{eq:mu_1p_diff}
\abs{\mu_1 - \mu_{1,p}} \le \frac{\polylog p}{p} \quad \text{and} \quad \abs{\mu_2 - \mu_{2,p}} \le \frac{\polylog p}{\sqrt{p}}.
\end{equation}
\begin{proof}
By construction, $\kernel(x) = \widehat{\kernel}_p(x)$ and  $\kernel'(x) = \widehat{\kernel}'_p(x)$ for $\abs{x} < \tau_p$. Let $z \sim \mathcal{N}(0, 1)$. We then have
\begin{align}
\abs{\mu_1 - \mu_{1, p}} &\le \E \big[\abs{\kernel'(z) - \widehat{\kernel}'_p(z)} \charfn_{\abs{z} \ge \tau_p}\big]\nonumber\\
&\le \sqrt{\E (\kernel'(z) - \widehat{\kernel}'_p(z))^2} \sqrt{\P(\abs{z} \ge \tau_p)}\nonumber\\
&\le \frac{\polylog p}{p},\label{eq:mu_1p_bnd}
\end{align}
where the last step uses the Gaussian tail bound $\P(\abs{z} \ge \tau_p) \le 2/p^2$ for $\tau_p \ge 2 \sqrt{\log p}$. The same truncation techniques will also give us
\[
\abs{\E \kernel^2(z) - \E \widehat{\kernel}_p^2(z)} \le \frac{\polylog p}{p}.
\]
Combining this bound with \eref{mu_1p_bnd} and recall the definitions of $\mu_2$ and $\mu_{2,p}$, we have $\abs{\mu_2^2 - \mu_{2,p}^2} \le ({\polylog p})/{p}$. Finally, the second bound in \eref{mu_1p_diff} can be obtained from the following inequality: $\abs{\sqrt{x} - \sqrt{y}} \le \sqrt{\abs{x-y}}$ for any two nonnegative numbers $x$ and $y$.
\end{proof}
\end{lem}

%% file: appendix-concentration.tex
\subsection{Some Concentration Results}

\subsubsection{\label{appendix:Concentration-of-GaussianVec}Concentration of Gaussian
Vectors}
%
\begin{lem}\label{lem:A1_high_prob}
Let $\mathcal{A}_1$ be the event defined in \eref{def_setA1}. There exists a constant $c>0$ such that
\[
\P(\mathcal{A}_{1})\geq1-c\exp\left(-(\log p)^{2}/c\right).
\]
\end{lem}
\begin{proof}
\textcolor[rgb]{0.00,0.00,0.00}{We start by stating the following simple result: for $\fvec_{1},\fvec_{2}\iid\calN\Big(0,\tfrac{1}{d}\mI_{d}\Big)$,
there exists positive constants $c$ and $K$ such that for any $\veps\geq 0$
\begin{equation}
\P\big(|\fvec_{1}\tran\fvec_{2}|\geq\veps\big)\leq2\exp\big[-cd\min\big\{ \tfrac{\veps^{2}}{K^{2}},\tfrac{\veps}{K}\big\} \big].\label{eq:fvec_weakcorrelation}
\end{equation}
and
\begin{equation}
\P\big(|\|\fvec_{1}\|^2-1|\geq\veps\big)\leq2\exp\big[-cd\min\big\{ \tfrac{\veps^{2}}{K^{2}},\tfrac{\veps}{K}\big\} \big].\label{eq:fnorm_concentrate_lbd}
\end{equation}
Indeed, for any $i\in[d]$, $f_{1,i}$ and $f_{2,i}$ are both sub-Gaussian random variables with sub-Gaussian norm bounded by $\frac{C}{\sqrt{d}}$, for some $C>0$  \cite[Example 2.5.8]{vershynin2018high}, so $f_{1,i}f_{2,i}$ is a sub-exponential
random variable with sub-exponential norm $\frac{C^2}{{d}}$ \cite[Lemma 2.7.7]{vershynin2018high}. Then we can apply Bernstein's inequality \cite[Corollary 2.8.3]{vershynin2018high}
to get (\ref{eq:fvec_weakcorrelation}). Also, \eqref{eq:fnorm_concentrate_lbd} can be proved in the same way.
Then we can let $\veps=\tfrac{(\log p)^{2}}{\sqrt{p}}$ in \eqref{eq:fvec_weakcorrelation} and \eqref{eq:fnorm_concentrate_lbd}
and use union bound to get for any $p$,
\begin{equation}
\P\Big(\max_{1\leq i<j\leq p}\big|\fvec_{i}\tran\fvec_{j}\big|\geq\tfrac{(\log p)^{2}}{\sqrt{p}}\Big)\leq c\exp\left(-(\log p)^{2}/c\right)\label{eq:fvec_weak_correlation_bd}
\end{equation}
and
\begin{equation}
\P\Big(\max_{1\leq i\leq p}\Big|\|\fvec_{i}\|^{2}-1\Big|\geq\tfrac{(\log p)^{2}}{\sqrt{p}}\Big)\leq c\exp\left(-(\log p)^{2}/c\right).\label{eq:fvec_norm_concentrate}
\end{equation}
where $c>0$ is some constant.}

\textcolor[rgb]{0.00,0.00,0.00}{Finally, we just need to verify that
\begin{equation}
\P\Big(\max_{1\leq i\leq p}\Big|\fvec_{i}\tran\sgl\Big|\geq\tfrac{(\log p)^{2}}{\sqrt{p}}\Big)\leq c\exp\left(-(\log p)^{2}/c\right).\label{eq:A1_high_prob_2}
\end{equation}
For any $i\in[p]$, we have $\fvec_{i}\tran\sgl\sim\mathcal{N}(0,\tfrac{1}{d})$. Thus, for any $\veps\geq0$, the standard Gaussian tail bound gives us
\[
\P\big(\big|\fvec_{i}\tran\sgl\big|\geq\veps\big)\leq2e^{-d\veps^{2}/2}.
\]
By setting $\veps=\tfrac{(\log p)^{2}}{\sqrt{p}}$ and applying union
bound, we can obtain (\ref{eq:A1_high_prob_2}). Recall the definition of $\mathcal{A}_{1}$ in (\ref{eq:def_setA1}). Combining \eref{fvec_weak_correlation_bd}, \eref{fvec_norm_concentrate} and \eref{A1_high_prob_2},  we complete the proof.}
\end{proof}

\subsubsection{\label{appendix:Concentration_Lipschitz}Concentration of Lipschitz
Functions of Gaussian Vectors}

The results presented in this section are all consequences of
the following well-known theorem about the concentration of Lipschitz
functions of independent Gaussian random variables. See \eg, \cite[Theorem 1.3.4]{Talagrand2010meanfieldv1} for a proof.
\begin{thm}
Let $X\sim\calN\left(\boldsymbol{0},\mI_{p}\right)$. For any $\kappa$-Lipschitz function $f\left(\vx\right)$
on $\R^{p}$ and any $\veps\geq0$,
\begin{equation}
\P\left(\left|f\left(X\right)-\E f\left(X\right)\right|\geq\veps\right)\leq2\exp\left(-\frac{\veps^{2}}{4\kappa^{2}}\right).\label{eq:Lipschitz_concentration}
\end{equation}
\end{thm}
We will also use the integral identity $\E \abs{X} = \int_{0}^{\infty}\P(\abs{X}>t)dt$ to control the moments of concentrated random variables.
If a random variable $X$ satisfies $\P\left(|X|>v\right)\leq ce^{-Cv}$ for some $C,c > 0$, then for any $\epn\in\mathbb{Z}^{+}$, it holds that
\begin{equation}
\E|X|^{\epn}\leq c\epn C^{-\epn}\int_{0}^{\infty}e^{-v}v^{\epn-1}dv = c (\epn!) C^{-\epn}.\label{eq:X_bd_moments_expo}
\end{equation}
Similarly, if $\P\left(|X|>v\right)\leq ce^{-Cv^{2}}$ for some $C,c>0$, then
\begin{align}
\E|X|^{\epn}\leq2c(\epn!)C^{-\frac{\epn}{2}}.\label{eq:X_bd_moments_gauss}
\end{align}

In what follows, we will consider probabilistic and moment bounds involving the regressors $\va_\didx$ and $\vb_\didx$ in \eref{a_vec} and \eref{b_vec}, for a \emph{fixed} feature matrix $\fmtx$. Correspondingly, the notation $\Pcnd{\fmtx}$ (resp. $\Ecnd{\fmtx}$) refer to the conditional probability (resp. expectation) for a given $\fmtx$.

\begin{lem}
\label{lem:linear_concentrate} Let $\mSig = \E[\vb_\didx \vb_\didx\tran]$. There exists $c>0$ such that
\begin{equation}
\Pcnd{\fmtx} \left(\left|\tfrac{1}{\sqrt{p}}\va_{t}\tran\wvv\right|\geq\veps\right)\leq2\exp\Big(-\tfrac{p\veps^{2}}{c\|\wvv\|^{2}\|\fmtx\|^{2}\norm{\kernel'}_\infty^{2}}\Big)\label{eq:atw_concentrate}
\end{equation}
and
\begin{equation}
\Pcnd{\fmtx}\left(\left|\tfrac{1}{\sqrt{p}}\vb_{t}\tran\wvv\right|\geq\veps\right)\leq2\exp\Big(-\tfrac{p\veps^{2}}{c\|\wvv\|^{2}\|\mSig\|}\Big),\label{eq:btw_concentrate}
\end{equation}
for any fixed vector $\wvv \in \R^p$ and $\veps\geq0$. Correspondingly, there exists $C>0$ such that
any $\epn\in\mathbb{Z}^{+}$,
\begin{equation}
\Ecnd{\fmtx}\Big(\left|\tfrac{1}{\sqrt{p}}\va_{t}\tran\wvv\right|^{\epn}\Big)\leq \epn!\left(\tfrac{C\|\wvv\|^{2}\|\mF\|^{2}\norm{\kernel'}_\infty^{2}}{p}\right)^{\frac{\epn}{2}}\label{eq:akw_bd_moments}
\end{equation}
and
\begin{equation}
\Ecnd{\fmtx}\Big(\left|\tfrac{1}{\sqrt{p}}\vb_{t}\tran\wvv\right|^{\epn}\Big)\leq \epn!\left(\tfrac{C\|\wvv\|^{2}\|\mSig\|}{p}\right)^{\frac{\epn}{2}}.\label{eq:bkw_bd_moments}
\end{equation}
\end{lem}
\begin{proof}
As a mapping from $\R^{d}$ to $\R^{p}$, $\cvec\mapsto\kernel\big(\fmtx\tran\cvec\big)$
is $(\norm{\kernel'}_\infty\|\fmtx\|)$-Lipschitz continuous. Indeed, for any $\cvec_{1},\cvec_{2}\in\R^{d}$, it is easy to verify that
\begin{align*}
\|\sigma\big(\fmtx\tran\cvec_{1}\big)-\sigma\big(\fmtx\tran\cvec_{2}\big)\|^{2} 
 & \leq\norm{\kernel'}_\infty^{2}\|\fmtx\|^{2}\|\cvec_{1}-\cvec_{2}\|^{2}.
\end{align*}
It follows that the function $f(\cvec) = \frac{1}{\sqrt{p}}\sigma\left(\cvec\tran\fmtx\right)\wvv$
is $\frac{\norm{\kernel'}_\infty\|\wvv\|\|\fmtx\|}{\sqrt{p}}$-Lipschitz
continuous. Therefore, using (\ref{eq:Lipschitz_concentration}) we
have
\begin{equation}
\Pcnd{\fmtx}\Big(\big|\tfrac{1}{\sqrt{p}}\va_{t}\tran\wvv-\Ecnd{\fmtx}\big(\tfrac{1}{\sqrt{p}}\va_{t}\tran\wvv\big)\big|\geq\veps\Big)\leq2\exp\Big(-\tfrac{p\veps^{2}}{c\|\wvv\|^{2}\|\fmtx\|^{2}\norm{\kernel'}_\infty^{2}}\Big).\label{eq:atw_concentrate_nonzeromean}
\end{equation}
Since $\kernel(x)$ is an odd function, we have $\Ecnd{\fmtx}\big(\tfrac{1}{\sqrt{p}}\va_{t}\tran\wvv\big)=0$ and thus (\ref{eq:atw_concentrate}).

To establish \eref{btw_concentrate}, we observe that $\vb_\didx$ can be represented as $\vb_\didx=\mSig^{1/2}\t{\vb}$,
where $\t{\vb}\sim\calN\left(\boldsymbol{0},\mI_{p}\right)$. It follows that $\tfrac{1}{\sqrt{p}}\vb_{t}\tran\wvv$
can also be seen as a Lipschitz function of a standard normal vector,
with a Lipschitz constant equal to $\frac{\|\wvv\|\|\mSig^{1/2}\|}{\sqrt{p}}$.
Therefore (\ref{eq:btw_concentrate}) is again a consequence of (\ref{eq:Lipschitz_concentration}). Finally, the moment bounds in (\ref{eq:akw_bd_moments}) and (\ref{eq:bkw_bd_moments})  can be obtained by applying \eqref{eq:X_bd_moments_gauss}.
\end{proof}
\begin{lem}
\label{lem:at_bt_norm_bd}There exists $c>0$ such that for any $t\in[n]$ and $s\geq\sqrt{\tfrac{4d}{p}}\norm{\kernel'}_\infty\|\mF\|$,
\begin{equation}\label{eq:norm_at_concentrate}
\Pcnd{\fmtx}\left(\scp\left\Vert \va_{t}\right\Vert \geq s\right)  \leq c\exp\big(-\tfrac{ps^{2}}{c\norm{\kernel'}_\infty^{2}\|\mF\|^{2}}\big).
\end{equation}
Similarly, for any $s\geq2\sqrt{\|\mSig\|}$, we have
\begin{equation}\label{eq:bnorm_prob_bnd}
\Pcnd{\fmtx}\left(\scp\left\Vert \vb_{t}\right\Vert \geq s\right)  \leq c\exp\big(-\tfrac{ps^{2}}{c\|\mSig\|}\big).
\end{equation}
Correspondingly, there exists $C>0$ such that
\begin{align}
\Ecnd{\fmtx}\left[\big(\scp\left\Vert \va_{t}\right\Vert \big)^{\epn}\right] & \leq\left(\sqrt{\tfrac{4d}{p}}\norm{\kernel'}_\infty\|\mF\|\right)^{\epn}+\epn!\left(\tfrac{C\norm{\kernel'}_\infty\|\mF\|}{\sqrt{p}}\right)^{\epn},\label{eq:a_nrm_bnd}\\
\Ecnd{\fmtx}\left[\big(\scp\left\Vert \vb_{t}\right\Vert \big)^{\epn}\right] & \leq\left(2\sqrt{\|\mSig\|}\right)^{\epn}+\epn!\left(C\sqrt{\tfrac{\|\mSig\|}{p}}\right)^{\epn},\label{eq:b_nrm_bnd}
\end{align}
for any $t\in[n]$ and $\epn\in\mathbb{Z}^{+}$.
\end{lem}
\begin{proof}
Recall that $\va_{t}=\sigma(\mF\tran\cvec_{t})$ and $\cvec\mapsto\sigma(\mF\tran\cvec)$
is a $(\norm{\kernel'}_\infty\|\mF\|)$-Lipschitz continuous mapping. It follows that
$\cvec\mapsto\|\sigma(\mF\tran\cvec)\| ~(= \|\va\|)$ is a $(\norm{\kernel'}_\infty\|\mF\|)$-Lipschitz continuous function.
From
(\ref{eq:Lipschitz_concentration}), there exists $c>0$ such that for
any $s>0$,
\begin{equation}
\Pcnd{\fmtx}\left(\scp\big|\left\Vert \va_{t}\right\Vert -\Ecnd{\fmtx}\left\Vert \va_{t}\right\Vert \big|>s\right)\leq c\exp\Big(-\tfrac{ps^{2}}{c\norm{\kernel'}_\infty^{2}\|\mF\|^{2}}\Big).\label{eq:at_norm_concentrate}
\end{equation}
On the other hand,
\begin{align}
\Ecnd{\fmtx}\Big(\scp\left\Vert \va_{t}\right\Vert \Big) & \overset{{(a)}}{=}\scp\Ecnd{\fmtx}\|\sigma(\mF\tran\cvec_{t})-\sigma(\mF\tran\boldsymbol{0}_p)\|\nonumber \\
 & \tleq{{(b)}}\scp\norm{\kernel'}_\infty\norm{\fmtx} \, \E\|\cvec_{t}\| \nonumber \\
 & \leq\sqrt{\tfrac{d}{p}}\norm{\kernel'}_\infty\|\mF\|.\label{eq:E_at_norm}
\end{align}
In step (a), we use the assumption that $\sigma(\cdot)$ is an odd function and thus $\sigma(0) = 0$; step (b) follows from the Lipschitz continuity of the mapping $\vg \mapsto \sigma(\mF\tran\vg)$; to reach the last inequality, we have used the Holder's inequality to get $\E \norm{\vg_t} \le \sqrt{\E \norm{\vg_t}^2} = \sqrt{d}$.

For any $s\geq\sqrt{\tfrac{4d}{p}}\norm{\kernel'}_\infty\|\mF\|$, we can use (\ref{eq:E_at_norm}) and (\ref{eq:at_norm_concentrate}) to deduce that
\begin{align}
\Pcnd{\fmtx}\left(\scp\left\Vert \va_{t}\right\Vert \geq s\right) &\le \Pcnd{\fmtx}\left(\scp\left\Vert \va_{t}\right\Vert -\scp\E\left\Vert \va_{t}\right\Vert \geq\tfrac{s}{2}\right)\nonumber \\
 &\le c\exp\big(-\tfrac{ps^{2}}{c\norm{\kernel'}_\infty^{2}\|\mF\|^{2}}\big).\nonumber
\end{align}

The proof of \eref{bnorm_prob_bnd} is analogous. We write $\vb_{t}=\mSig^{\frac{1}{2}}\widetilde{\vb}_{t}$, where
$\widetilde{\vb}_{t}\sim\calN\left(\boldsymbol{0},\mI_{p}\right)$.
Therefore, similar to what we did to reach \eqref{eq:at_norm_concentrate}, we can show there exists $c>0$ such that for any $s\geq0$,
\begin{align*}
&\Pcnd{\fmtx}\left(\scp\big|\left\Vert \vb_{t}\right\Vert -\Ecnd{\fmtx}\left\Vert \vb_{t}\right\Vert \big|\geq s\right)\\
=&\Pcnd{\fmtx}\left(\scp\big|\lVert\mSig^{\frac{1}{2}}\widetilde{\vb}_{t}\rVert-\Ecnd{\fmtx}\lVert\mSig^{\frac{1}{2}}\widetilde{\vb}_{t}\rVert\big|\geq s\right)\\
\leq& c\exp\big(-\tfrac{ps^{2}}{c\|\mSig\|}\big),
\end{align*}
where the last step follows from the fact that $\norm{\mSig^{\frac{1}{2}}\widetilde{\vb}_{t}}$
is a $\lVert\mSig^{1/2}\rVert$-Lipschitz function of $\widetilde{\vb}_{t}$. Meanwhile,
\begin{align*}
\Ecnd{\fmtx}\left(\scp\lVert{\vb}_{t}\rVert\right) \leq\sqrt{\tfrac{1}{p}\Ecnd{\fmtx}\lVert\mSig^{\frac{1}{2}}\widetilde{\vb}_{t}\rVert^{2}} \leq\sqrt{\|\mSig\|}.
\end{align*}
It follows that, for any $s\geq2\sqrt{\|\mSig\|}$,
\begin{align*}
\Pcnd{\fmtx}\left(\scp\left\Vert \vb_{t}\right\Vert \geq s\right) & \leq\Pcnd{\fmtx}\left(\scp\left\Vert \vb_{t}\right\Vert -\scp\Ecnd{\fmtx}\left\Vert \vb_{t}\right\Vert \geq\tfrac{s}{2}\right)\\
 & \leq c\exp\big(-\tfrac{ps^{2}}{c\|\mSig\|}\big).
\end{align*}

The bounds for the moments $\E\big[\scp\left\Vert \va_{t}\right\Vert\big]^{\epn}$
and $\E\big[\scp\left\Vert \vb_{t}\right\Vert \big]^{\epn}$ then
directly follow from the probabilistic bounds obtained above and \eqref{eq:X_bd_moments_gauss}.
\end{proof}
\begin{lem}
\label{lem:quadratic_asymmetry_concentrate} Let $\mathcal{A}$ be the admissible set of feature matrices defined in \eref{def_setA}, and $\mH_{\bs k}$ the leave-one-out Hessian matrix defined in \eref{looH}. There exists $c>0$ such
that, for every $k\in[n]$, $t\neq k$ and $\veps\geq 0$,
\begin{align}
\Pcnd{\fmtx}\big(|\va_{t}\tran\mH_{\bs k}^{-1}\va_{k}/p|\geq\veps\big) & \leq c\exp\big(-({p}/{c})\min\{\veps^2, \veps\}),\label{eq:ak_Hk_ak_concentrate}\\
\Pcnd{\fmtx}\big(|\va_{t}\tran\mH_{\bs k}^{-1}\vb_{k}/p|\geq\veps\big) & \leq c\exp\big(-({p}/{c})\min\{\veps^2, \veps\}),\label{eq:at_Hk_bk_concentrate}\\
\Pcnd{\fmtx}\big(|\vb_{t}\tran\mH_{\bs k}^{-1}\vb_{k}/p|\geq\veps\big) & \leq c\exp\big(-({p}/{c})\min\{\veps^2, \veps\}),\label{eq:bt_Hk_bk_concentrate}\\
\Pcnd{\fmtx}\big(|\vb_{t}\tran\mH_{\bs k}^{-1}\va_{k}/p|\geq\veps\big) & \leq c\exp\big(-({p}/{c})\min\{\veps^2, \veps\}).\label{eq:bt_Hk_ak_concentrate}
\end{align}
\end{lem}
\begin{proof}
Note that, conditioned on $\fmtx$, $\va_k$ is independent of $\va_{t}\tran\mH_{\bs k}^{-1}$ for $t\neq k$. For any $s\geq\sqrt{\tfrac{4d}{p}}\norm{\kernel'}_\infty\|\mF\|$,
\begin{align}
&\Pcnd{\fmtx}\Big(|\va_{t}\tran\mH_{\bs k}^{-1}\va_{k}/p|\geq\veps\Big)\nonumber \\
\leq& \Pcnd{\fmtx}\Big(\Big|\tfrac{\va_{t}\tran\mH_{\bs k}^{-1}}{\|\va_{t}\tran\mH_{\bs k}^{-1}\|}\tfrac{\va_{k}}{\sqrt{p}}\Big|\geq\tfrac{\sqrt{p}\veps}{\|\va_{t}\tran\mH_{\bs k}^{-1}\|},\scp\left\Vert \va_{t}\right\Vert <s\Big) \nonumber \\&\hspace{2em}+\Pcnd{\fmtx}\Big(\scp\left\Vert \va_{t}\right\Vert \geq s\Big)\nonumber \\
   \tleq{{(a)}}& \Pcnd{\fmtx}\Big(\Big|\tfrac{\va_{t}\tran\mH_{\bs k}^{-1}}{\|\va_{t}\tran\mH_{\bs k}^{-1}\|}\tfrac{\va_{k}}{\sqrt{p}}\Big|\geq\tfrac{\lambda\veps}{2s}\Big)+\Pcnd{\fmtx}\Big(\scp\left\Vert \va_{t}\right\Vert \geq s\Big)\nonumber \\
 \tleq{{(b)}}& c\exp\Big(-\tfrac{p\lambda^{2}\veps^{2}}{cs^{2}\norm{\kernel'}_\infty^{2}\|\mF\|^{2}}\Big)+c\exp\Big(-\tfrac{ps^{2}}{c\norm{\kernel'}_\infty^{2}\|\mF\|^{2}}\Big),\label{eq:concentration_at_Hk_ak}
\end{align}
where step (a) follows from the fact that $\mH_{\bs k}\succeq\frac{\lambda}{2}\mI_{p}$
for $\fmtx\in{\cal A}$ (see Remark~\ref{rem:convex}) and hence $\|\mH_{\bs k}^{-1}\va_{t}\|\leq2\lambda^{-1}\|\va_{t}\|$
and step (b) follows from the concentration inequalities in \eref{atw_concentrate} and \eqref{eq:norm_at_concentrate}.

To optimize the bound on the right-hand side of \eref{concentration_at_Hk_ak}, we choose different values of $s$ according to $\veps$. For $\veps\leq\tfrac{4d\norm{\kernel'}_\infty^{2}}{\lambda p}\|\mF\|^{2}$, we let $s=\sqrt{\tfrac{4d}{p}}\norm{\kernel'}_\infty\|\mF\|$ and get
\[
\Pcnd{\fmtx}\left(|\va_{t}\tran\mH_{\bs k}^{-1}\va_{k}/p|\geq\veps\right)\leq2c\exp\Big(-\tfrac{p^{2}\lambda^{2}\veps^{2}}{4cd\norm{\kernel'}_\infty^4\|\mF\|^{4}}\Big).
\]
For $\veps>\tfrac{4d\norm{\kernel'}_\infty^{2}}{\lambda p}\|\mF\|^{2}$, we let $s=\sqrt{\lambda\veps}$, which gives us
\[
\Pcnd{\fmtx}\left(|\va_{t}\tran\mH_{\bs k}^{-1}\va_{k}/p|\geq\veps\right)\leq2c\exp\Big(-\tfrac{p\lambda\veps}{c\norm{\kernel'}_\infty^{2}\|\mF\|^{2}}\Big).
\]
Combining these two inequalities and using Assumption \ref{a:kernel} that $\norm{\kernel'}_\infty < \infty$ and the fact that $\norm{\fmtx} \le 1 + 2 \sqrt{\eta}$ for $\fmtx \in \mathcal{A}$, we get (\ref{eq:ak_Hk_ak_concentrate}). The proofs of (\ref{eq:at_Hk_bk_concentrate})-(\ref{eq:bt_Hk_ak_concentrate}) follow exactly the same procedure, and we omit them.
\end{proof}

\subsubsection{\label{appendix:matrix}The Spectral Norm of Random
Matrices}

We first recall a well-known result on the spectral norm of Gaussian
random matrices, the proof of which can be found in \cite[Corollary 7.3.3]{vershynin2018high}.
\begin{lem}
\label{lem:Wishart_spectral_norm}For a random matrix $\fmtx\in\R^{d\times p}$
with $F_{ij}\iid\calN\left(0,\frac{1}{d}\right)$, there exists $c>0$
such that for any $t\geq0$,
\begin{equation}
\P\left(\left\Vert \fmtx\right\Vert \geq1+\sqrt{p/d}+t\right)\leq2e^{-cdt^{2}}.\label{eq:spectral_norm_F}
\end{equation}
In particular, choosing $t=\sqrt{p/d}$ gives us
\begin{equation}
\P\left(\left\Vert \fmtx\right\Vert \geq1+2\sqrt{p/d}\right)\leq2e^{-cp}.\label{eq:spectral_norm_Wishart}
\end{equation}
\end{lem}


Recall the definitions of $\va_\didx$ and $\vb_\didx$ in \eref{a_vec} and \eref{b_vec}, respectively. Next, we show that the spectral norms of $\big\Vert \frac{1}{p}\sum_{t=1}^{n}\va_{t}\va_{t}\tran\big\Vert $
and $\big\Vert \frac{1}{p}\sum_{t=1}^{n}\vb_{t}\vb_{t}\tran\big\Vert $
are bounded with high probability.
\begin{lem}
\label{lem:empi_spectral_norm_bd} There exists some positive constant $c$ such that, for any fixed $\fmtx$, the following holds.
\begin{equation}
\Pcnd{\fmtx}\Big(\Big.\Big\Vert\tfrac{1}{p}\sum_{t=1}^{n}\va_{t}\va_{t}\tran \Big\Vert\geq t  \Big)\leq2\exp\big(-\tfrac{pt}{4c\norm{\kernel'}_\infty^{2}\|\fmtx\|^{2}}\big)\label{eq:AtA_spectral_norm_bd}
\end{equation}
for any $t \ge 3c(1+n/p) \norm{\fmtx}^2\norm{\kernel'}_\infty^2$, and
\begin{equation}
\Pcnd{\fmtx}\Big(\Big.\Big\Vert\tfrac{1}{p}\sum_{t=1}^{n}\vb_{t}\vb_{t}\tran\Big\Vert \geq t \Big)\leq2\exp\big(-\tfrac{pt}{4c\|\mSig\|}\big),\label{eq:BtB_spectral_norm_bd}
\end{equation}
for any $t \ge 3c(1+n/p) \norm{\mSig}$.
\end{lem}
\begin{proof}
Let $\vx \in\calS^{p-1}$ and $\vu \in\calS^{n-1}$ be two \emph{fixed} vectors with unit norms. For any $\veps \geq 0$, we have from (\ref{eq:atw_concentrate}) that
\[
\Pcnd{\fmtx}\left(|\va_{t}\tran\vx|\geq\veps\right)\leq2e^{-\frac{\veps^{2}}{c\norm{\kernel'}_\infty^{2}\|\mF\|^{2}}},
\]
and thus $\va_{t}\tran\vx$ is a sub-Gaussian random variable. Then by the
independence of $\left\{ \va_{t}\right\} $,
\begin{align}
\Pcnd{\fmtx}\Big(\left|\scp\vu\tran\mA\vx\right|\geq\veps\Big) & =\Pcnd{\fmtx}\Big(\Big|\scp\sum_{t=1}^{n}u_{t}\va_{t}\tran\vx\Big|\geq\veps\Big)\nonumber \\
 & \le 2e^{-\frac{p\veps^{2}}{c\norm{\kernel'}_\infty^{2}\|\mF\|^{2}}},\label{eq:ut_A_x_concentrate}
\end{align}
where the last step follows from Hoeffding's inequality for sub-Gaussian random variables \cite[Theorem 2.6.3]{vershynin2018high}.

Next, we construct two $\veps$-nets: $\calN_{p}$ on $\calS^{p-1}$ and $\calN_{n}$ on $\calS^{n-1}$, with $\veps = 1/4$. It can be shown \cite[Corollary 4.2.13]{vershynin2018high}
that the cardinality of $\calN_{p}$ and $\calN_{n}$ satisfies: $\left|\calN_{p}\right|\leq 9^{p}$
and $\left|\calN_{n}\right|\leq 9^{n}$. Let $\mA$ be the matrix defined in \eref{mat_AB}. Its operator norm can be bounded as follows \cite[Lemma 4.4.1]{vershynin2018high}:
\begin{equation}
\frac{1}{\sqrt{p}}\|\mA\|\leq 2\max_{\vx\in\calN_{p}}\max_{\vu\in\calN_{n}}\frac{1}{\sqrt{p}}\vu\tran\mA\vx.\label{eq:spectral_norm_A_bd_epsilon_net}
\end{equation}
It follows that
\[
\begin{aligned}
    \Pcnd{\fmtx}\Big(\scp \norm{\mA} \geq \sqrt{t} \Big)  &\leq2\left|\calN_{p}\right|\left|\calN_{n}\right|e^{-\frac{pt}{c\norm{\kernel'}_\infty^{2}\|\mF\|^{2}}} \\
    &\leq2\cdot9^{n+p}e^{-\frac{pt}{c\norm{\kernel'}_\infty^{2}\|\mF\|^{2}}},
\end{aligned}
\]
where to reach the second inequality we have used \eref{ut_A_x_concentrate}. Since  $\norm{\tfrac{1}{p}\sum_{t=1}^{n}\va_{t}\va_{t}\tran} = \norm{\scp \mA}^2$, the desired inequality in \eref{AtA_spectral_norm_bd} immediately follows if we choose $t \ge 3c(1+n/p) \norm{\fmtx}^2\norm{\kernel'}_\infty^2$. We omit the proof of \eref{BtB_spectral_norm_bd} as it is completely analogous.
%
\end{proof}

\subsubsection{Concentration of Quadratic Forms}
%
Recall the quadratic form $\gamma_{k}(\vr)=({\vr\tran\mH_{\bs k}^{-1}\vr})/{p}$ defined in \eref{gamma_k}. In what follows, we derive some concentration inequalities for $\gamma_{k}(\vr)=({\vr\tran\mH_{\bs k}^{-1}\vr})/{p}$ with $\vr=\va_{k}$ or $\vb_{k}$. We shall use the notation $\P_k$ (resp. $\E_k$) to denote the conditional probability (resp. expectation) over $\va_k$ and $\vb_k$, with all other random variables, namely, $\set{\va_t, \vb_t}_{t \neq k}$ and $\mF$, fixed.

\begin{lem}
\label{lem:gamma_concentrate}There exists $c>0$, such that
\begin{equation}
\P_{k}\left(\left|\gamma_{k}(\va_{k})-\E_{k}\gamma_{k}(\va_{k})\right|\geq\veps\right)\leq c\exp\big(-(p/c)\min\{\veps^2, \veps\}\big)\label{eq:gamma_a_concentrate}
\end{equation}
for every $\fmtx \in \mathcal{A}$, $k\in[n]$ and $\veps\geq0$. Correspondingly, there exists $C>0$ such that
\begin{equation}
\E_{k}\big[\left|\gamma_{k}(\va_{k})-\E_{k}\gamma_{k}(\va_{k})\right|^{\epn}\big]\leq m!(C/p)^{m/2}.\label{eq:gamma_a_bd_moments}
\end{equation}
Similarly, there exists $c>0$ and $C > 0$, such that
\begin{equation}
\P_{k}\left(\left|\gamma_{k}\left(\vb_{k}\right)-\E_{k}\gamma_{k}\left(\vb_{k}\right)\right|\geq\veps\right)\leq2\exp\big(-cp\min\{\veps^2, \veps\}\big)\label{eq:gamma_b_concentrate}
\end{equation}
and
\begin{equation}
\E_{k}\big[\left|\gamma_{k}\left(\vb_{k}\right)-\E_{k}\gamma_{k}\left(\vb_{k}\right)\right|^{\epn}\big]\leq  m!(C/p)^{m/2} \label{eq:gamma_b_bd_moments}
\end{equation}
for every $\fmtx \in \mathcal{A}$, $k\in[n]$, $\veps\geq0$, and
$\epn\in\mathbb{Z}^{+}$,
\end{lem}
\begin{proof}
We first recall the definition of $\mH_{\bs k}$ in \eqref{eq:looH}. Since $\regu(x)$ is $\lambda$-strongly convex, and for $\fmtx\in{\cal A}$, $\tau_{1}\mSig\preceq\frac{\lambda}{2}\mI_{p}$, we must have $\mH_{\bs k}\succeq\frac{\lambda}{2}\mI_{p}$ and thus $\norm{\mH_{\bs k}^{-1}}\le\frac{2}{\lambda}$. (See Remark~\ref{rem:convex} for additional details.)

The concentration inequality (\ref{eq:gamma_a_concentrate}) then directly follows from \cite[Lemma 1]{louart2018random} and the fact that $\norm{\fmtx} \le 1 + 2\sqrt{\eta} < \infty$ for $\fmtx \in \mathcal{A}$. To show (\ref{eq:gamma_b_concentrate}), we note that $\vb_{k}\sim\calN\left(\boldsymbol{0},\mSig\right)$. Thus,
$\vb_{k}$ can be represented as $\vb_{k}=\mSig^{\frac{1}{2}}\vz_{k}$,
where $\vz_{k}\sim\calN\left(\boldsymbol{0},\mI_{p}\right)$. It follows that $\gamma_{k}\left(\vb_{k}\right)=\frac{\vz_{k}\tran\mSig^{\frac{1}{2}}\mH_{\bs k}^{-1}\mSig^{\frac{1}{2}}\vz_{k}}{p}$.
Since $\mH_{\bs k}^{-1}\preceq\frac{2}{\lambda}\mI_{p}$,
we have $\norm{\mSig^{\frac{1}{2}}\mH_{\bs k}^{-1}\mSig^{\frac{1}{2}}} \le \frac{2}{\lambda}\norm{\mSig} \le \frac{2}{\lambda} (\mu_1^2 \norm{\fmtx}^2 + \mu_2^2) < \infty$ for $\fmtx \in \mathcal{A}$. Applying the Hanson-Wright inequality (see, e.g., \cite[Theorem 6.2.1]{vershynin2018high}) then gives us the concentration inequality in \eref{gamma_b_concentrate}.

By applying the inequalities in \eref{X_bd_moments_expo} and \eref{X_bd_moments_gauss}, we can obtain the moment bounds \eref{gamma_a_bd_moments} and \eref{gamma_b_bd_moments} from \eref{gamma_a_concentrate} and \eref{gamma_b_concentrate}, respectively.
\end{proof}
\begin{lem}\label{lem:gamma_moments_bd1}
There exists a function
$B(\epn)$, $\epn\in\mathbb{Z}^{+}$ such that
\begin{equation}
\label{eq:gamma_moments_bd_goodF}
\begin{aligned}
    &\sup_{\fmtx \in \mathcal{A}, k \in [n]} \Eone{k}\left[\gamma_{k}^{\epn}\left(\va_k\right)\right]\leq B(\epn) \quad\text{and}\quad \\
    &\sup_{\fmtx \in \mathcal{A}, k \in [n]} \Eone{k}\left[\gamma_{k}^{\epn}\left(\vb_k\right)\right]\leq B(\epn).
\end{aligned}
\end{equation}
\end{lem}
\begin{proof}
Let $\vr=\va_{k}$ or $\vb_{k}$. We first show there exists $C>0$ such that
\begin{equation}
\label{eq:E_gamma_bd}
\Eone{k}\left[\gamma_{k}\left(\vr\right)\right]\leq C,
\end{equation}
for any $k\in[n]$ and $\fmtx\in\mathcal{A}$. By definition, $\Eone{k}[\gamma_{k}\left(\vr\right)]$ can be bounded as follows:
\begin{equation}
\begin{aligned}
\label{eq:ErrT_genbd}
\Eone{k}[\gamma_{k}\left(\vr\right)] & =\frac{1}{p}\Tr\left[\mH_{\bs k}^{-1}\Eone{k}\left( \vr \vr\tran \right)\right] \leq \frac{1}{p}\|\mH_{\bs k}^{-1}\|_\text{F}\, \| \Eone{k}\left(\vr\vr\tran\right)\|_\text{F}\\
&\leq \|\mH_{\bs k}^{-1}\|\cdot \|\Eone{k}\left(\vr\vr\tran\right)\|.
\end{aligned}
\end{equation}
For $\vr=\vb_{k}$, recall that
$\E(\vb_{k}\vb_{k}\tran)=\mu_{1}^{2}\fmtx\tran\fmtx+\mu_{2}^{2}\mI_{p}$. Moreover, for $\fmtx\in{\cal A}$, $\|\fmtx\|\leq1+2\sqrt{\eta}$ [see \eref{def_setA2}] and $\|\mH_{\bs k}^{-1}\|\leq\frac{2}{\lambda}$.
Therefore, from \eqref{eq:ErrT_genbd}, there exists $C>0$ such that
$\Eone{k}[\gamma_{k}\left(\vb\right)] \leq C,$ for every $\fmtx \in \mathcal{A}$ and $k\in[n]$.
For $\vr=\va_{k}$, we can first write $\|\E(\va_{k} \va_{k}\tran)\|$ as:
\begin{equation}\label{eq:EaaT}
\|\E(\va_{k} \va_{k}\tran)\| =\max_{\|\vx\|=1}\vx\tran\E(\va_{k} \va_{k}\tran)\vx =\max_{\|\vx\|=1}\E\left(\va_{k}\tran\vx\right)^{2}.
\end{equation}
As is shown in (\ref{eq:atw_concentrate}), for any $\vx\in\calS^{p-1}$,
$\va_{k}\tran\vx$ is a sub-Gaussian variable, with a sub-Gaussian norm
proportional to $\norm{\kernel'}_\infty\|\mF\|$. It follows from (\ref{eq:X_bd_moments_gauss}) that
\[
\E\left(\va_{k}\tran\vx\right)^{2}\leq c\norm{\kernel'}_\infty^{2}\|\mF\|^{2} \le  c\norm{\kernel'}_\infty^{2} (1 + 2 \sqrt{\eta})
\]
for some $c > 0$, where the last step is due to \eref{def_setA2}. Substituting this inequality into \eref{EaaT} and \eref{ErrT_genbd}, we have verified \eref{E_gamma_bd} for $\vr=\va_{k}$.

To show \eqref{eq:gamma_moments_bd_goodF}, we use the following simple inequality due to convexity: $(x+y)^m \le 2^{m-1} (x^m + y^m)$ for $x, y > 0$ and $m \in \mathbb{Z}^+$. This allows us to write
\begin{align*}
\E_{k}[\gamma_{k}^\epn\left(\vr\right)] & =\E_{k}\left|\gamma_{k}\left(\vr\right)-\E_{k}\gamma_{k}\left(\vr\right)+\E_{k}\gamma_{k}\left(\vr\right)\right|^{\epn}\\
 &\le 2^{\epn-1}\big(\E_{k}\left|\gamma_{k}\left(\vr\right)-\E_{k}\gamma_{k}\left(\vr\right)\right|^{\epn}+\left|\E_{k}\gamma_{k}\left(\vr\right)\right|^{\epn}\big).
\end{align*}
Applying (\ref{eq:gamma_a_bd_moments}),
(\ref{eq:gamma_b_bd_moments}) and \eqref{eq:E_gamma_bd}, we reach the desired bounds in \eref{gamma_moments_bd_goodF}.
\end{proof}

%% file: appendix-boundedness.tex

\subsection{\label{appendix:opt_properties}Characterizations of the Optimization Problems}

In this appendix, we collect some useful properties of the optimization
problems that we encounter when constructing and analyzing the interpolation path based
on Lindeberg's method.

For each $k\in[n]$, define
\begin{equation}
R_{\bs k}(\vw)\bydef\sum_{t\neq k}\loss\big(\tfrac{\vr_{t}\tran\vw}{\sqrt{p}};y_{t}\big)+\sum_{j=1}^{p}\regu(w_{j})+Q(\vw),\label{eq:loo_objective_func_1}
\end{equation}
where $Q(\vw)$ is the function defined in \eref{Q_def}, $\vr_t = \vb_t$ for $1 \le t \le k-1$, and $\vr_t = \va_t$ for $k+1 \le t \le n$.  Let
\begin{equation}
R_{k}(\vw;\vr)\bydef R_{\bs k}(\vw)+\loss\big(\tfrac{\vr{\tran}\vw}{\sqrt{p}};y_{k}\big)\label{eq:original_objective_func_1},
\end{equation}
and
\begin{equation}
S_{k}(\vw;\vr)\bydef\Phi_{\bs k}+\frac{1}{2}(\vw-\loowt{\bs k}){\tran}\mH_{\bs k}(\vw-\loowt{\bs k})+\loss(\tfrac{\vr{\tran}\vw}{\sqrt{p}};y_{k}),\label{eq:quadratic_approx_objctive_function_1}
\end{equation}
where $\mH_{\bs k}$ is the Hessian matrix defined in \eref{looH}, and $\Phi_{\bs k}=\min_{\vw\in\R^{p}}R_{\bs k}(\vw)$. We will be studying the following three related optimization problems:
\begin{align}
\label{eq:originprob_def}
\Phi_{k}(\vr)&=\min_{\vw\in\R^{p}}R_{k}(\vw;\vr),\hspace{3em}\wt_k(\vr)=\argmin{\vw\in\R^{p}}R_{k}(\vw;\vr),\\
\label{eq:looprob_def}
\Phi_{\bs k}&=\min_{\vw\in\R^{p}}R_{\bs k}(\vw),\hspace{4.6em}\loowt{\bs k}=\argmin{\vw\in\R^{p}}R_{\bs k}(\vw),\\
\label{eq:quadprob_def}
\Psi_{k}(\vr)&=\min_{\vw\in\R^{p}}S_{k}(\vw;\vr),\hspace{3.2em}\qwt_k(\vr)=\argmin{\vw\in\R^{p}}S_{k}(\vw;\vr).
\end{align}
As explained in \sref{Lindeberg}, the optimization problems formulated in \eqref{eq:originprob_def}-\eqref{eq:quadprob_def} can be referred to as the ``original problem'', the ``leave-one-out problem'' and the ``quadratic approximation problem'', respectively.

\subsubsection{Deterministic Characterizations}

We first show that the quadratic approximation
problem \eqref{eq:quadprob_def} allows for convenient closed-form solutions.
\begin{lem}
\label{lem:quadapprox_optsol}
For every $k \in [n]$, it holds that
\begin{equation}
\Psi_{k}(\vr)=\Phi_{\bs k}+\calM_{k}\left(\scp\vr{\tran}\loowt{\bs k};\gamma_{k}\left(\vr\right)\right),\label{eq:Rk_cr_decomp_general_2}
\end{equation}
where $\calM_{k}(z;\gamma)$ is the \emph{Moreau envelope} of $\loss\left(x;y_{k}\right)$ as defined in \eqref{eq:Moreau_envelope}, and $\gamma_k(\vr)$ is the quadratic term defined in \eref{gamma_k}. Moreover,
\begin{equation}
\qwt_{k}(\vr)=\loowt{\bs k}-\ell'\big(\scp{\vr{\tran}\qwt_{k}(\vr)};y_{k}\big)\tfrac{\mH_{\bs k}^{-1}\vr}{\sqrt{p}}\label{eq:wtilde_loowt_diff}
\end{equation}
and
\begin{equation}
\scp{\vr{\tran}\qwt_{k}(\vr)}=\prox_k\left(\scp\vr{\tran}\loowt{\bs k};\gamma_{k}\left(\vr\right)\right),\label{eq:scalar_observation}
\end{equation}
where $\prox_k\left(z;\gamma\right)$ denotes the \emph{proximal
operator} of $\loss\left(x;y_{k}\right)$, i.e.,
\[
\prox_k\left(z;\gamma\right)\bydef\argmin x\loss\left(x;y_{k}\right)+\frac{(x-z)^{2}}{2\gamma}.
\]
\end{lem}
\begin{proof}
We have
\begin{align}
&\Psi_{k}(\vr) \nonumber\\
=& \Phi_{\bs k}+\min_{\vw\in\R^{p}}\Big\{\frac{1}{2}(\vw-\loowt{\bs k}){\tran}\mH_{\bs k}(\vw-\loowt{\bs k})\nonumber\\
&\hspace{15em}+\loss\left(\scp{\vr{\tran}\vw};y_{k}\right)\Big\}\nonumber \\
 =&\Phi_{\bs k}+\min_{\tau}\min_{\scp{\vr{\tran}(\vw-\loowt{\bs k})}=\tau}\Big\{\frac{1}{2}(\vw-\loowt{\bs k}){\tran}\mH_{\bs k}(\vw-\loowt{\bs k})\nonumber\\
 &\hspace{13em}+\loss\big(\scp{\vr{\tran}\loowt{\bs k}}+\tau;y_{k}\big)\Big\}\nonumber \\
 & =\Phi_{\bs k}+\min_{\tau}\set{\tfrac{\tau^{2}}{2\gamma_{k}(\vr)}+\loss\big(\scp{\vr{\tran}\loowt{\bs k}}+\tau;y_{k}\big)}.\label{eq:quadratic_t}
\end{align}
By the definition of Moreau envelopes, we immediately get (\ref{eq:Rk_cr_decomp_general_2}). Besides, the
optimal solution $\tau^{*}$ of \eref{quadratic_t} is
\begin{equation}
\tau^{*}=\prox_k\big(\scp{\vr{\tran}\loowt{\bs k}};\gamma_{k}(\vr)\big)-\scp{\vr{\tran}\loowt{\bs k}}.\label{eq:t_opt}
\end{equation}
Since $\scp{\vr{\tran}[\qwt_{k}(\vr)-\loowt{\bs k}]}=\tau^{*}$, we then get (\ref{eq:scalar_observation}). Finally, by using the first order optimality condition $\nabla S_{k}(\vw;\vr)=\boldsymbol{0}$,
we can directly get (\ref{eq:wtilde_loowt_diff}).
\end{proof}
The next result is a deterministic bound for $\left\Vert \wt_k-\qwt_k\right\Vert $,
$\ie$, the distance between the true optimal solution and the solution to the quadratic
approximation problem.
\begin{lem}
\label{lem:qwt_wt_deterministic_bd}For any $\fmtx\in \mathcal{A}$ and $k\in[n]$, there exists $C>0$ such that
\begin{equation}\label{eq:xstar_xtilde_diff}
\begin{aligned}
 &\left\Vert \wt_{k}(\vr)-\qwt_{k}(\vr)\right\Vert \\ 
 \le& C\left|\loss_{k}'\right|^{2}\lsbd\Big(\sup_{t\neq k}\big\{ |\vr_{t}\tran\mH_{\bs k}^{-1}\vr/p|\big\} \big\lVert\tfrac{1}{p}\sum_{t\neq k}\vr_{t}\vr_{t}\tran\big\rVert \cdot {\|\scp\mH_{\bs k}^{-1}\vr\|}{} \\
 &\hspace{5em}+\tfrac{1}{p}\big[\sum_{i=1}^{p}(\vh\tran_{\bs k,i}\vr)^{4}\big]^{\tfrac{1}{2}}\Big),
 \end{aligned}
\end{equation}
where $\loss_{k}'\bydef\loss'\big(\scp{\vr{\tran}\qwt_{k}(\vr)};y_{k}\big)$, $\vh_{\bs k,i}$ denotes the $i$th column of $\mH_{\bs k}^{-1}$, and
\begin{equation}\label{eq:Ls}
\lsbd\bydef\sup_{t\in[n]}\{1+|\cvec_t\tran\sgl|^{K_1}\}.
\end{equation}
Here, $K_1\in\mathbb{Z}^+$ is the constant defined in Assumption \ref{a:loss}.
\end{lem}
\begin{proof}
We follow the proof technique of \cite[Proposition 3.4]{el2018impact}. For
notational simplicity, we write $\wt := \wt_{k}(\vr)$ and $\qwt := \qwt_{k}(\vr)$ in the proof. We start by noting that, since $R_{k}(\vw;\vr)$
is $\tfrac{\lambda}{2}$-strongly convex for $\fmtx\in\mathcal{A}$, we have
\begin{align}
\left\Vert \wt-\qwt\right\Vert  & \leq\frac{2}{\lambda}\left\Vert \nabla R_{k}(\wt;\vr)-\nabla R_{k}(\qwt;\vr)\right\Vert \nonumber \\
 & =\frac{2}{\lambda}\left\Vert \nabla R_{k}(\qwt;\vr)\right\Vert,\label{eq:grad_bd_diff}
\end{align}
where the first inequality is a property of strongly-convex functions (see, e.g., \cite[pp.\ 112--113]{HL2001fundamentals}), and the last equality is due to the optimality condition $\nabla R_{k}(\wt;\vr)=\boldsymbol{0}$. Therefore, to prove \eref{xstar_xtilde_diff}, it suffices to control $\left\Vert \nabla R_{k}(\qwt;\vr)\right\Vert$.

To that end, we note that $\loowt{\bs k}=\argmin{\vw\in\R^{p}}R_{\bs k}(\vw)$ and thus $\nabla R_{\bs k}(\loowt{\bs k})=\boldsymbol{0}$. This allows us to write
\begin{align}
&\nabla R_{k}(\qwt;\vr) \nonumber \\
=& \nabla R_{k}(\qwt;\vr)-\nabla R_{\bs k}(\loowt{\bs k})\nonumber \\
=& \sum_{t\neq k}\loss'\big(\tfrac{\vr_{t}\tran\qwt}{\sqrt{p}};y_{t}\big)\tfrac{\vr_t}{\sqrt{p}}+\nabla h\left(\qwt\right)+\nabla Q(\qwt)+\loss'\big(\tfrac{\vr{\tran}\qwt}{\sqrt{p}};y_{k}\big)\tfrac{\vr}{\sqrt{p}}\nonumber \\
 &\qquad-\Big[\sum_{t\neq k}\loss'\big(\tfrac{\vr_{t}\tran\loowt{\bs k}}{\sqrt{p}};y_{t}\big)\tfrac{\vr_t}{\sqrt{p}}+\nabla h(\loowt{\bs k})+\nabla Q(\loowt{\bs k})\Big]\nonumber \\
=& \Big[\tfrac{1}{p}\sum_{t\neq k}\loss''\left(v_{t};y_{t}\right)\vr_{t}\vr_{t}\tran+\nabla^{2}Q(\loowt{\bs k})\Big](\qwt-\loowt{\bs k})\nonumber \\
 &\qquad +\loss'\big(\tfrac{\vr{\tran}\qwt}{\sqrt{p}};y_{k}\big)\tfrac{\vr}{\sqrt{p}} +\nabla h\left(\qwt\right)-\nabla h(\loowt{\bs k}),\label{eq:grad_R}
\end{align}
where in reaching the last step we have used the intermediate value theorem, with $v_{t}$ being some number that lies between $\tfrac{\vr_{t}\tran\loowt{\bs k}}{\sqrt{p}}$
and $\tfrac{\vr_{t}\tran\qwt}{\sqrt{p}}$. From (\ref{eq:wtilde_loowt_diff})
and the definition of $\mH_{\bs k}$ in (\ref{eq:looH}), we have
\begin{equation}
\begin{aligned}
    &\Big[\tfrac{1}{p}\sum_{t\neq k}\loss''\big(\tfrac{\vr_{t}\tran\loowt{\bs k}}{\sqrt{p}};y_{t}\big)\vr_{t}\vr_{t}\tran \\
    &\hspace{5em}+\diag\{ \regu''(w_{\bs k,i}^{*})\} +\nabla^{2}Q(\loowt{\bs k})\Big]\times(\qwt-\loowt{\bs k})\\
    &\hspace{3em}+\ell'\left(\tfrac{\vr{\tran}\qwt}{\sqrt{p}};y_{k}\right)\tfrac{\vr}{\sqrt{p}}=\boldsymbol{0}.\label{eq:quadapprox_1st_optcond}
\end{aligned}
\end{equation}
Substituting this inequality into (\ref{eq:grad_R}) then gives us
\begin{align}
&\nabla R_{k}(\qwt;\vr) \nonumber \\
=& \tfrac{1}{p}\sum_{t\neq k}\Big[\loss''\left(v_{t};y_{t}\right)-\loss''\big(\tfrac{\vr_{t}\tran\loowt{\bs k}}{\sqrt{p}};y_{t}\big)\Big]\vr_{t}\vr_{t}\tran(\qwt-\loowt{\bs k})\nonumber \\
 &\qquad +\nabla h\left(\qwt\right)-\nabla h(\loowt{\bs k})-\diag\{ \regu''(w_{\bs k,i}^{*})\} (\qwt-\loowt{\bs k})\nonumber \\
=& -  \tfrac{1}{p}\sum_{t\neq k}\Big[\loss''\left(v_{t};y_{t}\right)-\loss''\big(\tfrac{\vr_{t}\tran\loowt{\bs k}}{\sqrt{p}};y_{t}\big)\Big]\vr_{t}\vr_{t}\tran\bigg(\tfrac{\loss_{k}'\mH_{\bs k}^{-1}\vr}{\sqrt{p}}\bigg)\nonumber \\
 &\qquad +\nabla h\left(\qwt\right)-\nabla h(\loowt{\bs k})-\diag\{ \regu''(w_{\bs k,i}^{*})\} (\qwt-\loowt{\bs k}),\label{eq:grad_Rk_qwt_2}
\end{align}
where in the last step, $\loss_{k}'=\ell'\left(\tfrac{\vr{\tran}\qwt}{\sqrt{p}};y_{k}\right)$,
and we have used (\ref{eq:wtilde_loowt_diff}). By the intermediate value theorem,
\begin{equation}
\label{eq:lsbd_bd_2}
\begin{aligned}
&\Big\Vert\tfrac{1}{p}\sum_{t\neq k}\Big[\loss''\left(v_{t};y_{t}\right)-\loss''\big(\tfrac{\vr_{t}\tran\loowt{\bs k}}{\sqrt{p}};y_{t}\big)\Big]\vr_{t}\vr_{t}\tran\Big\Vert \\
\leq & \sup_{t\neq k}\big\{\big|\loss'''(u_{t};y_{t})\scp\vr_{t}\tran(\qwt-\loowt{\bs k})\big|\big\}\Big\Vert\tfrac{1}{p}\sum_{t\neq k}\vr_{t}\vr_{t}\tran\Big\Vert\\
\leq& C\underbrace{\sup_{t\in[n]}\{1+|\cvec_t\tran\sgl|^{K_1}\}}_{=\lsbd} \sup_{t\neq k}\left\{ \left|\scp\vr_{t}\tran(\qwt-\loowt{\bs k})\right|\right\}\Big\Vert\tfrac{1}{p}\sum_{t\neq k}\vr_{t}\vr_{t}\tran\Big\Vert,
\end{aligned}
\end{equation}
where $u_t$ is some number lying between $v_t$ and $\tfrac{\vr_{t}\tran\loowt{\bs k}}{\sqrt{p}}$, and the last step follows from Assumption \ref{a:loss}.
From (\ref{eq:grad_Rk_qwt_2}) and \eqref{eq:lsbd_bd_2}, there exists $C>0$,
\begin{align*}
&\left\Vert \nabla R_{k}(\qwt;\vr)\right\Vert \\
\leq & C\lsbd\left|\loss_{k}'\right|\sup_{t\neq k}\left\{ \left|\scp\vr_{t}\tran(\qwt-\loowt{\bs k})\right|\right\} \Big\Vert\tfrac{1}{p}\sum_{t\neq k}\vr_{t}\vr_{t}\tran\Big\Vert\cdot\left\Vert \scp \mH_{\bs k}^{-1}\vr\right\Vert \nonumber \\
 & \hspace{5em}+\tfrac{\|\regu'''\|_{\infty}}{2}\Big[\sum_{i=1}^{p}(\tw_{i}-w_{\bs k,i}^{*})^{4}\Big]^{1/2}\nonumber \\
= & C\lsbd\left|\loss_{k}'\right|^{2}\sup_{t\neq k}\left\{ \left|\vr_{t}\tran\mH_{\bs k}^{-1}\vr/p\right|\right\} \Big\Vert\tfrac{1}{p}\sum_{t\neq k}\vr_{t}\vr_{t}\tran\Big\Vert\cdot\left\Vert \scp\mH_{\bs k}^{-1}\vr\right\Vert \nonumber \\
 & \hspace{5em}+\tfrac{C\left|\loss_{k}'\right|^{2}}{p}\Big[\sum_{i=1}^{p}(\vh\tran_{\bs k,i}\vr)^{4}\Big]^{1/2},
\end{align*}
where in the last step, we have used (\ref{eq:wtilde_loowt_diff}) and the assumption that $\|\regu'''\|_{\infty}<\infty$. Substituting this inequality into \eref{grad_bd_diff} and using the fact that $L_S \ge 1$, we conclude the proof.
\end{proof}

\subsubsection{Bounding $\|\bf{w}_k^*\|$}

We will introduce a function $G(\vw)$ to wrap up
all the terms in (\ref{eq:original_objective_func_1}), except the
loss function, $\ie$,
\begin{equation}
G(\vw)=\sum_{j=1}^{p}\regu\left(w_{j}\right)+Q\left(\vw\right),\label{eq:Gw_def_2}
\end{equation}
where $Q(\vw)$ is defined in \eqref{eq:Q_def}.
%
\begin{lem}
Let $\wt_k(\vr)$ denote either $\wt_k(\va_k)$ or $\wt_k(\vb_k)$, and $\loowt{\bs k}$ be the leave-one-out solution in \eref{looprob_def}. There exists $C,c>0$ such that for every
$k \in [n]$,
\begin{equation}
\P\left(\scp\left\Vert \wt_{k}(\vr)\right\Vert \geq C\right)\leq ce^{-(\log p)^2/c}\label{eq:wt_norm_prob_bd}
\end{equation}
and
\begin{equation}
\P\left(\scp\Vert \loowt{\bs k}\Vert \geq C\right)\leq ce^{-(\log p)^2/c}.\label{eq:wt_leave_norm_prob_bd}
\end{equation}
\end{lem}
\begin{proof}
Recall the definition of the set $\mathcal{A}_2$ in \eref{def_setA2}. We start by noting that
\begin{align}
&\P\left(\scp\left\Vert \wt_{k}(\vr)\right\Vert \geq C\right) \nonumber \\
\le& \P\left(\scp\left\Vert \wt_{k}(\vr)\right\Vert \geq C \cap \fmtx \in \mathcal{A}_2\right) + \P(\fmtx \in \mathcal{A}^c_2)\nonumber\\
\le& \sup_{\fmtx \in \mathcal{A}_2} \Pcnd{\fmtx}\left(\scp\left\Vert \wt_{k}(\vr)\right\Vert \geq C\right) + 2e^{-cp},\label{eq:wt_norm_prob_bd_cond}
\end{align}
where the last inequality is due to \eref{A3_p}. Therefore, to show \eref{wt_norm_prob_bd}, it suffices to bound the conditional probability $\Pcnd{\fmtx}\left(\scp\left\Vert \wt_{k}(\vr)\right\Vert \geq C\right)$ for any fixed $\fmtx \in \mathcal{A}_2$.

On the one hand, since $\loss(x;y)\geq0$,
we have
\begin{align*}
G(\wt_{k}(\vr)) & \leq\sum_{t=1}^{n}\loss\left(\scp\vr_{t}\tran\wt_{k}(\vr);y_{t}\right)+G(\wt_{k}(\vr))\\
 & \le\sum_{t=1}^{n}\loss\left(0;y_{t}\right)+G(\boldsymbol{0}),
\end{align*}
where the last step is due to the fact that $\wt_{k}(\vr)$ is the optimal solution.
On the other hand, for $\fmtx\in\mathcal{A}_2$, $G(\vw)$
is $\frac{\lambda}{2}$-strongly convex. This then gives us
\begin{align*}
G(\wt_{k}(\vr)) & \geq G(\boldsymbol{0})+\nabla{\tran}G(\boldsymbol{0})\wt_{k}(\vr)+\frac{\lambda}{4}\|\wt_{k}(\vr)\|^{2}\\
 & \geq G(\boldsymbol{0})-\|\nabla G(\boldsymbol{0})\|\|\wt_{k}(\vr)\|+\frac{\lambda}{4}\|\wt_{k}(\vr)\|^{2}.
\end{align*}
Combining the above upper and lower bounds for $G(\wt_k)$, we have
\[
\frac{\lambda}{4}\|\wt_{k}(\vr)\|^{2}-\|\nabla G(\boldsymbol{0})\|\|\wt_{k}(\vr)\|\leq\sum_{t=1}^{n}\loss\left(0;y_{t}\right)
\]
and thus
\begin{align}
\frac{\|\wt_{k}(\vr)\|}{\sqrt{p}} & \leq\frac{2\|\nabla G(\boldsymbol{0})\|+2\sqrt{\|\nabla G(\boldsymbol{0})\|^{2}+\lambda\sum_{t=1}^{n}\loss\left(0;y_{t}\right)}}{\lambda\sqrt{p}}\nonumber \\
 & \leq\frac{2}{\lambda\sqrt{p}}\big[2\|\nabla G(\boldsymbol{0})\|+\big(\lambda\textstyle\sum_{t=1}^{n}\loss\left(0;y_{t}\right)\big)^{1/2}\big].\label{eq:wt_norm_bd}
\end{align}
By its definition in (\ref{eq:Gw_def_2}), $\nabla G(\boldsymbol{0})=\regu'({0})\boldsymbol{1}_{p}+\tau_{2}\mu_{1}\sqrt{p}\fmtx{\tran}\sgl$, and thus
\[
\norm{\nabla G(\boldsymbol{0})} \le C_1\sqrt{p},
\]
where $C_1 = \abs{h'(0)} + \tau_2 \mu_1 (1+2\sqrt{\eta})$ and we have used \eref{def_setA2}. It then follows from (\ref{eq:wt_norm_bd}) that
\begin{equation}\label{eq:wt_norm_bd_1}
\tfrac{\|\wt_k(\vr)\|}{\sqrt{p}}\le \tfrac{2}{\lambda}\Big[2C_1 +\big(\tfrac{\lambda}{p}\textstyle\sum_{t=1}^{n}\loss\left(0;y_{t}\right)\big)^{1/2}\Big].
\end{equation}

From Assumption \ref{a:loss}, we know there exists some $ C_{2},C_{2}'>0$ such that for any $B>0$
\begin{align}
&\P\big(\tfrac{1}{p}\textstyle\sum_{t=1}^{n}\loss\left(0;y_{t}\right)\geq C_2\big) \\
\leq& \P\big(\tfrac{1}{n}\textstyle\sum_{t=1}^{n}|s_t|^{K_1}\geq 2C_{2}'\big)\nonumber\\
\tleq{}& \P\big(\tfrac{1}{n}\textstyle\sum_{t=1}^{n}|s_t|^{K_1}\charfn_{\{|s_t|\leq B\}}\geq 2C_{2}'\big)
+\P(\max_{t\in[n]}|s_t| > B)\nonumber\\
\tleq{(a)}& \P\big(\tfrac{1}{n}\textstyle\sum_{t=1}^{n}(|s_t|^{K_1}\charfn_{\{|s_t|\leq B\}}-e_B)\geq C_{2}'\big)\nonumber\\
&\hspace{2em}+\P(\max_{t\in[n]}|s_t| > B)\nonumber\\
\tleq{(b)}& \exp\big(-\tfrac{2n(C_{2}')^2}{B^{2K_1}}\big)+2n\exp(-B^2/2),\label{eq:concentration_aver_l_y}
\end{align}
where $s_t\iid \mathcal{N}(0,1)$, $e_B:=\E \big(|s_t|^{K_1}\charfn_{\{|s_t|\leq B\}}\big)$, (a) follows from the fact $e_B\leq \E|s_t|^{K_1}<\infty$ and (b) follows from Hoeffding's inequality for bounded random variables \cite[Theorem 2.2.6]{vershynin2018high} and the tail bound for standard Gaussian:  $\P(\abs{s} \ge t) \le 2 e^{-t^2/2}$. Letting $B=1+\log p$ in \eqref{eq:concentration_aver_l_y}, we have
\begin{equation}
\label{eq:empi_mean_l0y_prbd}
\P\Big(\tfrac{1}{p}\textstyle\sum_{t=1}^{n}\loss\left(0;y_{t}\right)\geq C_2\Big)\leq C_{3}e^{-(\log p)^2/C_{3}},
\end{equation}
for some $C_3>0$. Combining \eqref{eq:wt_norm_bd_1} and \eqref{eq:empi_mean_l0y_prbd} and choosing $C_4 = (2/\lambda)(2C_1 + \sqrt{\lambda}\sqrt{C_2})$, we get
\[
\Pcnd{\fmtx}\left(\scp\left\Vert \wt_{k}(\vr)\right\Vert \geq C_4\right) \le C_{3}e^{-(\log p)^2/C_{3}}.
\]
As this holds uniformly over all $\mathcal{F} \in \mathcal{A}_2$, we get \eref{wt_norm_prob_bd} from \eref{wt_norm_prob_bd_cond}. The proof of \eref{wt_leave_norm_prob_bd} follows exactly the same steps, and we omit it.
\end{proof}
\begin{lem}
\label{lem:norm_optwt_Expec_bd_A2}Let $\wt_k(\vr)$ denote either $\wt_k(\va_k)$ or $\wt_k(\vb_k)$, and $\loowt{\bs k}$ be the leave-one-out solution in \eref{looprob_def}. There exists a function $B(\epn)$
of $\epn\in\mathbb{Z}^{+}$ such that for any $\fmtx\in{\cal A}_2$, $p\geq 2$ and $k \in [n]$,
\begin{equation}
\Ecnd{\fmtx}\left(\scp\|\wt_{k}(\vr)\|\right)^{\epn}\leq B(\epn)(\log p)^{\epn K_1/2}\label{eq:norm_optwt_Expec_bd_goodF}
\end{equation}
and
\begin{equation}
\Ecnd{\fmtx}\left(\scp\|\loowt{\bs k}\|\right)^{\epn}\leq B(\epn)(\log p)^{\epn K_1/2},\label{eq:norm_loowt_Expec_bd_goodF}
\end{equation}
where $K_1$ is the constant defined in Assumption \ref{a:loss}.
\end{lem}
\begin{proof}
Using the simple inequality $\sqrt{x} < 1 + x$ for $x \ge 0$, we can deduce from \eref{wt_norm_bd_1} that
\[
\scp\|\wt_k(\vr)\| \le C \big[1 + \tfrac{1}{n}\textstyle\sum_{t=1}^{n}\loss\left(0;y_{t}\right)\big].
\]
It follows that
\begin{equation}
\label{eq:wk_norm_bd}
\left(\scp\|\wt_k(\vr)\|\right)^{\epn} \leq (2C)^{\epn} \big[1+\big(\tfrac{1}{n}\textstyle\sum_{t=1}^{n}\loss\left(0;y_{t}\right)\big)^{\epn}\big].
\end{equation}
According to Assumption \ref{a:loss}, there exists $c>0$ such that for any $\veps \geq 0$,
\begin{align}
\P\Big[\big(\tfrac{1}{n}\textstyle\sum_{t=1}^{n}\loss\left(0;y_{t}\right)\big)^{\epn}\geq \veps\Big] \leq& n \P\big(\loss\left(0;y_{t}\right)\geq \veps^{1/\epn}\big) \nonumber \\
\leq& cp\exp\big(-\veps^{\tfrac{2}{\epn K_1}}/c\big).
\end{align}
Then by the integral identity $\E \abs{X} = \int_{0}^{\infty}\P(\abs{X}>t)dt$, there exists some $C>0$ such that for $p\geq 2$,
\begin{equation}
\label{eq:ly_mom_bd}
\E\big[\tfrac{1}{n}\textstyle\sum_{t=1}^{n}\loss\left(0;y_{t}\right)\big]^{\epn}\leq \epn !(C\log p)^{\epn K_1/2}.
\end{equation}
Combining \eqref{eq:wk_norm_bd} and \eqref{eq:ly_mom_bd} gives us (\ref{eq:norm_optwt_Expec_bd_goodF}). The proof of \eref{norm_loowt_Expec_bd_goodF} follows the same steps, and we omit it.
\end{proof}

\subsubsection{Bounding $|\scp{\mathbf{r}}{\tran}\widetilde{\mathbf{w}}_{k}(\mathbf{r})|$}
\begin{lem}\label{lem:lnresponse_bd_2}
Let $\qwt_{k}(\vr)$ be the optimal solution to the quadratic optimization problem as defined in \eqref{eq:quadprob_def}. There exists
$c>0$ such that for any $k\in[n]$ and $\veps \geq 0$
\begin{equation}
\P\left(|\scp\vr{\tran}\qwt_{k}(\vr)|\geq\veps\right)\leq c\exp\big(-\veps^{\tfrac{4}{K_1+2}}/c\big) + ce^{-(\log p)^2/c},\label{eq:lnresponse_quadratic_bd}
\end{equation}
where $\vr = \va_{k}$ or $\vb_{k}$ and $K_1\in\mathbb{Z}^+$ is the constant defined in Assumption \ref{a:loss}.
\end{lem}
\begin{proof}
We first show there exists $c>0$ such that for any $\veps \geq 0$,
\begin{equation}
\P\left(|\scp\vr{\tran}\loowt{\bs k}|\geq\veps\right)\leq ce^{-\veps^{2}/c}+ce^{-(\log p)^2/c},\label{eq:lnresponse_indep_bd}
\end{equation}
where $\loowt{\bs k}$ is the leave-one-out solution defined in \eref{looprob_def}.

Note that $\loowt{\bs k}$ is independent of $\va_k$. From (\ref{eq:atw_concentrate}), there exists $c>0$ such that for any $\veps\geq 0$, when conditioned on $\fmtx$ and $\loowt{\bs k}$,
\begin{equation}
\P\left(|\tfrac{1}{\sqrt{p}}\va_{k}{\tran}\loowt{\bs k}|\geq\veps\mid\fmtx,\loowt{\bs k}\right)\leq ce^{-\frac{p\veps^{2}}{c\|\loowt{\bs k}\|^{2}\|\fmtx\|^{2}\norm{\kernel'}_\infty^{2}}}.\label{eq:axk_concentrate1}
\end{equation}
Define the following event:
\[
\mathcal{E} \bydef \big\{ \scp\norm{\loowt{\bs k}}\leq C,~\norm{\fmtx}\leq 1+2\sqrt{\eta} \big\},
\]
where $C$ is the same constant as the one in \eqref{eq:wt_leave_norm_prob_bd}. Then it holds that
\begin{align*}
&\P\left(|\tfrac{1}{\sqrt{p}}\va_{k}{\tran}\loowt{\bs k}|\geq\veps\right)\\
=& \Eone{\fmtx,\loowt{\bs k}}\P\left(|\tfrac{1}{\sqrt{p}}\va_{k}{\tran}\loowt{\bs k}|\geq\veps\mid\fmtx,\loowt{\bs k}\right)\\
\leq& \Eone{\fmtx,\loowt{\bs k}} \Big[\charfn_{\mathcal{E}} \P\big(|\tfrac{1}{\sqrt{p}}\va_{k}{\tran}\loowt{\bs k}|\geq\veps\mid\fmtx,\loowt{\bs k}\big)\Big]+\P(\mathcal{E}^c).
\end{align*}
It then follows from (\ref{eq:spectral_norm_F}),  \eqref{eq:wt_leave_norm_prob_bd}, (\ref{eq:axk_concentrate1}) and Assumption \ref{a:kernel} that there exists $c>0$ such that
\begin{align}
\label{eq:at_wt_prob_bd_2}
\P\left(|\tfrac{1}{\sqrt{p}}\va_{k}{\tran}\loowt{\bs k}|\geq\veps\right)\leq ce^{-\veps^{2}/c}+ce^{-(\log p)^2/c},
\end{align}
for every $k \in [n]$ and $\veps\geq 0$. The case of $\vr = \vb_{k}$ for (\ref{eq:lnresponse_indep_bd}) can be
proved in the same way and we omit its proof.

Next, we show (\ref{eq:lnresponse_quadratic_bd}) by using the characterization in (\ref{eq:scalar_observation}). Since
\begin{align}
\label{eq:prox_0_def}
\prox_{k}\big(0;\gamma_{k}\left(\vr\right)\big)=\argmin{x}\Big\{\frac{x^{2}}{2}+\gamma_{k}(\vr)\loss\left(x;y_{k}\right)\Big\}
\end{align}
and $\gamma_{k}(\vr)$, $\loss\left(x;y_{k}\right)\geq0$, we can
get
\begin{align*}
&\frac{1}{2}\prox_{k}\big(0;\gamma_{k}\left(\vr\right)\big)^{2}\\
\leq& \frac{1}{2}\prox_{k}\big(0;\gamma_{k}\left(\vr\right)\big)^{2}+\gamma_{k}(\vr)\loss\big(\prox_{k}\left(0;\gamma_{k}\left(\vr\right)\right);y_{k}\big)\\
\leq& \gamma_{k}(\vr)\loss\left(0;y_{k}\right),
\end{align*}
where in the last step, we substitute $x=0$ in the right-hand side of \eqref{eq:prox_0_def} and use the optimality of $\prox_{k}\big(0;\gamma_{k}\left(\vr\right)\big)$. This gives us $\left|\prox_{k}\left(0;\gamma_{k}\left(\vr\right)\right)\right|\leq\sqrt{2\gamma_{k}\left(\vr\right)\loss\left(0;y_{k}\right)}$.
By the non-expansiveness of proximal operators, we can get
\begin{align}
\abs{\scp{\vr{\tran}\qwt_{k}(\vr)}} &= \left|\prox_{k}\left(\scp\vr{\tran}\loowt{\bs k};\gamma_{k}(\vr)\right)\right|\nonumber\\
&\leq\sqrt{2\gamma_{k}\left(\vr\right)\loss\left(0;y_{k}\right)}+\left|\scp\vr{\tran}\loowt{\bs k}\right|.\label{eq:prox_bd}
\end{align}
From Assumption \ref{a:loss}, $\loss(0;y_{k})\leq C(1+|\cvec_k\tran \sgl|^{K_1})$, with $\cvec_k\tran \sgl\sim\mathcal{N}(0,1)$, so by standard Gaussian concentration bound, there exists $c>0$ such that for any $\veps \geq 0$,
\begin{equation}
\label{eq:l0y_bd}
\P\big(\loss(0;y_{k})\geq \veps\big) \leq c\exp(-\veps^{2/K_1}/c).
\end{equation}
On the other hand, from Lemma \ref{lem:gamma_concentrate} and Lemma \ref{lem:gamma_moments_bd1}, there exists $c>0$ such that for any $\veps\geq 0$
\begin{equation}
\label{eq:gammakr_bd}
\P(\gamma_k(\vr) \ge \veps) \leq c\exp(-\veps/c).
\end{equation}
Then it follows from \eqref{eq:prox_bd}, \eqref{eq:l0y_bd}, \eqref{eq:gammakr_bd} and \eqref{eq:lnresponse_indep_bd} that there exists $c>0$ such that for any $\veps\geq0$,
\begin{align}
&\P\left(|\scp\vr{\tran}\qwt_{k}(\vr)|\geq\veps\right)\nonumber \\ 
\le& \P\Big[\gamma_k(\vr) \ge \big(\tfrac{\veps}{2\sqrt{2}}\big)^{\tfrac{2K_1}{K_1+2}}\Big] + \P\Big[\loss(0, y_k) \ge \big(\tfrac{\veps}{2\sqrt{2}}\big)^{\tfrac{4}{K_1+2}} \Big] \nonumber \\
&\hspace{2em}+ \P\left(|\tfrac{1}{\sqrt{p}}\vr{\tran}\loowt{\bs k}| \ge \veps/2\right)\nonumber\\
\leq& c\exp\big(-\veps^{\tfrac{4}{K_1+2}}/c\big)+ce^{-(\log p)^2/c}.
\end{align}
This concludes our proof.
\end{proof}
\begin{lem}
\label{lem:lk_moments_bd_goodF}There exists a function $B(m)$, $m\in\mathbb{Z}^{+}$,
such that for every $\fmtx\in{\cal A}$ and $p\geq 2$,
\begin{equation}
\Ecnd{\fmtx}\left|\loss'(\scp\vr{\tran}\qwt_{k}(\vr);y_{k})\right|^{m}\leq B(m)(\log p)^{\epn K_1},\label{eq:lk_moments_bd_goodF}
\end{equation}
where $\vr=\va_{k}$ or $\vb_{k}$ and $K_1\in\mathbb{Z}^+$ is the constant defined in Assumption \ref{a:loss}.
\end{lem}
\begin{proof}
We start by showing that $\Ecnd{\fmtx}\big|\scp\vr{\tran}\loowt{\bs k}\big|^{\epn}$ is bounded. Indeed, from the independence of $\loowt{\bs k}$ and $\vr$, we can apply (\ref{eq:akw_bd_moments}), (\ref{eq:bkw_bd_moments}), and \eref{norm_loowt_Expec_bd_goodF} to get for $p\geq 2$,
\begin{equation}
\label{eq:lnresponse_indep_bd_goodF_2}
\begin{aligned}
    \Ecnd{\fmtx}\left|\scp\vr{\tran}\loowt{\bs k}\right|^{\epn} &\le B_{1}(\epn)\Ecnd{\fmtx}\big(\scp\|\loowt{\bs k}\|\big)^{\epn} \\
    &\le B_{2}(\epn)(\log p)^{\epn K_1/2},
\end{aligned}
\end{equation}
where $B_1(\epn)$ and $B_2(\epn)$ are two constants that depend on $\epn$. Using \eref{prox_bd}, we have for $p\geq 2$,
\begin{align}
&\Ecnd{\fmtx}\left|\scp\vr{\tran}\qwt_{k}(\vr)\right|^{\epn} \nonumber \\ 
\le& \Ecnd{\fmtx}\Big(\sqrt{2\gamma_{k}\left(\vr\right)\loss\left(0;y_{k}\right)}+\left|\scp\vr{\tran}\loowt{\bs k}\right|\Big)^{\epn}\nonumber\\
\le& 3^{\epn-1}\Ecnd{\fmtx}\left([\gamma_{k}\left(\vr\right)]^{\epn}+[\loss\left(0;y_{k}\right)]^{\epn}+\left|\scp\vr{\tran}\loowt{\bs k}\right|^{\epn}\right)\nonumber\\
\le& B_{3}(\epn)(\log p)^{\epn K_1/2},\label{eq:lnresponse_quadratic_bd_goodF}
\end{align}
where $B_3(\epn)$ is a constant that depend on $\epn$ and the last step follows from (\ref{eq:gamma_moments_bd_goodF}), (\ref{eq:lnresponse_indep_bd_goodF_2}) and Assumption \ref{a:loss}.

Now we are ready to obtain \eqref{eq:lk_moments_bd_goodF}.
By Assumption \ref{a:loss}, there exists $C, C_1>0$ such that for any $x$,
\begin{align}
&\abs{\loss'\left(x;y_k\right)}\nonumber\\
\leq& \abs{\loss'\left(0;y_k\right)}+\int_{-|x|}^{|x|}\abs{\loss''\left(t;y_k\right)}dt\nonumber\\
\leq& \abs{\loss'\left(0;y_k\right)} +2\abs{\loss''\left(0;y_k\right)}|x| +\int_{-|x|}^{|x|} \int_{-|t|}^{|t|}\abs{\loss'''\left(u;y_k\right)}dudt\nonumber\\
\leq& C \big(\left| \vg_k\tran \sgl\right|^{K_{1}}+1\big) (1+2|x|+2{|x|^2})\nonumber\\
\leq& C_1(|x|^{2}+1)(\left| \vg_k\tran \sgl\right|^{K_{1}}+1),
\label{eq:loss_derivative_bd_2}
\end{align}
so we can get
\begin{align*}
&\Ecnd{\fmtx}\left|\loss'(\scp\vr{\tran}\qwt_{k}(\vr);y_{k})\right|^{m} \\
\leq& C_1(m)\Ecnd{\fmtx}\Big(\left|\scp{\vr{\tran}\qwt_{k}(\vr)}\right|^{2}+1\Big)^{m}\Big(|\cvec_{k}{\tran}\sgl|^{K_{1}}+1\Big)^{m}\\
\leq& C_2(m)\sqrt{\Ecnd{\fmtx}\Big(\left|\scp{\vr{\tran}\qwt_{k}(\vr)}\right|^{4m}+1\Big)}\sqrt{\E\Big(|\cvec_{k}{\tran}\sgl|^{2mK_{1}}+1\Big)},
\end{align*}
where $C_1(m),C_2(m)>0$ are two constants that depend on $m$. Then \eqref{eq:lk_moments_bd_goodF} can be proved by using (\ref{eq:lnresponse_quadratic_bd_goodF}) and standard moment bounds for $\cvec_{k}{\tran}\sgl\sim\mathcal{N}(0,1)$.
\end{proof}

\subsubsection{Bounding $\left\Vert \protect\mathbf{w}^{*}_{k}(\mathbf{r})-\tilde{\mathbf{w}}_{k}(\mathbf{r})\right\Vert $}
\label{appendix:quaddiff_prob_bd}
\begin{lem}
\label{lem:optwt_qwt_prob_bd}
There exists $c>0$ such that for every
$k\in[n]$ and $\veps \geq 0$,
\begin{equation}
\label{eq:optwt_qwt_prob_bd}
\begin{aligned}
&\P\big(\| \wt_{k}(\vr)-\qwt_{k}(\vr) \| \geq\veps \big)\\
\leq& cp\exp\Big[ -\min\big\{(\sqrt{p}\veps)^{\tfrac{2}{5K_1+6}}, (\sqrt{p}\veps)^{\tfrac{4}{5K_1+6}}, (\log p)^2\big\} / c \Big],
\end{aligned}
\end{equation}
where $\vr=\va_{k}$ or $\vb_{k}$ and $K_1\in\mathbb{Z}^+$ is the constant defined in Assumptions \ref{a:loss}.
\end{lem}
\begin{proof}
For notational simplicity, we write $\wt:=\wt_{k}(\vr)$ and $\qwt:=\qwt_{k}(\vr)$ in the proof. $C > 0$ and $c > 0$ denote constants whose values can change from one line to the other. From (\ref{eq:xstar_xtilde_diff}), we have that
\begin{equation}
\begin{aligned}
    &\left\Vert \wt-\qwt\right\Vert \\
    \leq&  C\left|\loss_{k}'\right|^{2}\lsbd\Big(\sup_{t\neq k}\big\{ |\vr_{t}\tran\mH_{\bs k}^{-1}\vr/p|\big\} \big\lVert\tfrac{1}{p}\textstyle\sum_{t\neq k}\vr_{t}\vr_{t}\tran\big\rVert \cdot {\|\scp\mH_{\bs k}^{-1}\vr\|}{}\\
    &\hspace{8em}+\tfrac{1}{p}\big[\textstyle\sum_{i=1}^{p}(\vh\tran_{\bs k,i}\vr)^{4}\big]^{\tfrac{1}{2}}\Big),\label{eq:optwt_qwt_diff_bd_2}
\end{aligned}
\end{equation}
where $\vr=\va_{k}$ or $\vb_{k}$ and $\vh_{\bs k,i}$ denotes the $i$th column of $\mH_{\bs k}^{-1}$.
Therefore, to show \eref{optwt_qwt_prob_bd}, it suffices to control each term on the right-hand side of (\ref{eq:optwt_qwt_diff_bd_2}).

(I) $\tfrac{1}{p}\big[\sum_{i=1}^{p}(\vh\tran_{\bs k,i}\vr)^{4}\big]^{\tfrac{1}{2}}$.
Conditioned on $\fmtx\in{\cal A}_{2}$, $\|\mH_{\bs k}^{-1}\|\leq\frac{2}{\lambda}$
and hence $\norm{\vh_{\bs k,i}}\leq\frac{2}{\lambda}$, for any $i\in[p]$.
Applying (\ref{eq:atw_concentrate}) and (\ref{eq:btw_concentrate})
and taking into account the independence between $\vh_{\bs k,i}$
and $\vr$, we can find a constant $c>0$ such that for any $\veps \geq 0$, $i\in[p]$ and $\fmtx \in \mathcal{A}_2$,
\begin{equation}
\label{eq:optwt_qwt_diff_bd_2_aux_10}
\Pcnd{\fmtx}\big(|\vh\tran_{\bs k,i}\vr|\geq\veps \big)\leq ce^{-\veps^{2}/c}.
\end{equation}
By \eref{spectral_norm_Wishart}, $\P\left({\cal A}_{2}\right)\geq1-c e^{-p/c}$. It then follows that
\[
\P\big(|\vh\tran_{\bs k,i}\vr|\geq\veps\big)\leq ce^{-{\veps^{2}}/{c}}+ce^{-p/c}.
\]
Applying the union bound then gives us
\begin{equation}
\begin{aligned}
\label{eq:optwt_qwt_diff_bd_2_aux_1}
\P\Big(\tfrac{1}{p}\big[\textstyle\sum_{i=1}^{p}(\vh\tran_{\bs k,i}\vr)^{4}\big]^{\tfrac{1}{2}}\geq \veps \Big)&\leq \sum_{i=1}^{p} \P\big(|\vh\tran_{\bs k,i}\vr|>p^{\tfrac{1}{4}}\veps^{\tfrac{1}{2}}\big)\\
&\leq cpe^{-{\sqrt{p}\veps}/{c}}+cpe^{-p/c}.
\end{aligned}
\end{equation}

(II) $\left|\loss_{k}'\right|^2\lsbd$.
Recall from (\ref{eq:xstar_xtilde_diff}) that $\loss_{k}'=\loss'(\frac{\vr{\tran}\qwt}{\sqrt{p}};y_{k})$ and $\lsbd\bydef\sup_{t\in[n]}\{1+|\cvec_t\tran\sgl|^{K_1}\}$.
From \eqref{eq:loss_derivative_bd_2}, we know there exists $C>0$ such that for any $x$,
$\abs{\loss'\left(x;y_k\right)}^2\leq C(|x|^{4}+1)\big(\left| \vg_k\tran \sgl\right|^{2K_{1}}+1\big)$ and thus
\begin{equation}\label{eq:loss_derivative_LS_bd_3}
\abs{\loss'\left(x;y_k\right)}^2 \lsbd \le C_1(\abs{x}^4+1) \big(\textstyle\sup_{t\in[n]} \abs{\vg_t\tran \sgl}^{3K_1}+1\big),
\end{equation}
for some $C_1 > 0$.
Therefore, there exists $c>0$ such that for any
sufficiently large $\veps>0$,
\begin{align}
&\P\left(|\loss_k'|^2 \lsbd\geq C_1\veps\right) \nonumber \\
\le&\P\Big(|\scp\vr{\tran}\qwt|^{4}+1\geq2\left(\tfrac{\veps}{4}\right)^{\tfrac{2K_1+4}{5K_{1}+4}}\Big) \nonumber \\
&\hspace{6em}+\P\Big(\sup_{t \in [n]}\left| \vg_t\tran \sgl \right|^{3K_{1}}+1\geq 2\left(\tfrac{\veps}{4}\right)^{\tfrac{3K_{1}}{5K_{1}+4}}\Big)\nonumber \\ \leq&\P\Big(|\scp\vr{\tran}\qwt|^{4}\geq\left(\tfrac{\veps}{4}\right)^{\tfrac{2K_1+4}{5K_{1}+4}}\Big)\nonumber \\
&\hspace{6em}+\P\Big(\sup_{t \in [n]}\left| \vg_t\tran \sgl\right|^{3K_{1}}\geq\left(\tfrac{\veps}{4}\right)^{\tfrac{3K_{1}}{5K_{1}+4}}\Big)\nonumber \\
 \le& cp\exp\big[-\min\{\veps^{\tfrac{2}{5K_{1}+4}}, (\log p)^2\}/c\big],\label{eq:loss_derivative_LS_prob_bd_5}
\end{align}
where to reach the last step we have used (\ref{eq:lnresponse_quadratic_bd}) and the standard tails bound for Gaussian random variables $\vg_t\tran \sgl$, together with union bound.
Then, by choosing a large enough $c$, we can make \eqref{eq:loss_derivative_LS_prob_bd_5} hold for any $\veps\geq 0$.

(III) $\lVert\tfrac{1}{p}\sum_{t\neq k}\vr_{t}\vr_{t}\tran\rVert\cdot\scp\|\mH_{\bs k}^{-1}\vr\|$. Notice that
\begin{equation}\label{eq:rrt_ab}
\textstyle\lVert\tfrac{1}{p}\sum_{t\neq k}\vr_{t}\vr_{t}\tran\rVert \le \lVert\tfrac{1}{p}\sum_{1\le t \le n}\va_{t}\va_{t}\tran\rVert + \lVert\tfrac{1}{p}\sum_{1\le t \le n}\vb_{t}\vb_{t}\tran\rVert.
\end{equation}
From Lemma \ref{lem:empi_spectral_norm_bd}, we can then find two constants $C > 0$ and $c > 0$ such that
\begin{equation}\label{eq:rrt_bnd}
\sup_{\fmtx \in \mathcal{A}_2} \Pcnd{\fmtx}(\textstyle\lVert\tfrac{1}{p}\sum_{t\neq k}\vr_{t}\vr_{t}\tran\rVert \geq C) \le c e^{-p/c}.
\end{equation}
Moreover, as $\|\mH_{\bs k}^{-1}\| \le 2 / \lambda$ when $\fmtx \in \mathcal{A}_2$, we have from Lemma~\ref{lem:at_bt_norm_bd} that
\begin{equation}\label{eq:hr_bnd}
\sup_{\fmtx \in \mathcal{A}_2} \Pcnd{\fmtx}(\scp\|\mH_{\bs k}^{-1}\vr\| \geq C) \le c e^{-p/c},
\end{equation}
for some $C, c>0$. Combining \eref{rrt_bnd}, \eref{hr_bnd}, and using Lemma \ref{lem:Wishart_spectral_norm}, we have
\begin{equation}
\label{eq:optwt_qwt_diff_bd_2_aux_3}
\P\Big(\lVert\tfrac{1}{p}\sum_{t\neq k}\vr_{t}\vr_{t}\tran\rVert\cdot\scp\|\mH_{\bs k}^{-1}\vr\|\geq C\Big)\leq ce^{-p/c}.
\end{equation}

(IV) $\sup_{t\neq k}\{ |\vr_{t}\tran\mH_{\bs k}^{-1}\vr/p|\} .$
By Lemmas \ref{lem:Wishart_spectral_norm}, \ref{lem:quadratic_asymmetry_concentrate} and the union bound,
we have, for every $\veps\geq0$,
\begin{equation}
\label{eq:optwt_qwt_diff_bd_2_aux_4}
\begin{aligned}
\P\Big(\sup_{t\neq k}\{ |\vr_{t}\tran\mH_{\bs k}^{-1}\vr/p|\} \geq \veps\Big) &\leq cpe^{-p\veps^2/c}+cpe^{-p/c}.\\
\end{aligned}
\end{equation}
Substituting the bounds \eqref{eq:optwt_qwt_diff_bd_2_aux_1}, \eqref{eq:loss_derivative_LS_prob_bd_5}, \eqref{eq:optwt_qwt_diff_bd_2_aux_3} and \eqref{eq:optwt_qwt_diff_bd_2_aux_4} into (\ref{eq:optwt_qwt_diff_bd_2}),
we have for any $\veps\geq0$,
\begin{align}
&\P\left(\left\Vert \wt-\qwt\right\Vert \geq C \veps \right) \nonumber\\
\leq&
\P\Big(\sup_{t\neq k}\big\{ |\vr_{t}\tran\mH_{\bs k}^{-1}\vr/p|\big\} \big\lVert\tfrac{1}{p}\sum_{t\neq k}\vr_{t}\vr_{t}\tran\big\rVert\tfrac{\|\mH_{\bs k}^{-1}\vr\|}{\sqrt{p}}\geq\tfrac{(\sqrt{p}\veps)^{\tfrac{2}{5K_1+6}}}{2\sqrt{p}}\Big)\nonumber\\
&~~+\P\Big(|\loss_{k}'|^2\lsbd \geq (\sqrt{p}\veps)^{\frac{5K_1+4}{5K_1+6}}\Big) \nonumber\\
&~~+\P\Big[\tfrac{1}{p}\big(\sum_{i=1}^{p}(\vh\tran_{\bs k,i}\vr)^{4}\big)^{\tfrac{1}{2}} \geq\tfrac{(\sqrt{p}\veps)^{\tfrac{2}{5K_1+6}}}{2\sqrt{p}} \Big]\nonumber\\
\leq& cp\exp\Big[ -\min\big\{(\sqrt{p}\veps)^{\tfrac{2}{5K_1+6}}, (\sqrt{p}\veps)^{\tfrac{4}{5K_1+6}}, (\log p)^2\big\} / c \Big],
\label{eq:wdiff_bd}
\end{align}
where constant $c$ does not depend on $k$ and $p$. This completes our proof.
\end{proof}
\begin{lem}\label{lem:wk_qwtk_diff_moments_bd}
There exists a function $B(m)$, $m\in\mathbb{Z}^{+}$ such that for every $\fmtx\in{\cal A}$, $p\geq 2$ and $k\in [n]$,
\begin{equation}
\label{eq:wk_qwtk_diff_moments_bd}
\Ecnd{\fmtx}\left\Vert \wt_{k}(\vr)-\qwt_{k}(\vr)\right\Vert ^{m}\leq B(m)  \frac{(\log p)^{({2.5K_1}+3)m}}{p^{m/2}},
\end{equation}
where $\vr=\va_k$ or $\vb_k$ and $K_1\in\mathbb{Z}^+$ is the constant defined in Assumption \ref{a:loss}.
\end{lem}
\begin{proof}
From (\ref{eq:xstar_xtilde_diff}), there exists a function $B_{1}(m)$ such that
\begin{align}
 &\left\Vert \wt_{k}(\vr)-\qwt_{k}(\vr)\right\Vert ^{m}\nonumber\\
 \leq& B_{1}(m)(\lsbd\left|\loss_{k}'\right|^2)^{m} \nonumber\\
 &\hspace{1em}\times\Big(\sup_{t\neq k}\{ |\vr_{t}\tran\mH_{\bs k}^{-1}\vr/p|\} ^{m}
 \big\Vert\tfrac{1}{p}\sum_{t\neq k}\vr_{t}\vr_{t}\tran\big\Vert^{m}\big(\scp{\|\mH_{\bs k}^{-1}\vr\|}\big)^{m} \nonumber\\
 &\hspace{6em}+\big[\tfrac{1}{p^{2}}\sum_{i=1}^{p}(\vh\tran_{\bs k,i}\vr)^{4}\big]^{m/2}\Big),\label{eq:wk_qwtk_diff_expo_bd}
\end{align}
where $\vr=\va_{k}$ or $\vb_{k}$. It follows that
\begin{align}
\label{eq:quad_approx_sol_mombd}
&\Ecnd{\fmtx}\left\Vert \wt_{k}(\vr)-\qwt_{k}(\vr)\right\Vert ^{m}\nonumber\\
\le& B_{1}(m)\Big[\Ecnd{\fmtx}(\lsbd\left|\loss_{k}'\right|^{2})^{4m} \Ecnd{\fmtx}\sup_{t\neq k}\big\{ |\vr_{t}\tran\mH_{\bs k}^{-1}\vr/p|\big\}^{4m} \nonumber\\
&\hspace{4em}\times\Ecnd{\fmtx}\big\Vert\tfrac{1}{p}\sum_{t\neq k}\vr_{t}\vr_{t}\tran\big\Vert^{4m} \Ecnd{\fmtx}\big(\scp{\|\mH_{\bs k}^{-1}\vr\|}\big)^{4m}\Big]^{1/4}\nonumber\\
 & +B_{1}(m)\Big[\Ecnd{\fmtx}(\lsbd\left|\loss_{k}'\right|^{2})^{2m} \Ecnd{\fmtx}\big(\tfrac{1}{p^{2}}\sum_{i=1}^{p}(\vh\tran_{\bs k,i}\vr)^{4}\big)^{m}\Big]^{1/2},
\end{align}
where we have used the following generalized H\"{o}lder's inequality: for random variables $X_{1},\ldots,X_{4}\geq0$, $\E\left(X_{1}X_{2}\cdots X_{4}\right)\leq\prod_{i=1}^{4}\left(\E X_{i}^{4}\right)^{1/4}$. Therefore, to show \eref{wk_qwtk_diff_moments_bd}, it suffices to bound each term on the right-hand side of \eqref{eq:quad_approx_sol_mombd}. Following the same steps leading towards \eqref{eq:loss_derivative_LS_prob_bd_5}, we get there exists $c>0$ such that for any $\veps\geq 0$ and $\fmtx\in{\cal A}$,
\[
\Pcnd{\fmtx}\left(\lsbd\big|\loss'_k\big|^2\geq\veps\right) \le cp\exp\big(-\veps^{\tfrac{2}{5K_{1}+4}}/c\big).
\]
Applying the integral identity $\E \abs{X} = \int_{0}^{\infty}\P(\abs{X}>t)dt$, we can then show that for $p\geq2$, $ \Ecnd{\fmtx}(\lsbd\left|\loss_{k}'\right|^{2})^{4m} \le (\log p)^{(10K_1+8)m}B_2(m)$ and $\Ecnd{\fmtx}(\lsbd\left|\loss_{k}'\right|^{2})^{2m} \le (\log p)^{(5K_1+4)m}B_2(m)$ for some function $B_2(m)$. Similarly, combining \eref{rrt_ab} and Lemma \ref{lem:empi_spectral_norm_bd} gives us $\Ecnd{\fmtx}\big\Vert\tfrac{1}{p}\sum_{t\neq k}\vr_{t}\vr_{t}\tran\big\Vert^{4m} \le B_3(m)$. Since $\|\mH_{\bs k}^{-1}\| \le 2/\lambda$ for $\fmtx\in{\cal A}$, we have $\Ecnd{\fmtx}\big(\scp{\|\mH_{\bs k}^{-1}\vr\|}\big)^{4m} \le C\Ecnd{\fmtx}(\scp \norm{\vr})^{4m} \le B_4(m)$, where the last step is due to \eref{a_nrm_bnd} and \eref{b_nrm_bnd}.

Next, we consider $\Ecnd{\fmtx}\sup_{t\neq k}\big\{ |\vr_{t}\tran\mH_{\bs k}^{-1}\vr/p|\big\}^{4m}$. Applying Lemma~\ref{lem:quadratic_asymmetry_concentrate} and the union bound gives us
\[
\Pcnd{\fmtx}\Big(\sup_{t\neq k}\{ |\vr_{t}\tran\mH_{\bs k}^{-1}\vr/p|\} \geq \veps\Big) \leq cpe^{-p\veps^2/c}.
\]
We can then show that when $p\geq 2$, $\Ecnd{\fmtx}\sup_{t\neq k}\big\{ |\vr_{t}\tran\mH_{\bs k}^{-1}\vr/p|\big\}^{4m} \le B(m) (\log p / p)^{2m}$ for some function $B(m)$. Similarly as \eqref{eq:optwt_qwt_diff_bd_2_aux_1}, we can get $\Pcnd{\fmtx}\big[\big(\tfrac{1}{p^2}\sum_{i=1}^{p}(\vh\tran_{\bs k,i}\vr)^{4}\big)^{\tfrac{1}{2}} \geq \veps \big]\leq cpe^{-{\sqrt{p}\veps}/{c}}$
and then it can be verified that $\Ecnd{\fmtx}\big(\tfrac{1}{p^{2}}\sum_{i=1}^{p}(\vh\tran_{\bs k,i}\vr)^{4}\big)^{m}  \le B(m) (\log p / \sqrt{p})^{2m}$. Substituting these bounds in \eqref{eq:quad_approx_sol_mombd}, we reach the desired inequality in \eref{wk_qwtk_diff_moments_bd}.
\end{proof}

\subsubsection{\label{appendix:linfty_boundedness}The $\ell_{\infty}$ Boundedness
of Optimal Solutions}
\begin{lem}\label{lem:loowt_l_infty_bd}
Let $\wt_k$ be the optimal solution to the optimization problem defined in \eref{interpolation_k_wt}. There exists some $c_{\infty}>0$ such that for every $p$ and $0\leq k\leq n$,
\begin{equation}
\P\left(\left\Vert \wt_{k}\right\Vert _{\infty}\geq\left(\log p\right)^{3+2K_{1}}\right)\leq c_{\infty}\exp\left[-c_{\infty}^{-1}\left(\log p\right)^{2}\right],\label{eq:loowt_l_infty_bd}
\end{equation}
where $K_1\in \mathbb{Z}^+$ is the constant in Assumptions~\ref{a:loss}.
\end{lem}
\begin{proof}
The general strategy of our proof is as follows. To bound $\norm{\wt_k}_\infty$, we just need to show that any given coordinate of $\wt_k$, e.g., its last entry, is bounded with high probability. By symmetry, all the coordinates have the same marginal distribution. Consequently, each
coordinate of $\wt_k$ can be analyzed in the same way and $\left\Vert \wt_k\right\Vert _{\infty}$ can then be controlled by using the union bound.

Recall that $\wt_k \in \R^p$. To simplify the notation, we will instead study a $(p+1)$-dimensional version of the problem in \eref{interpolation_k} and focus on, without loss of generality, the last coordinate of the optimal solution, denoted by $u^\ast$. Let $\vf_{p+1}$ be the new column added to the feature matrix. Also define a vector \\
\begin{equation}\label{eq:et}
\ve \bydef [b_{1} ~ \cdots ~ b_{k} ~~ a_{k+1} ~ \cdots ~ a_{n}]{\tran},
\end{equation}
where $b_t = \mu_1 \vf_{p+1}\tran \vg_t + \mu_2 z_t$ (with $z_t \sim \mathcal{N}(0, 1)$, independent of $\vg_t$) and $a_t = \kernel(\vf_{p+1}\tran \vg_t )$. From \eref{interpolation_k}, the $(p+1)$th coordinate $u^\ast$ can be expressed as
\begin{align}
u^\ast= & \argmin{u}\min_{\vw}\sum_{t=1}^{n}\loss\left(\scp\left(\vr_{t}\tran\vw+e_{t}u\right);y_{t}\right)+G\left(\vw\right)\nonumber\\
&\hspace{7em}+(2\tau_{1}\mu_{1}^{2}\fvec_{p+1}\tran\fmtx\vw) u +h\left(u\right) \nonumber \\
 & \hspace{7em}+\tau_{1}\Big(\mu_{1}^{2}\left\Vert \fvec_{p+1}\right\Vert ^{2} +\mu_{2}^{2}\Big)u^{2} \nonumber\\
 & \hspace{7em}+(\tau_{2}\mu_{1}\sqrt{p}\sgl{\tran}\fvec_{p+1})u,\label{eq:w0_opt}
\end{align}
where
\[
G\left(\vw\right)=Q(\vw)+\sum_{i=1}^{p}h(w_{i}).
\]
The rest of the proof consists of two steps. First, we will show
\begin{equation}
\begin{aligned}
    \left|u^\ast\right|\leq&\tfrac{4}{\lambda}\Big|h'(0)+\scp\sum_{t=1}^{n}\loss'(\scp\vr_{t}\tran\wt_k;y_{t})e_{t}+2\tau_{1}\mu_{1}^{2}\fvec_{p+1}\tran\fmtx\wt_k\\    &~~~+\tau_{2}\mu_{1}\sqrt{p}\sgl{\tran}\fvec_{p+1}\Big|,
\end{aligned}
\label{eq:w0_bd_deterministic}
\end{equation}
provided that
\begin{equation}\label{eq:bd_cond}
\|\fmtx\|\leq 1+2\sqrt{\eta} \quad \text{and} \quad \left\Vert \fvec_{p+1}\right\Vert \leq1+2\sqrt{\eta}.
\end{equation}
Second, we show that \eref{bd_cond} holds with high probability and that each term on the right-hand side of \eref{w0_bd_deterministic} is also bounded with high probability.

We start by proving the bound in \eref{w0_bd_deterministic}. Let $\calL(u)$ denote the objective function of $u$ in \eref{w0_opt}, i.e., $u^\ast = \arg\,\min_{u} \calL(u)$. We first derive a lower bound for $\calL(u)$. To that end, we note from the convexity of the loss function that
\[
\begin{aligned}
    \loss(\scp\left(\vr_{t}\tran\vw+e_{t}u\right);y_{t}) \ge& \loss\left(\scp\vr_{t}\tran\wt_k;y_{t}\right) + \loss'\left(\scp\vr_{t}\tran\wt_k;y_{t}\right)\\
    &\times\left[\scp\vr_{t}\tran(\vw-\wt_k)+\scp e_{t}u\right].
\end{aligned}
\]
Moreover, recall from Remark~\ref{rem:convex} that $G(\w)$ is $\tfrac{\lambda}{2}$-strongly convex when $\fmtx$ satisfies $\|\fmtx\|\leq 1+2\sqrt{\eta}$. It follows that $G(\vw) \ge G(\wt_k) +\nabla{\tran}G(\wt_k)(\vw-\wt_k) + \frac{\lambda}{4} \norm{\vw-\wt_k}^2$. Furthermore, $h(u)$ being $\lambda$-strongly convex gives us $h(u) \ge h(0) + h'(0)u + \frac{\lambda}{2} u^2$. Substituting these inequalities into \eref{w0_opt} and using the first-order optimality condition of $\wt_k$, we have
\begin{align}
\calL(u) &\geq \calL(0) + \min_{\vw}\Big\{2\tau_{1}\mu_{1}^{2}\fvec_{p+1}\tran\fmtx(\vw-\wt_k)u+\frac{\lambda}{4}\left\Vert \vw-\wt_k\right\Vert ^{2}\Big\} \nonumber \\
&\hspace{3em}+ \chi u + \frac{\lambda}{2} u^{2}\nonumber \\
&= \calL(0) + \chi u + \big(\frac{\lambda}{2}-\frac{4}{\lambda}\tau_{1}^{2}\mu_{1}^{4}\fvec_{p+1}\tran\fmtx\fmtx{\tran}\fvec_{p+1}\big)u^{2}\nonumber \\
&\ge \calL(0)+\chi u + \frac{\lambda}{4} u^2,\label{eq:Lw0_lbd_2}
\end{align}
where $\chi = h'(0)+\scp\sum_{t=1}^{n}\loss'(\scp\vr_{t}\tran\wt_k;y_{t})e_{t}+2\tau_{1}\mu_{1}^{2}\fvec_{p+1}\tran\fmtx\wt_k+\tau_{2}\mu_{1}\sqrt{p}\sgl{\tran}\fvec_{p+1}$. To reach \eref{Lw0_lbd_2}, we have used \eref{bd_cond} and the constraint \eref{tau12} on the magnitude of $\tau_1$. In the meanwhile, we must have $\min_u \calL(u) \le \calL(0)$. It follows that $\abs{u^\ast} = \abs{\arg\,\min_u \calL(u)} \le \frac{4 \abs{\chi}}{\lambda}$ and thus \eref{w0_bd_deterministic}.

From \eref{spectral_norm_Wishart} and \eref{fnorm_concentrate_lbd}, the conditions in \eref{bd_cond} hold with probability greater than $1-2e^{-cp}$, for some $c>0$. Thus, to complete the proof, we just need to bound the following three terms on the right-hand side of \eref{w0_bd_deterministic}: (I) $\sqrt{p}\sgl{\tran}\fvec_{p+1}$; (II) $\fvec_{p+1}\tran\fmtx\wt_k$; and (III) $\scp\sum_{t=1}^{n}\loss'(\scp\vr_{t}\tran\wt_k;y_{t})e_{t}$.

(I) Since $\sqrt{p}\sgl{\tran}\fvec_{p+1}\sim \mathcal{N}(0,p/d)$, there exists $c>0$ such that
\begin{equation}
\P\left(\left|\sqrt{p}\sgl{\tran}\fvec_{p+1}\right| \ge \log p\right)\le c e^{-(\log p)^2 / c}.\label{eq:f0mu_bd}
\end{equation}

(II) $\fvec_{p+1}\tran\fmtx\wt_k$. Note that $\fvec_{p+1}$ is independent
of $\fmtx\wt_k$. Given $\fmtx\wt_k$, the conditional distribution of $\fvec_{p+1}\tran\fmtx\wt_k$ is $\mathcal{N}(0, \norm{\fmtx \wt_k}^2/d)$. From (\ref{eq:spectral_norm_Wishart}) and (\ref{eq:wt_norm_prob_bd}),
there exists $C,c>0$ such that $\P\big(\tfrac{1}{\sqrt{d}}\left\Vert \fmtx\wt_k\right\Vert \geq C\big)\leq ce^{-(\log p)^2/c}$. It then follows that for some $c>0$,
\begin{equation}
\label{eq:f0Fw_bd}
\P\big(\big|\fvec_{p+1}\tran\fmtx\wt_k\big|\geq \log p\big)<ce^{-(\log p)^2 / c}.
\end{equation}

(III) $\scp\sum_{t=1}^{n}\loss'(\scp\vr_{t}\tran\wt_k;y_{t})e_{t}$. To simplify the notation, let $\theta_{t}^{*}=\loss'(\scp\vr_{t}\tran\wt_k;y_{t})$. We first show that $\theta_t^\ast$ is bounded with high probability.
From inequality \eqref{eq:loss_derivative_bd_2}, there exists a constant $C > 0$ such that
\begin{equation}
\label{eq:linresponse_original_5}
\begin{aligned}
    &\P\left(\big|\theta_t^\ast\big|\geq C(\log p)^{2+2K_1}\right) \\
    \leq& \P\big(|\scp\vr_{t}\tran\wt_k|\geq 2(\log p)^{(K_1+2)/2} \big)+\P\left(\left| \vg_t\tran \sgl\right|\geq\log p\right).
\end{aligned}
\end{equation}
To bound $\P\big(|\scp\vr_{t}\tran\wt_k|\geq2(\log p)^{(K_1+2)/2} \big)$ in \eqref{eq:linresponse_original_5}, we consider different $t$. When $1\leq t \leq k$, we have
\begin{align}
&\P\Big(|\scp\vr_{t}\tran\wt_k|\geq 2(\log p)^{\tfrac{K_1+2}{2}} \Big)&\nonumber\\
=& \P\Big(|\scp\vb_{t}\tran\wt_k|\geq 2(\log p)^{\tfrac{K_1+2}{2}} \Big)\nonumber\\
=& \P\Big(|\scp\vb_{k}\tran\wt_k|\geq 2(\log p)^{\tfrac{K_1+2}{2}}\Big)\nonumber\\
\leq& \P\Big(|\scp\vb_{k}\tran\qwt_{k}(\vb_{k})|\geq(\log p)^{\tfrac{K_1+2}{2}}\Big) \nonumber \\
&\hspace{2.5em}+\P\Big(\scp\|\vb_{k}\|\|\wt_k-\qwt_{k}(\vb_{k})\|\geq (\log p)^{\tfrac{K_1+2}{2}}\Big)\nonumber\\
=& \P\Big(|\scp\vb_{k}\tran\qwt_{k}(\vb_{k})|\geq(\log p)^{\tfrac{K_1+2}{2}}\Big) \nonumber\\
&\hspace{2.5em}+\P\Big(\scp\|\vb_{k}\|\|\wt_k(\vb_{k})-\qwt_{k}(\vb_{k})\|\geq (\log p)^{\tfrac{K_1+2}{2}}\Big),\label{eq:linresponse_original_6}
\end{align}
where the second equality is due to symmetry of $\vb_{t}, 1\leq t \leq k$. Similarly, when $k< t\leq n$, we can get
\begin{equation}
\label{eq:linresponse_original_7}
\begin{aligned}
&\P\Big(|\scp\vr_{t}\tran\wt_k|\geq 2(\log p)^{\tfrac{K_1+2}{2}}\Big)\\
\leq& \P\Big(|\scp\va_{k+1}\tran\qwt_{k+1}(\va_{k+1})|\geq (\log p)^{\tfrac{K_1+2}{2}}\Big)+ \\
&\hspace{-1em}\P\Big(\scp\|\va_{k+1}\|\|\wt_{k+1}(\va_{k+1})-\qwt_{k+1}(\va_{k+1})\|\geq (\log p)^{\tfrac{K_1+2}{2}}\Big).
\end{aligned}
\end{equation}
By Lemma~\ref{lem:lnresponse_bd_2}, there exists $c>0$ such that
\[
\P\Big(|\scp\vr_{k}\tran\qwt_{k}(\vr_{k})|\geq(\log p)^{\tfrac{K_1+2}{2}}\Big) \le c e^{-(\log p)^2 / c},
\]
for $\vr_k = \va_a$ or $\vb_k$. Moreover, there exists $C,c>0$ such that
\begin{equation*}
\begin{aligned}
&\P\left(\scp\|\vr_{k}\|\|\wt_k(\vr_{k})-\qwt_{k}(\vr_{k})\|\geq (\log p)^{\tfrac{K_1+2}{2}} \right) \\
 \leq& \P\left(\scp\|\vr_{k}\|\geq C\right)+\P\Big(\|\wt_k(\vr_{k})-\qwt_{k}(\vr_{k})\|\geq \tfrac{(\log p)^{\tfrac{K_1+2}{2}}}{C}\Big)\\
 \leq& c e^{-(\log p)^2 / c},
\end{aligned}
\end{equation*}
where in reaching the last step we have used (\ref{eq:optwt_qwt_prob_bd}), Lemma \ref{lem:Wishart_spectral_norm} and Lemma \ref{lem:at_bt_norm_bd}. Substituting these two bounds into  \eqref{eq:linresponse_original_6} and \eqref{eq:linresponse_original_7}, we get there exists $c>0$ such that for every $1\leq t \leq k$,
\begin{equation}
\label{eq:linresponse_original_8}
\P\Big(|\scp\vr_{t}\tran\wt_k|\geq 2(\log p)^{\tfrac{K_1+2}{2}}\Big) \leq c e^{-(\log p)^2 / c}.
\end{equation}
On the other hand, since $\vg_t\tran \sgl \sim \mathcal{N}(0, 1)$, we have
\begin{equation}
\label{eq:linresponse_original_9}
\P\left(\left| \vg_t\tran \sgl \right|\geq\log p\right) \le c e^{-(\log p)^2 / c}.
\end{equation}
Therefore, from \eqref{eq:linresponse_original_5}, \eqref{eq:linresponse_original_8} and \eqref{eq:linresponse_original_9}, we get there exists $c>0$ such that for any $p\in\mathbb{Z}^{+}$,
\begin{equation}
\label{eq:linresponse_prob_bd2}
\P\left(\big|\theta_t^\ast\big|\geq C(\log p)^{2+2K_{1}}\right) \le c e^{-(\log p)^2 / c}.
\end{equation}

Recall the definition of $e_t$ in \eref{et}. We have
\begin{equation}\label{eq:Uz}
\textstyle\scp \sum_t \theta_t^\ast e_t = U(\sqrt{d} \vf_{p+1}) + \tfrac{\mu_2}{\sqrt{p}}\sum_{t \le k} \theta_t^\ast z_t,
\end{equation}
where $U: \R^d \mapsto \R$ is a function defined as
\begin{equation}\label{eq:Ux}
\textstyle U(\vx) \bydef \scp \sum_{t \le k} \theta_t^\ast \mu_1 \tfrac{1}{\sqrt{d}}\vg_t\tran \vx + \scp \sum_{k+1 \le t \le n} \theta_t^\ast \kernel(\tfrac{1}{\sqrt{d}}\vg_t\tran \vx).
\end{equation}
Let us consider the following event
\[
E=\left\{ \scp\left\Vert \vtheta^{*}\right\Vert \leq C(\log p)^{2+2K_{1}},\tfrac{1}{\sqrt{d}}\left\Vert \cmtx\right\Vert \leq K\right\},
\]
where $\vtheta^\ast = [\theta_1^\ast, \theta_1^\ast, \ldots, \theta_n^\ast]\tran$, $C$ is the constant in \eref{linresponse_prob_bd2}, $\cmtx=[\cvec_1~\cvec_2~\dots~\cvec_n]{\tran}$ is the matrix of the latent input vectors in Assumption~\ref{a:Gaussian}, and $K$ is some sufficiently large constant. Notice that $E$ is a high probability event. Indeed,  from (\ref{eq:spectral_norm_F}) and \eqref{eq:linresponse_prob_bd2}, there exists $c>0$ such that for every $K$ large enough,
\begin{equation}
\label{eq:EProb_bded}
\begin{aligned}
\P(E^C) &\leq \P(\scp\left\Vert \vtheta^{*}\right\Vert > C(\log p)^{2+2K_{1}}) + \P(\scp\left\Vert \cmtx\right\Vert > K)\\
&\leq c e^{-(\log p)^2 / c}.
\end{aligned}
\end{equation}
Conditioned on any $\cmtx$ and $\fmtx$ in $E$, the two terms of the right-hand side of \eref{Uz} can be easily bounded. Specifically, let
\begin{equation}\label{eq:J}
J = \tfrac{\lambda}{32}\left(\log p\right)^{3+2K_{1}}.
\end{equation}
Since $\set{z_t}$ is a set of i.i.d. standard normal random variables independent of $\theta_t^\ast$, we have
\begin{align}
&\textstyle\P\Big(\abs{\tfrac{\mu_2}{\sqrt{p}} \sum_{t \le k} \theta_t^\ast z_t} \geq J\Big) \nonumber\\
\le& \textstyle\P\Big(\set{\abs{\tfrac{\mu_2}{\sqrt{p}} \sum_{t \le k} \theta_t^\ast z_t} \geq J} \cap E\Big) + \P(E^c)\nonumber\\
\le& c e^{-(\log p)^2 / c},\label{eq:bd_z}
\end{align}
where in reach the last inequality we have used the standard tail bound for Gaussian random variables. To bound the first term on the right-hand side of \eref{Uz}, we note that, given any $\cmtx$ and $\fmtx$ in $E$, the function $U(\vx)$ in \eref{Ux} is a Lipschitz continuous mapping with a Lipschitz constant $C K (\log p)^{2+2K_{1}}$ for some constant $C>0$. Since $\sqrt{d} \fvec_{p+1}$ is a standard Gaussian vector and $\Ecnd{\cmtx,\mF}\left[U(\sqrt{d} \fvec_{p+1})\right] =0$, we can apply \eqref{eq:Lipschitz_concentration} to get
\[
\begin{aligned}
&\P\Big(\abs{U(\sqrt{d} \fvec_{p+1})} \geq J\Big)\\
\le& \P\Big(\set{\abs{U(\sqrt{d} \fvec_{p+1})} \geq J} \cap E\Big) + \P(E^c)\\
\le& c e^{-(\log p)^2 / c},
\end{aligned}
\]
where in the last step we have used the specific value of $J$ in \eref{J}. Combining this inequality and \eref{bd_z}, we can then get from \eref{Uz} that $\P\big(\big|\textstyle\scp \sum_t \theta_t^\ast e_t\big| \geq \tfrac{\lambda}{16}\left(\log p\right)^{3+2K_{1}}\big) < c e^{-(\log p)^2 / c}$. Finally, substituting this bound, \eref{f0mu_bd}, and \eref{f0Fw_bd} into \eref{w0_bd_deterministic}, we have
\[
\P[\abs{u^\ast} \geq \left(\log p\right)^{3+2K_{1}}] \le c e^{-(\log p)^2 / c}.
\]
Since $u^\ast$ is the last coordinate of the optimal weight vector, and since all the coordinates have the same distribution by symmetry, we get from the union bound that
\[
\P[\abs{\wt_k} \geq  \left(\log p\right)^{3+2K_{1}})] \le cp e^{-(\log p)^2 / c}.
\]
Note that there exists $p_0$ such that for any $p\geq p_0$, $cp e^{-\left(\log p\right)^{2}/c} \leq 2ce^{-\left(\log p\right)^{2}/(2c)}$. We can get (\ref{eq:loowt_l_infty_bd}) by choosing $c_{\infty}$ to be the smallest number satisfying $c_{\infty}\geq 2c$ and $c_{\infty}e^{-\left(\log p_0\right)^{2}/c_{\infty}}\geq 1$.
\end{proof}

\subsubsection{Proof of Proposition \ref{prop:wt_bnd}}

We write ${\cal A}_{3}$ as ${\cal A}_{3}=\cap_{k=1}^{n}{\cal A}_{3,k}$, where
\[
{\cal A}_{3,k}\bydef\left\{ \fmtx:\Ecnd{\fmtx}(\|\loowt k\|_{\infty}^{2})\leq \left(\log p\right)^{7+4K_{1}}\right\}.
\]
To show \eref{A4_p}, it suffices to show that each ${\cal A}_{3,k}$ has high probability. Consider the following set of $\fmtx$:
\begin{equation}
\begin{aligned}
    {\cal B}_{k}\bydef&\Big\{ \fmtx:\Pcnd{\fmtx}\Big(\linf{\loowt k}\leq\big(\log p\big)^{3+2K_{1}}\Big) \\
    &\hspace{10em}\geq 1-c_{\infty}e^{-\left(\log p\right)^{2}/(2c_{\infty})}\Big\} ,
\end{aligned}
\label{eq:def_Bk_set}
\end{equation}
where $c_{\infty}$ is the constant in (\ref{eq:loowt_l_infty_bd}). From
(\ref{eq:loowt_l_infty_bd}), we have
\begin{align*}
&1-c_{\infty}e^{-\left(\log p\right)^{2}/c_{\infty}}\\ \leq&  \P\left(\linf{\loowt k} \leq \left(\log p\right)^{3+2K_{1}}\right)\\
=&  \E_{\fmtx}\left[\charfn_{{\cal B}_{k}}\Pcnd{\fmtx}\left(\linf{\loowt k}\leq\left(\log p\right)^{3+2K_{1}}\right)\right]\\
 & ~~+\E_{\fmtx}\left[\charfn_{{\cal B}_{k}^{C}}\Pcnd{\fmtx}\left(\linf{\loowt k}\leq\left(\log p\right)^{3+2K_{1}}\right)\right]\\
 \leq& \P\left({\cal B}_{k}\right)+\left[1-\P\left({\cal B}_{k}\right)\right]\Big[1-c_{\infty}e^{-\left(\log p\right)^{2}/(2c_{\infty})}\Big],
\end{align*}
which indicates that
\[
\P\left(\calB_{k}\right)\geq1-e^{-\left(\log p\right)^{2}/(2c_{\infty})}.
\]
Let $\mathcal{A}_2$ be the set defined in \eref{def_setA2}. From Lemma~\ref{lem:norm_optwt_Expec_bd_A2}, we know there
exists $c>0$ such that, for every $\fmtx\in{\cal A}_2$, $p\geq 2$ and $0\leq k \leq n$,
\begin{equation}
\label{eq:wk_8moments_bd}
\Ecnd{\fmtx}(\|\loowt k\|^{4})\leq cp^{2}(\log p)^{2K_1}.
\end{equation}
Therefore, for every $\fmtx\in{\cal A}_2\cap{\cal B}_{k}$,
it holds that for $p\geq 2$,
\begin{align}
&\Ecnd{\fmtx}\linf{\loowt k}^{2} \nonumber\\
=&\Ecnd{\fmtx}(\charfn_{\linf{\loowt k}\leq\left(\log p\right)^{3+2K_{1}}}\linf{\loowt k}^{2}) \nonumber \\
&\hspace{1em}+\Ecnd{\fmtx}(\charfn_{\linf{\loowt k}>\left(\log p\right)^{3+2K_{1}}}\linf{\loowt k}^{2})\nonumber\\
 \leq&\left(\log p\right)^{6+4K_{1}}+\Ecnd{\fmtx}(\charfn_{\linf{\loowt k}>\left(\log p\right)^{3+2K_{1}}}\|\loowt k\|^{2})\nonumber\\
 \le&\left(\log p\right)^{6+4K_{1}}+\sqrt{\Ecnd{\fmtx}\big(\|\loowt k\|^{4}\big)}\times\nonumber\\
 &\hspace{9em}\sqrt{\Pcnd{\fmtx}\big(\linf{\loowt k}>\left(\log p\right)^{3+2K_{1}}\big)}\nonumber\\
 \le&\left(\log p\right)^{6+4K_{1}}+cp(\log p)^{K_1} e^{-\left(\log p\right)^{2}/c},\label{eq:w_infnorm_mom2bd}
\end{align}
where $c>0$ is some constant, and we have used \eqref{eq:wk_8moments_bd} and \eqref{eq:def_Bk_set} in reaching the last step. There exists a constant $p_0$ such that for any $p\geq p_0$, the right-hand side of \eqref{eq:w_infnorm_mom2bd} is bounded by $\left(\log p\right)^{7+4K_{1}}$
and in that case, ${\cal A}_{2}\cap{\cal B}_{k}\subset{\cal A}_{3,k}.$
Since there exists $c_1>0$ such that $\P\left({\cal A}_{2}\right)\geq1-c_1e^{-p/c}$
and $\P\left(\calB_{k}\right)\geq1-e^{-\left(\log p\right)^{2}/(c_1)}$,
we know there exists some $c_{2}>0$ such that
$\P\left({\cal A}_{3,k}\right)\geq1-c_{2}e^{-\left(\log p\right)^{2}/c_{2}}$ for every $p\geq p_0$ and $0\leq k \leq n$. Choose a large enough constant $c$ satisfying $c \ge c_2$ and $ce^{-\left(\log p_0\right)^{2}/c}\geq 1$, we then have
\[
\P\left({\cal A}_{3,k}\right)\geq1-ce^{-\left(\log p\right)^{2}/c},
\]
for every $p$ and $0\leq k \leq n$. Finally, (\ref{eq:A4_p}) can be obtained by applying the union bound.

%% file: appendix-quadapprox.tex

\subsubsection{\label{appendix:quad_approx}Proof of Lemma \ref{lem:quad_approx}}
Recall the definitions of $\Phi_{k}(\vr)$ and $\Psi_{k}(\vr)$ in \eqref{eq:originprob_def} and \eqref{eq:quadprob_def} of Appendix~\ref{appendix:opt_properties}.
The corresponding optimal solutions $\qwt_k(\vr)$ and $\wt_k(\vr)$ are also defined in \eqref{eq:originprob_def} and \eqref{eq:quadprob_def}, respectively. We first show \eref{quad_approx_psi_phi_k}. Let $\vr = \va_k$ or $\vb_k$. It follows from (\ref{eq:Rk_cr_decomp_general_2}) that
\begin{align*}
\Ecnd{\fmtx}\left(\Psi_{k}(\vr)-\Phi_{\bs k}\right)^{2} &=\Ecnd{\fmtx}\calM_{k}\Big(\scp\vr\tran\loowt{\bs k};\gamma_{k}(\vr)\Big)^{2}\\
 & \leq\Ecnd{\fmtx}\loss\Big(\scp\vr\tran\loowt{\bs k};y_{k}\Big)^{2}\\
 & \tleq{{(a)}}\Ecnd{\fmtx} Q\Big(\scp\|{\loowt{\bs k}}\|\Big)\\
 & \tleq{{(b)}}\polylog p,
\end{align*}
where $Q(x)$ in step (a) is some finite degree polynomial. To reach (a), we have used
\eqref{eq:loss_derivative_bd_2} and Lemma \ref{lem:linear_concentrate} and (b) follows from (\ref{eq:norm_loowt_Expec_bd_goodF}).

We now move on to showing \eref{quad_approx_phi_psi}.  By applying Taylor expansion, $R_{k}(\vw;\vr)$ in \eqref{eq:original_objective_func_1} can be written as
\begin{align}
R_{k}(\vw;\vr)= & \Phi_{\bs k}+\frac{1}{2}(\vw-\loowt{\bs k})\tran\mH_{\bs k}(\vw-\loowt{\bs k})\nonumber \\
 & +\frac{1}{6}\sum_{t=1}^{k-1}\loss'''(\nu_{t};y_{t})\left[\scp\vb_{t}\tran(\vw-\loowt{\bs k})\right]^{3}\nonumber\\
 & +\frac{1}{6}\sum_{t=k+1}^{n}\loss'''(\nu_{t};y_{t})\left[\scp\va_{t}\tran(\vw-\loowt{\bs k})\right]^{3}\nonumber \\
 & +\frac{1}{6}\sum_{i=1}^{p}\regu'''\left(w_{i}'\right)(w_{i}-w_{\bs k,i}^{*})^{3}+\loss\big(\scp{\vr\tran\vw};y_{k}\big),\label{eq:loo_objective_function_Taylor}
\end{align}
where $\mH_{\bs k}$ is the Hessian matrix defined in \eqref{eq:looH}, $\nu_{t}$ denotes some point that lies between $\frac{\vr_t\tran\vw}{\sqrt{p}}$ and
$\tfrac{\vr_t\tran\loowt{\bs k}}{\sqrt{p}}$, with $\vr_t = \va_t\text{ or }\vb_t$, $t\neq k$ and $w_{i}'$ denotes some point that lies between
$w_{i}$ and $w_{\bs k,i}^{*}$. By recalling the definition of $S_{k}(\vw;\vr)$ in \eqref{eq:quadratic_approx_objctive_function_1} and that of $\lsbd$ in \eref{Ls}, we have
\begin{align}
&\left|R_{k}(\vw;\vr)-S_{k}(\vw;\vr)\right| \nonumber\\
\leq & C_1\lsbd\Big(\sum_{t=1}^{k-1}\left|\scp\vb_{t}\tran(\vw-\loowt{\bs k})\right|^{3}+\sum_{t=k+1}^{n}\left|\scp\va_{t}\tran(\vw-\loowt{\bs k})\right|^{3}\Big)\nonumber \\
 & +C_1 \sum_{i=1}^{p}|w_{i}-w_{\bs k,i}^{*}|^{3}\nonumber \\
\leq & C_2\lsbd\Big(\sum_{t=1}^{k-1}\left|\scp\vb_{t}\tran\left(\vw-\qwt_{k}(\vr)\right)\right|^{3} \nonumber\\
&\hspace{8em}+\sum_{t=k+1}^{n}\left|\scp\va_{t}\tran\left(\vw-\qwt_{k}(\vr)\right)\right|^{3}\Big)\nonumber \\
 & +C_2\lsbd\Big(\sum_{t=1}^{k-1}\left|\scp\vb_{t}\tran(\qwt_{k}(\vr)-\loowt{\bs k})\right|^{3}\nonumber\\
 &\hspace{8em}+\sum_{t=k+1}^{n}\left|\scp\va_{t}\tran(\qwt_{k}(\vr)-\loowt{\bs k})\right|^{3}\Big)\nonumber \\
 & +C_2\sum_{i=1}^{p}\big(\left|w_{i}-\tw_{i}(\vr)\right|^{3}+|\tw_{i}(\vr)-w_{\bs k,i}^{*}|^{3}\big),\label{eq:quadapprox_bd_objfunc}
\end{align}
for some constants $C_1, C_2 > 0$, where the first step is obtained similar as \eqref{eq:lsbd_bd_2}.

Let  $\calB=\{\wt_k(\vr)\} \cup \set{\qwt_k(\vr)}$. It is easy to verify that
\begin{align}
\left|\Phi_{k}(\vr)-\Psi_{k}(\vr)\right| & =|\min_{\vw\in\calB}R_{k}(\vw;\vr)-\min_{\vw\in{\cal B}}S_{k}(\vw;\vr)|\nonumber \\
 & \leq\max_{\vw\in\calB}|R_{k}(\vw;\vr)-S_{k}(\vw;\vr)|.\label{eq:quadapprox_bd_optvalue}
\end{align}
This then allows us to apply (\ref{eq:quadapprox_bd_objfunc}) to get
\begin{align}
&\left|\Phi_{k}(\vr)-\Psi_{k}(\vr)\right| \nonumber \\
\leq& C\lsbd\left\Vert \vw_{k}^{*}(\vr)-\qwt_{k}(\vr)\right\Vert ^{3}\bigg[\sum_{t=1}^{k-1}\left(\scp\left\Vert \vb_{t}\right\Vert \right)^{3}\nonumber\\
&\hspace{13em}+\sum_{t=k+1}^{n}\left(\scp\left\Vert \va_{t}\right\Vert \right)^{3}\bigg]\nonumber \\
 & +C\lsbd\left|\loss_{k}'\right|^{3}\Big(\sum_{t=1}^{k-1}\left|\tfrac{1}{p}\vb_{t}\tran\mH_{\bs k}^{-1}\vr\right|^{3}+\sum_{t=k+1}^{n}\left|\tfrac{1}{p}\va_{t}\tran\mH_{\bs k}^{-1}\vr\right|^{3}\Big)\nonumber \\
 & +C\sum_{i=1}^{p}\Big(\left\Vert \vw_{k}^{*}(\vr)-\qwt_{k}(\vr)\right\Vert ^{3}+\left|\loss_{k}'\right|^{3}\left|\scp\vh_{\bs k,i}\tran\vr\right|^{3}\Big),\label{eq:quadapprox_objfunc_diff_bd-1}
\end{align}
where $C>0$, $\loss_{k}'\bydef\loss'\left(\frac{\vr\tran\qwt_{k}(\vr)}{\sqrt{p}};y_{k}\right)$,
$\vh_{\bs k,i}$ is the $i$th column of $\mH_{\bs k}^{-1}$ and we have
used (\ref{eq:wtilde_loowt_diff}). Using the simple inequality $(\sum_{i=1}^n \abs{a_i})^2 \le n \sum_{i=1}^n a_i^2$, we then have
\begin{align}
\label{eq:quadapprox_objfunc_diff_bd-2}
&\left|\Phi_{k}(\vr)-\Psi_{k}(\vr)\right|^2 \nonumber\\
\leq& C p \lsbd^2 \left\Vert \vw_{k}^{*}(\vr)-\qwt_{k}(\vr)\right\Vert^{6} \nonumber\\
&\hspace{7em}\times\sum_{t=1}^{n} \Big[ \big(\scp\left\Vert \vb_{t}\right\Vert \big)^{6} + \big(\scp\left\Vert \va_{t}\right\Vert \big)^{6}+1\Big]\nonumber\\
& ~~~+Cp\lsbd^2\left|\loss_{k}'\right|^{6}\sum_{t\neq k}\Big(\left|\tfrac{1}{p}\vb_{t}\tran\mH_{\bs k}^{-1}\vr\right|^{6}+\left|\tfrac{1}{p}\va_{t}\tran\mH_{\bs k}^{-1}\vr\right|^{6}\Big)\nonumber\\
& ~~~+Cp\left|\loss_{k}'\right|^{6}\sum_{i=1}^{p}\left|\scp\vh_{\bs k,i}\tran\vr\right|^{6}.
\end{align}
Therefore, it suffices to control the expectation of each term on the right-hand side of \eqref{eq:quadapprox_objfunc_diff_bd-2}, which can be done as follows.
Similar to what we did in reaching \eqref{eq:loss_derivative_LS_prob_bd_5}, we can get there exists $c>0$ such that for any $\veps>0$, $\P(\lsbd\geq\veps)\leq cp \exp({-\veps^{{2}/{K_1}}/c})$, which implies $\Ecnd{\fmtx}\lsbd^{8}\leq\polylog p$ by the identity $\E \abs{X} = \int_{0}^{\infty}\P(\abs{X}>t)dt$. Also
from (\ref{eq:wk_qwtk_diff_moments_bd}), we have $\Ecnd{\fmtx}\left\Vert \vw_{k}^{*}(\vr)-\qwt_{k}(\vr)\right\Vert ^{24}\leq \frac{\polylog p}{p^{12}}$. Hence $\Ecnd{\fmtx}\lsbd^4\left\Vert \vw_{k}^{*}(\vr)-\qwt_{k}(\vr)\right\Vert ^{12}\leq \frac{\polylog p}{p^6}$. From Lemma \ref{lem:at_bt_norm_bd}, we have $\Ecnd{\fmtx}\big(\scp\left\Vert \va_{t}\right\Vert \big)^{12}\leq C$ and $\Ecnd{\fmtx}\big(\scp\left\Vert \vb_{t}\right\Vert \big)^{12}\leq C$. It follows from H\"{o}lder's inequality that
\[
\begin{aligned}
    &\Ecnd{\fmtx} \Big[\lsbd^2\left\Vert \vw_{k}^{*}(\vr)-\qwt_{k}(\vr)\right\Vert^{6} \\ 
    &\hspace{6em}\times\sum_{t=1}^{n} \Big( \big(\scp\left\Vert \vb_{t}\right\Vert \big)^{6} + \big(\scp\left\Vert \va_{t}\right\Vert \big)^{6}+1\Big)\Big] \\
    \le& \tfrac{\polylog p}{p^2}.
\end{aligned}
\]
The other terms in \eref{quadapprox_objfunc_diff_bd-2} can be bounded similarly. From \eqref{eq:lk_moments_bd_goodF}, we have $\Ecnd{\fmtx}\left|\loss_{k}'\right|^{24}\leq \polylog p$. Also we have obtained $\Ecnd{\fmtx}\lsbd^8\leq\polylog p$. Therefore, $\Ecnd{\fmtx}\lsbd^4\left|\loss_{k}'\right|^{12}\leq\polylog p$. Combining Lemma \ref{lem:quadratic_asymmetry_concentrate} and \eqref{eq:X_bd_moments_gauss}, we can get $\Ecnd{\fmtx}\big|\tfrac{1}{p}\vr_{t}\tran\mH_{\bs k}^{-1}\vr\big|^{12}\leq \frac{C}{p^6}$ for $t \neq k$, with $\vr_{t} = \va_{t}$ or $\vb_{t}$. Finally, for $\fmtx \in \mathcal{A}$, we have $\norm{\mH_{\bs k}^{-1}} \le 2/\lambda$ and thus $\norm{\vh_{\bs k,i}} \le 2/\lambda$, ($\vh_{\bs k,i}$ is the $i$th column of $\mH_{\bs k}^{-1}$). We can then apply Lemma \ref{lem:linear_concentrate} to get $\Ecnd{\fmtx}\big|\scp\vh_{\bs k,i}\tran\vr\big|^{12}\leq \frac{C}{p^6}$. Substituting the above bounds into \eqref{eq:quadapprox_objfunc_diff_bd-2}, we reach the claim \eref{quad_approx_phi_psi} of the lemma.

\subsubsection{Two Auxiliary Lemmas for Proving Theorem~\ref{thm:get}}
\label{appendix:Thm2_satisfy}

\begin{lem}\label{lem:pathdiff_D1D2_bd}
Let $\Delta_1$ and $\Delta_2$ be the quantities defined in \eref{Phi_decomp_D1} and \eref{Phi_decomp_D2}, respectively. It holds that $\Delta_{1}\leq \frac{\polylog p}{\sqrt{p}}$ and $\Delta_{2}\leq \frac{\polylog p}{\sqrt{p}}$, uniformly over $\fmtx \in \mathcal{A}$ and $k \in [n]$.
\end{lem}
\begin{proof}
From (\ref{eq:Moreau_envelope}) we can get
\begin{equation}
\tfrac{\partial\calM_{k}(z;\gamma)}{\partial\gamma}=-\tfrac{1}{2}\loss'\big(\prox_{k}\left(z;\gamma\right);y_{k}\big)^{2}.\label{eq:Moreau_derivative}
\end{equation}
To bound the right-hand side of \eqref{eq:Moreau_derivative}, first note from (\ref{eq:prox_bd}) that
\begin{equation}
\left|\prox_{k}\left(z;\gamma\right)\right|  \leq\sqrt{2\gamma\loss\left(0;y_{k}\right)}+\left|z\right| \leq\gamma+\loss\left(0;y_{k}\right)+\left|z\right|.\label{eq:prox_bd_2}
\end{equation}
Thus, under Assumptions~\ref{a:loss}, there exists $C_{1},C_{2}>0$ such that
\begin{align}
&\loss'\big(\prox_{k}\left(z;\gamma\right);y_{k}\big) \nonumber\\
\leq& C_{1}\big(\left|\prox_{k}\left(z;\gamma\right)\right|^{2}+1\big)\big(\left|s_{k}\right|^{K_{1}}+1\big)\nonumber \\
\leq& C_{2}\big(\gamma^{2}+\loss\left(0;y_{k}\right)^{2}+\left|z\right|^{2}+1\big)\big(\left|s_{k}\right|^{K_{1}}+1\big),\label{eq:l_prox_bd_2}
\end{align}
where $y_k = \fteacher(s_k)$ and $s_{k}=\cvec_k\tran \sgl\sim{\cal N}(0,1)$ and the first inequality follows from \eqref{eq:loss_derivative_bd_2}. From (\ref{eq:Moreau_derivative}), there exists
$C>0$ such that for any $\gamma'$ between $\gamma_{k}(\vb_{k})$ and
$\gamma_{k}$,
\[
\left|\tfrac{\partial\calM_{k}(z;\gamma')}{\partial\gamma}\right|\leq C\big(\gamma_{k}^{4}+\gamma_{k}(\vb_{k})^{4}+\loss\left(0;y_{k}\right)^{4}+\left|z\right|^{4}+1\big)\big(\left|s_{k}\right|^{2K_{1}}+1\big).
\]
Then using \eqref{eq:gamma_moments_bd_goodF},
(\ref{eq:bkw_bd_moments}) and Assumption \ref{a:loss}, we can get
\begin{equation}
\E_{k}\Big(\tfrac{\partial\calM_{k}(\frac{1}{\sqrt{p}}\vb_{k}\tran\loowt{\bs k};\gamma)}{\partial\gamma}\Big)^{2}\leq Q\left(\scp\|{\loowt{\bs k}}\|\right),\label{eq:Moreau_derivative_bd_b}
\end{equation}
where $Q(x)$ is a finite degree polynomial. Therefore, for some $\gamma'$ between $\gamma_{k}(\vb_{k})$ and
$\gamma_{k}$,
\begin{align*}
\Delta_{1}\leq & \Ecnd{\fmtx}\E_{k}\left\{ \big|\tfrac{\partial\calM_{k}(\frac{1}{\sqrt{p}}\vb_{k}\tran\loowt{\bs k};\gamma')}{\partial\gamma}\big|\left|\gamma_{k}(\vb_{k})-\gamma_{k}\right|\right\} \\
\tleq{{(a)}} & \Ecnd{\fmtx}\left\{ \sqrt{Q\big(\scp\|{\loowt{\bs k}}\|\big)}\sqrt{\E_{k}\left[\gamma_{k}(\vb_{k})-\gamma_{k}\right]^{2}}\right\} \\
\tleq{{(b)}} & \tfrac{C_{1}}{\sqrt{p}}\sqrt{\Ecnd{\fmtx} Q\big(\scp\|{\loowt{\bs k}}\|\big)}\\
\tleq{{(c)}} & \tfrac{\polylog p}{\sqrt{p}},
\end{align*}
where $C_{1}>0$ is some constant. Here, (a) follows from \eqref{eq:Moreau_derivative_bd_b}; in (b), we use (\ref{eq:gamma_b_bd_moments}); in (c), we use (\ref{eq:norm_optwt_Expec_bd_goodF}).

The term $\Delta_2$ can be bounded similarly. Following the same steps as above, we can show there exists some polynomial $Q(x)$ such that
\begin{equation}
\label{eq:Moreau_derivative_bd_a}
\E_{k}\Big(\tfrac{\partial\calM_{k}(\frac{1}{\sqrt{p}}\va_{k}\tran\loowt{\bs k};\gamma)}{\partial\gamma}\Big)^{2}\leq Q\left(\scp\|{\loowt{\bs k}}\|\right)
\end{equation}
for any $\gamma'$ between $\gamma_{k}(\va_{k})$ and
$\gamma_{k}$. It follows that
\begin{align}
\Delta_{2} & \leq\Ecnd{\fmtx}\E_{k}\Big[\big|\tfrac{\partial\calM_{k}(\frac{1}{\sqrt{p}}\va_{k}\tran\loowt{\bs k};\gamma')}{\partial\gamma}\big|\left|\gamma_{k}(\va_{k})-\gamma_{k}\right|\Big]\nonumber\\
 & \leq\Ecnd{\fmtx}\E_{k}\Big[\big|\tfrac{\partial\calM_{k}(\frac{1}{\sqrt{p}}\va_{k}\tran\loowt{\bs k};\gamma')}{\partial\gamma}\big|\Big(\left|\gamma_{k}(\va_{k})-\E_{k}\gamma_{k}(\va_{k})\right|\nonumber\\
 &\hspace{13em}+\left|\E_{k}\gamma_{k}(\va_{k})-\gamma_{k}\right|\Big)\Big]\nonumber\\
 & \leq \sqrt{\Ecnd{\fmtx}\big[Q(\scp\|{\loowt{\bs k}}\|)\E_{k}\left|\gamma_{k}(\va_{k})-\E_{k}\gamma_{k}(\va_{k})\right|^2\big]} \nonumber\\
 &\hspace{1em}+\sqrt{\Ecnd{\fmtx}\big[Q(\scp\|{\loowt{\bs k}}\|)\left|\E_{k}\gamma_{k}(\va_{k})-\gamma_{k}\right|^2\big]},\label{eq:Delta2_bd}
\end{align}
where in the last step we use \eqref{eq:Moreau_derivative_bd_a} and H\"{o}lder's inequality. We need to bound the term $\left|\E_{k}\gamma_{k}(\va_{k})-\gamma_{k}\right|$ in \eqref{eq:Delta2_bd}. Recall that $\gamma_{k}=\E_{k}\gamma_{k}(\vb_{k})$. Thus,
\begin{align}
|\E_{k}\gamma_{k}(\va_{k})-\gamma_{k}| &= \tfrac{1}{p}\big|\E_{k}({\va_{k}\tran\mH_{\bs k}^{-1}\va_{k}-\vb_{k}\tran\mH_{\bs k}^{-1}\vb_{k}})\big|\nonumber\\
&= \tfrac{1}{p}\big|\Tr[\mH_{\bs k}^{-1}(\mSig_a - \mSig_b)]\big|\nonumber\\
&\leq \tfrac{\polylog p}{\sqrt{p}},\label{eq:gamma_condmean_diff_bd}
\end{align}
where in the last step,
we use Lemma \ref{lem:sigma_ab} and the fact that $\|\mH_{\bs k}^{-1}\|\leq \tfrac{2}{\lambda}$ for $\fmtx\in\cal{A}$. Plugging \eqref{eq:gamma_condmean_diff_bd}, \eqref{eq:gamma_a_bd_moments} and \eqref{eq:norm_loowt_Expec_bd_goodF} into \eqref{eq:Delta2_bd}, we conclude that $\Delta_2\leq \tfrac{\polylog p}{\sqrt{p}}$.
\end{proof}

\begin{lem}
\label{lem:Thm2_satisfy}
There exists a function $B(s)$ such that $\E B^4(Z)<\infty$ for $Z\sim\mathcal{N}(0,1)$ and for each $k\in[n]$,
\begin{equation}
\label{eq:moreau_satisfy}
\max\{\calM_{k}\big(x;\gamma_{k}\big), \calM_{k}'\big(x;\gamma_{k}\big)\} \leq B(\cvec_k\tran\sgl)(1+|x|^{3}).
\end{equation}
\end{lem}
\begin{proof}
From \eqref{eq:Moreau_envelope}, we can verify that
\begin{align}
\label{eq:Moreau_bd_5}
\calM_{k}\big(x;\gamma_{k}\big) \leq \loss(x;y_{k})
\end{align}
and
\begin{align}
\label{eq:Moreauderi_bd_5}
\calM_{k}'\big(x;\gamma_{k}\big) = \loss'\big(\prox_{k}(x;\gamma_{k}); y_{k}\big),
\end{align}
where $\prox_{k}(x;\gamma_{k})$ is the proximal operator of $\loss(x;y_k)$. Moreover, from \eqref{eq:prox_bd},
\begin{align}
\label{eq:prox_bd_5}
|\prox_{k}(x;\gamma_{k})| \leq \gamma_{k} + \loss(0;y_k) + |x|.
\end{align}
Combining \eqref{eq:Moreau_bd_5}, \eqref{eq:prox_bd_5} with Assumption \ref{a:loss} allows us to show that $\calM_{k}\big(x;\gamma_{k}\big)$ satisfies \eqref{eq:moreau_satisfy}. Indeed, similar as \eqref{eq:loss_derivative_bd_2}, we can get
\begin{equation}
\label{eq:loss_derivative_bd_8}
\loss(x;y_{k}) \leq C_1(|x|^{3}+1)(\left| \vg_k\tran \sgl\right|^{K_{1}}+1),
\end{equation}
for some $C_1>0$.
Then from \eqref{eq:Moreau_bd_5} and \eqref{eq:loss_derivative_bd_8}, there exists $C>0$ such that
\begin{equation}
\calM_{k}\big(x;\gamma_{k}\big) \le (|x|^{3}+1)
\underbrace{C(|\cvec\tran_{k}\sgl|^{K_1}+1)}_{B_1(\cvec_k\tran\sgl)}.\label{eq:Moreau_prob_bd_5}
\end{equation}
Similarly, there exist $C_1', C_2', C_3'>0$ such that
\begin{align}
|\calM_{k}'\big(x;\gamma_{k}\big)| &\tleq{{(a)}} C_1'(|\prox_{k}(x;\gamma_{k})|^{2}+1)(|\cvec\tran_{k}\sgl|^{K_1}+1)\nonumber\\
&\tleq{{(b)}} C_2'(\gamma_{k}^{2} + \loss(0;y_k)^{2} + |x|^{2}+1)(|\cvec\tran_{k}\sgl|^{K_1}+1)\nonumber\\
&\tleq{{(c)}}(|x|^{3}+1) \underbrace{C_3'(|\cvec\tran_{k}\sgl|^{3K_1}+1)}_{B_2(\cvec_k\tran\sgl)}.\label{eq:Moreauderi_prob_bd_5}
\end{align}
In (a), we use \eref{Moreauderi_bd_5} and \eqref{eq:loss_derivative_bd_2}; in (b), we use \eqref{eq:prox_bd_5}; in (c), we use \eqref{eq:gamma_moments_bd_goodF} and Assumption \ref{a:loss}. It is clear that $B_1(s)$ and $B_2(s)$ in \eqref{eq:Moreau_prob_bd_5} and \eqref{eq:Moreauderi_prob_bd_5} satisfy $\E B_1^4(Z), \E B_2^4(Z)<\infty$, for $Z\sim\mathcal{N}(0, 1)$. Choosing $B(s) = \max\{B_1(s),B_2(s)\}$ then gives us the desired result.
\end{proof}

%% file: get_kernel.bbl
\begin{thebibliography}{10}
\providecommand{\url}[1]{#1}
\csname url@samestyle\endcsname
\providecommand{\newblock}{\relax}
\providecommand{\bibinfo}[2]{#2}
\providecommand{\BIBentrySTDinterwordspacing}{\spaceskip=0pt\relax}
\providecommand{\BIBentryALTinterwordstretchfactor}{4}
\providecommand{\BIBentryALTinterwordspacing}{\spaceskip=\fontdimen2\font plus
\BIBentryALTinterwordstretchfactor\fontdimen3\font minus
  \fontdimen4\font\relax}
\providecommand{\BIBforeignlanguage}[2]{{%
\expandafter\ifx\csname l@#1\endcsname\relax
\typeout{** WARNING: IEEEtran.bst: No hyphenation pattern has been}%
\typeout{** loaded for the language `#1'. Using the pattern for}%
\typeout{** the default language instead.}%
\else
\language=\csname l@#1\endcsname
\fi
#2}}
\providecommand{\BIBdecl}{\relax}
\BIBdecl

\bibitem{rahimi2008random}
A.~Rahimi and B.~Recht, ``Random features for large-scale kernel machines,'' in
  \emph{Advances in neural information processing systems}, 2008, pp.
  1177--1184.

\bibitem{daniely2016toward}
A.~Daniely, R.~Frostig, and Y.~Singer, ``Toward deeper understanding of neural
  networks: The power of initialization and a dual view on expressivity,'' in
  \emph{Advances In Neural Information Processing Systems}, 2016, pp.
  2253--2261.

\bibitem{daniely2017sgd}
A.~Daniely, ``{SGD} learns the conjugate kernel class of the network,'' in
  \emph{Advances in Neural Information Processing Systems}, 2017, pp.
  2422--2430.

\bibitem{bach2017equivalence}
F.~Bach, ``On the equivalence between kernel quadrature rules and random
  feature expansions,'' \emph{The Journal of Machine Learning Research},
  vol.~18, no.~1, pp. 714--751, 2017.

\bibitem{jacot2018neural}
A.~Jacot, F.~Gabriel, and C.~Hongler, ``Neural tangent kernel: Convergence and
  generalization in neural networks,'' in \emph{Advances in neural information
  processing systems}, 2018, pp. 8571--8580.

\bibitem{belkin2018understand}
M.~Belkin, S.~Ma, and S.~Mandal, ``To understand deep learning we need to
  understand kernel learning,'' \emph{arXiv preprint arXiv:1802.01396}, 2018.

\bibitem{liu2020random}
F.~Liu, X.~Huang, Y.~Chen, and J.~A. Suykens, ``Random features for kernel
  approximation: A survey in algorithms, theory, and beyond,'' \emph{arXiv
  preprint arXiv:2004.11154}, 2020.

\bibitem{louart2018random}
C.~Louart, Z.~Liao, and R.~Couillet, ``A random matrix approach to neural
  networks,'' \emph{The Annals of Applied Probability}, vol.~28, no.~2, pp.
  1190--1248, 2018.

\bibitem{hastie2019surprises}
T.~Hastie, A.~Montanari, S.~Rosset, and R.~J. Tibshirani, ``Surprises in
  high-dimensional ridgeless least squares interpolation,'' \emph{arXiv
  preprint arXiv:1903.08560}, 2019.

\bibitem{mei2019generalization}
S.~Mei and A.~Montanari, ``The generalization error of random features
  regression: Precise asymptotics and double descent curve,'' \emph{arXiv
  preprint arXiv:1908.05355}, 2019.

\bibitem{montanari2019generalization}
A.~Montanari, F.~Ruan, Y.~Sohn, and J.~Yan, ``The generalization error of
  max-margin linear classifiers: High-dimensional asymptotics in the
  overparametrized regime,'' \emph{arXiv preprint arXiv:1911.01544}, 2019.

\bibitem{goldt2019modelling}
S.~Goldt, M.~M{\'e}zard, F.~Krzakala, and L.~Zdeborov{\'a}, ``Modelling the
  influence of data structure on learning in neural networks,'' \emph{arXiv
  preprint arXiv:1909.11500}, 2019.

\bibitem{gerace2020generalisation}
F.~Gerace, B.~Loureiro, F.~Krzakala, M.~M{\'e}zard, and L.~Zdeborov{\'a},
  ``Generalisation error in learning with random features and the hidden
  manifold model,'' \emph{arXiv preprint arXiv:2002.09339}, 2020.

\bibitem{Goldt2020Gaussian}
S.~Goldt, G.~Reeves, M.~M\'{e}zard, F.~Krzakala, and L.~Zdeborov\'{a}, ``The
  {Gaussian} equivalence of generative models for learning with two-layer
  neural networks,'' \emph{arXiv preprint arXiv:2006.14709}, 2020.

\bibitem{ghorbani2019linearized}
B.~Ghorbani, S.~Mei, T.~Misiakiewicz, and A.~Montanari, ``Linearized two-layers
  neural networks in high dimension,'' \emph{arXiv preprint arXiv:1904.12191},
  2019.

\bibitem{ba2019generalization}
J.~Ba, M.~Erdogdu, T.~Suzuki, D.~Wu, and T.~Zhang, ``Generalization of
  two-layer neural networks: An asymptotic viewpoint,'' in \emph{International
  Conference on Learning Representations}, 2019.

\bibitem{dhifallah2020precise}
O.~Dhifallah and Y.~M. Lu, ``A precise performance analysis of learning with
  random features,'' \emph{arXiv preprint arXiv:2008.11904}, 2020.

\bibitem{gordon1985some}
Y.~Gordon, ``Some inequalities for gaussian processes and applications,''
  \emph{Israel Journal of Mathematics}, vol.~50, no.~4, pp. 265--289, 1985.

\bibitem{thrampoulidis2018precise}
C.~Thrampoulidis, E.~Abbasi, and B.~Hassibi, ``Precise error analysis of
  regularized {$M$}-estimators in high-dimensions,'' \emph{IEEE Trans. Inf.
  Theory}, vol.~64, no.~8, pp. 5592--5628, 2018.

\bibitem{chandrasekaran2012convex}
V.~Chandrasekaran, B.~Recht, P.~A. Parrilo, and A.~S. Willsky, ``The convex
  geometry of linear inverse problems,'' \emph{Foundations of Computational
  Mathematics}, vol.~12, no.~6, pp. 805--849, 2012.

\bibitem{amelunxen2014living}
D.~Amelunxen, M.~Lotz, M.~B. McCoy, and J.~A. Tropp, ``Living on the edge:
  Phase transitions in convex programs with random data,'' \emph{Information
  and Inference: A Journal of the IMA}, vol.~3, no.~3, pp. 224--294, 2014.

\bibitem{seddik2020random}
M.~E.~A. Seddik, C.~Louart, M.~Tamaazousti, and R.~Couillet, ``Random matrix
  theory proves that deep learning representations of gan-data behave as
  gaussian mixtures,'' \emph{arXiv preprint arXiv:2001.08370}, 2020.

\bibitem{Dhifallah2021inherent}
O.~Dhifallah and Y.~M. Lu, ``On the inherent regularization effects of noise
  injection during training,'' \emph{arXiv:2102.07379}, 2021.

\bibitem{Loureiro2021capturing}
B.~Loureiro, C.~Gerbelot, H.~Cui, S.~Goldt, F.~Krzakala, and L.~Zdeborov{\'a},
  ``Capturing the learning curves of generic features maps for realistic data
  sets with a teacher-student model,'' \emph{arXiv:2102.08127}, 2021.

\bibitem{cheng2013spectrum}
X.~Cheng and A.~Singer, ``The spectrum of random inner-product kernel
  matrices,'' \emph{Random Matrices: Theory and Applications}, vol.~2, no.~04,
  p. 1350010, 2013.

\bibitem{pennington2017nonlinear}
J.~Pennington and P.~Worah, ``Nonlinear random matrix theory for deep
  learning,'' in \emph{Advances in Neural Information Processing Systems},
  2017, pp. 2637--2646.

\bibitem{mezard1987spin}
M.~M{\'e}zard, G.~Parisi, and M.~Virasoro, \emph{Spin glass theory and beyond:
  An Introduction to the Replica Method and Its Applications}.\hskip 1em plus
  0.5em minus 0.4em\relax World Scientific Publishing Company, 1987, vol.~9.

\bibitem{lindeberg1922neue}
J.~W. Lindeberg, ``Eine neue herleitung des exponentialgesetzes in der
  wahrscheinlichkeitsrechnung,'' \emph{Mathematische Zeitschrift}, vol.~15,
  no.~1, pp. 211--225, 1922.

\bibitem{el2018impact}
N.~El~Karoui, ``On the impact of predictor geometry on the performance on
  high-dimensional ridge-regularized generalized robust regression
  estimators,'' \emph{Probability Theory and Related Fields}, vol. 170, no.
  1-2, pp. 95--175, 2018.

\bibitem{oymak2018universality}
S.~Oymak and J.~A. Tropp, ``Universality laws for randomized dimension
  reduction, with applications,'' \emph{Information and Inference: A Journal of
  the IMA}, vol.~7, no.~3, pp. 337--446, 2018.

\bibitem{korada2011applications}
S.~B. Korada and A.~Montanari, ``Applications of the lindeberg principle in
  communications and statistical learning,'' \emph{IEEE transactions on
  information theory}, vol.~57, no.~4, pp. 2440--2450, 2011.

\bibitem{montanari2017universality}
A.~Montanari and P.-M. Nguyen, ``Universality of the elastic net error,'' in
  \emph{2017 IEEE International Symposium on Information Theory (ISIT)}.\hskip
  1em plus 0.5em minus 0.4em\relax IEEE, 2017, pp. 2338--2342.

\bibitem{panahi2017universal}
A.~Panahi and B.~Hassibi, ``A universal analysis of large-scale regularized
  least squares solutions,'' in \emph{Advances in Neural Information Processing
  Systems}, 2017, pp. 3381--3390.

\bibitem{abbasi2019universality}
E.~Abbasi, F.~Salehi, and B.~Hassibi, ``Universality in learning from linear
  measurements,'' in \emph{Advances in Neural Information Processing Systems},
  2019, pp. 12\,372--12\,382.

\bibitem{gerace2022gaussian}
F.~Gerace, F.~Krzakala, B.~Loureiro, L.~Stephan, and L.~Zdeborova, ``Gaussian
  universality of linear classifiers with random labels in high-dimension,''
  \emph{arXiv preprint arXiv:2205.13303}, 2022.

\bibitem{liang2020precise}
T.~Liang and P.~Sur, ``A precise high-dimensional asymptotic theory for
  boosting and minimum-$\ell_1$-norm interpolated classifiers,'' \emph{arXiv
  preprint arXiv:2002.01586}, 2020.

\bibitem{ba2022high}
J.~Ba, M.~A. Erdogdu, T.~Suzuki, Z.~Wang, D.~Wu, and G.~Yang,
  ``High-dimensional asymptotics of feature learning: How one gradient step
  improves the representation,'' \emph{arXiv preprint arXiv:2205.01445}, 2022.

\bibitem{montanari2022universality}
A.~Montanari and B.~Saeed, ``Universality of empirical risk minimization,''
  \emph{arXiv preprint arXiv:2202.08832}, 2022.

\bibitem{lu2022equivalence}
Y.~M. Lu and H.-T. Yau, ``An equivalence principle for the spectrum of random
  inner-product kernel matrices,'' \emph{arXiv preprint arXiv:2205.06308},
  2022.

\bibitem{misiakiewicz2022spectrum}
T.~Misiakiewicz, ``Spectrum of inner-product kernel matrices in the polynomial
  regime and multiple descent phenomenon in kernel ridge regression,''
  \emph{arXiv preprint arXiv:2204.10425}, 2022.

\bibitem{xiao2022precise}
L.~Xiao and J.~Pennington, ``Precise learning curves and higher-order scaling
  limits for dot product kernel regression,'' \emph{arXiv preprint
  arXiv:2205.14846}, 2022.

\bibitem{hu2022sharp}
H.~Hu and Y.~M. Lu, ``Sharp asymptotics of kernel ridge regression beyond the
  linear regime,'' \emph{arXiv preprint arXiv:2205.06798}, 2022.

\bibitem{stein1972bound}
C.~Stein, ``A bound for the error in the normal approximation to the
  distribution of a sum of dependent random variables,'' in \emph{Proceedings
  of the Sixth Berkeley Symposium on Mathematical Statistics and Probability,
  Volume 2: Probability Theory}.\hskip 1em plus 0.5em minus 0.4em\relax The
  Regents of the University of California, 1972.

\bibitem{barbour2005introduction}
A.~D. Barbour and L.~H.~Y. Chen, \emph{An introduction to Stein's
  method}.\hskip 1em plus 0.5em minus 0.4em\relax World Scientific, 2005,
  vol.~4.

\bibitem{Chen2011normal}
L.~H.~Y. Chen, L.~Goldstein, and Q.~Shao, \emph{Normal approximation by
  {Stein's} method}.\hskip 1em plus 0.5em minus 0.4em\relax New York: Springer,
  2011.

\bibitem{boucheron2013concentration}
S.~Boucheron, G.~Lugosi, and P.~Massart, \emph{Concentration inequalities: {A}
  nonasymptotic theory of independence}.\hskip 1em plus 0.5em minus 0.4em\relax
  Oxford university press, 2013.

\bibitem{vershynin2018high}
R.~Vershynin, \emph{High-dimensional Probability: {An} Introduction with
  Applications in Data Science}.\hskip 1em plus 0.5em minus 0.4em\relax
  Cambridge University Press, 2018.

\bibitem{Talagrand2010meanfieldv1}
M.~Talagrand, \emph{Mean Field Models for Spin Glasses}.\hskip 1em plus 0.5em
  minus 0.4em\relax Springer, 2010, vol.~1.

\bibitem{HL2001fundamentals}
J.-B. Hiriart-Urrut and C.~Lemar\'{e}chal, \emph{Fundamentals of convex
  analysis}.\hskip 1em plus 0.5em minus 0.4em\relax Berlin: Springer-Verlag,
  2001.

\end{thebibliography}
